UNIVERSIDAD DE CONCEPCION
FACULTAD DE INGENIERIA
DEPARTAMENTO DE INGENIERIA ELECTRICA

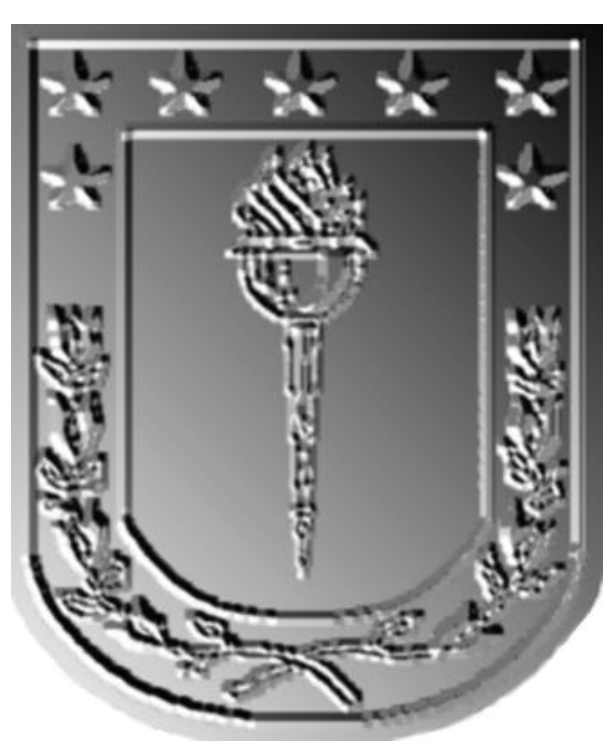

# APLICACIÓN GRÁFICA PARA EL ESTUDIO DE UN MODELO DE CELDA ELECTROLÍTICA USANDO TÉCNICAS DE VISUALIZACIÓN DE CAMPOS VECTORIALES

César A. Mena L.

**UNIVERSIDAD DE CONCEPCION**
**FACULTAD DE INGENIERIA**
**Departamento de Ingeniería Eléctrica**

**Profesor Patrocinante**
Sr. Ricardo Sánchez Schulz, Ph. D.

# *Aplicación Gráfica para el estudio de un Modelo de Celda Electrolí tica usando Técnicas de Visualización de Campos Vectoriales*

**César Antonio Mena Labraña**

Informe de Memoria de Título
Para optar al Título de

**Ingeniero Civil Electrónico**

Septiembre de 2000

*Deseo expresar mi más sincero agradecimiento a mi profesor patrocinante Sr. Ricardo Sánchez, especialmente por su motivación y paciencia.*

# *Sumario*

El uso de electrodos bipolares flotantes en celdas de electro-obtención de cobre constituye una tecnología no convencional que promete impactos económicos y operacionales para la industria. En el presente trabajo se desarrolla una herramienta computacional cuyo objetivo es contribuir al análisis y diseño de tales celdas, de modo de reducir el tiempo y costo de desarrollo, pues en el caso de la tecnología bipolar, el diseño ha estado basado en el uso de prototipos. El software desarrollado proporciona una interfaz gráfica que permite, en un entorno 3D interactivo; el diseño geométrico, la simulación y el análisis de los resultados. Se puede estudiar geometrías rectangulares arbitrarías de celdas con electrodos bipolares flotantes y celdas convencionales con electrodos unipolares. La simulación de la celda se sustenta en un modelo base que fue desarrollado en una investigación previa, la que motivó el presente trabajo.

Para que la tecnología de electrodos bipolares se pueda aplicar comercialmente se debe asegurar un producto de calidad, lo que implica garantizar un depósito de cobre homogéneo en toda la superficie de los cátodos. El modelo base se orienta a la predicción del depósito de cobre, lo que se logra mediante el cálculo de la distribución del campo de densidad de corriente sobre los electrodos, pues esta se relaciona directamente con la del depósito de cobre. Se usa un método de diferencia finita para obtener la distribución tridimensional de potencial eléctrico, lo que permite obtener el campo de densidad de corriente en todo el volumen de la celda.

Para el análisis interactivo de los datos se proporcionan varias herramientas de visualización. Como herramienta principal de análisis, se provee la capacidad para visualizar campos vectoriales 3D como líneas continuas de flujo, las que son generadas en forma automática. Esta técnica resulta muy apropiada para esta aplicación, permite visualizaciones significativas, locales o globales, del campo densidad de corriente. Para una adecuada percepción espacial de las líneas de flujo se crea efectos de sombreado, mediante un apropiado modelo de iluminación de líneas. Aunque el hardware gráfico convencional sólo proporciona soporte para la iluminación de primitivas de superficie, la iluminación de líneas se implementa convenientemente mediante mapeo de textura, característica optimizada en aceleradores gráficos. De este modo, todos los cálculos de iluminación se pueden efectuar eficientemente con el soporte de hardware gráfico. Además, las líneas de flujo incluyen sugestión de profundidad y se colorean para codificar la magnitud de campos escalares (potencial eléctrico, intensidad del campo densidad de corriente).

Desde la perspectiva de la Visualización científica, una aplicación de este tipo debe ser interactiva y precisa. Con objeto de soportar geometrías rectangulares arbitrarias, el algoritmo del modelo base debía ser reestructurado. En tal proceso, se concibió algunas mejoras al modelo y una importante corrección, lo que constituye un aporte adicional significativo al esfuerzo por modelar la celda de electro-obtención con mayor precisión. En particular, se desarrollo un mejor modelo para electrodos flotantes, el que calcula el potencial eléctrico del electrodo a partir de una deducción más consistente.

## CAPITULO 3

## Visualización Interactiva de Campos Vectoriales mediante Líneas de Flujo 26

## CAPÍTULO 4

# *Glosario y Abreviaciones*

---

**Visualización :** Un método para extraer información significativa a partir de conjuntos de datos complejos, a través del uso de imágenes y gráficos interactivos.

**Rendering :** Término general que describe el proceso completo para transformar una representación en base de datos de un objeto tridimensional, en una proyección bidimensional sombreada sobre una superficie de visualización (pantalla).

---

**EW** : *Electrowinning* o Electro-obtención.

**OOP** : *Object Oriented Programming* o Programación Orientada a Objetos.

**API** : *Application Programming Interface* o Interfaz de Programación de Aplicaciones.

**GUI** : *Graphical User Interface* o Interfaz Gráfica de Usuario.

---

# *Nomenclatura*

**Notación :**

- Caracteres en *cursiva* denotan variables o constantes escalares.
- Caracteres en **negrita** denotan vectores o matrices. Preferentemente se usa caracteres en minúscula para vectores y en mayúscula para matrices.

**Símbolos :**

La siguiente lista contiene los símbolos más importantes usados en este documento.

| | | |
|---|---|---|
| $V$ | [V] : | potencial eléctrico en el interior de la celda electrolítica (potencial en el electrolito) |
| $V_m$ | [V] : | *potencial metálico*. Potencial en el interior de un electrodo metálico |
| $DV$ | [V] : | *potencial de electrodo*. Diferencia de potencial aplicada a una interfase metal-electrolito ($V_m$- $V$) |
| $DV_{\mathrm{A}}$ | [V] : | potencial de electrodo en la región de un electrodo que actúa como Ánodo |
| $DV_{\mathrm{C}}$ | [V] : | potencial de electrodo en la región de un electrodo que actúa como Cátodo |

$e_0$ [V] : potencial de equilibrio de una reacción electroquímica en condiciones estándar (temperatura y concentración)

$e_R$ [V] : potencial de equilibrio de una reacción electroquímica en condiciones no estándar (temperatura y concentración)

$\mathbf{E}$ : vector campo eléctrico ($|\mathbf{E}|$ [V/m])

$\mathbf{J}$ : vector densidad de corriente eléctrica ($|\mathbf{J}|$ [A/m$^2$])

$\hat{\mathbf{J}}$ : vector **J** normalizado

$\sigma$ [$\Omega^{-1}$m$^{-1}$] : conductividad eléctrica

$\rho$ [C/m$^3$] : densidad volumétrica de carga eléctrica

$\varepsilon$ [F/m] : permitividad eléctrica

$\mathbf{x}$ : vector de coordenadas espaciales ($|\mathbf{x}|$ [m])

$\hat{\mathbf{x}}$, $\hat{\mathbf{y}}$, $\hat{\mathbf{z}}$ : vectores unitarios en coordenadas rectangulares. $\hat{\mathbf{x}}=(1,0,0)^T$, $\hat{\mathbf{y}}=(0,1,0)^T$, y $\hat{\mathbf{z}}=(0,0,1)^T$

$\hat{\mathbf{n}}$ : vector unitario normal a la superficie de una interfase del electrolito con otro medio, orientado hacia el electrolito

$dS$ [m$^2$] área de un elemento diferencial de superficie

$\Delta S$ [m$^2$] : aproximación finita de $dS$

$d\mathbf{s}$ : elemento diferencial orientable de superficie ($\hat{\mathbf{n}}\,dS$)

$N$ : número de elementos de superficie $\Delta S$ en los que se discretiza la superficie de un electrodo

$k_S$ : factor de ajuste de área para elementos de superficie no cuadrados ($\Delta S = k_S h^2$)

$h$ [m] : espaciamiento de la grilla de muestreo rectangular del dominio

$d$ [m] : longitud del paso adaptivo de integración de líneas de flujo

$d_{min}$, $d_{max}$ [m] : cotas inferior y superior para $d$

$|\Delta|$ [m] : error en la longitud de una línea de flujo correspondiente a un paso de integración

$d^*$ [m] : $d$ óptimo para lograr que $|\Delta|$ sea inferior a una cota $TOL$

$\mathbf{N}$ : vector unitario que localmente representa la dirección normal a una superficie

$\mathbf{L}$ : vector unitario que representa la dirección de una fuente de luz

$\mathbf{R}$ : vector unitario que representa la dirección de la luz reflejada

$\mathbf{V}$ : vector unitario que representa la dirección de vista (observador)

$\mathbf{T}$ : vector unitario que representa la dirección de una línea

$k_a$ : intensidad de la luz ambiente (modelo de iluminación Phong)

$k_d$ : coeficiente de reflexión difusa (modelo de iluminación Phong)

$k_s$ : coeficiente de reflexión especular (modelo de Phong)

$s$ : exponente de reflexión especular (modelo de Phong)

$p$ : exponente de reflexión difusa que compensa exceso de brillo

$\mathbf{M}$ : matriz de transformación de coordenadas de textura (4x4)

$\mathbf{t_0}$ : vector homogéneo de coordenadas de textura

$\rho_S$ [ptos./$h^2$] : densidad superficial de puntos

$I$ [$h$] : dimensión de electrodo rectangular en la dirección $x$

$J$ [$h$] : dimensión de electrodo rectangular en la dirección $y$

$K$ [$h$] : dimensión de electrodo rectangular en la dirección $z$

**Funciones y operadores :**

En las siguientes fórmulas, $\boxtimes$ denota "en coordenadas rectangulares $\mathbf{x} = (x, y, z)^T$". Además, se supone que $f = f(\mathbf{x})$ y $\mathbf{g} = \mathbf{g}(\mathbf{x}) = (g_x, g_y, g_z)$ tienen derivadas parciales.

$\nabla$ : operador Nabla. En $\boxtimes$, $\nabla = (\partial/\partial x)\,\hat{\mathbf{x}} + (\partial/\partial y)\,\hat{\mathbf{y}} + (\partial/\partial z)\,\hat{\mathbf{z}}$

$\nabla f$ : gradiente de $f$. En $\boxtimes$, $\nabla f = (\partial f/\partial x)\,\hat{\mathbf{x}} + (\partial f/\partial y)\,\hat{\mathbf{y}} + (\partial f/\partial z)\,\hat{\mathbf{z}}$

$\nabla\cdot\mathbf{g}$ : divergencia de $\mathbf{g}$. En $\boxtimes$, $\nabla\cdot\mathbf{g} = \partial g_x/\partial x + \partial g_y/\partial y + \partial g_z/\partial z$

$\nabla^2 f$ : laplaciano de $f$. En $\boxtimes$, $\nabla^2 f = \nabla\cdot(\nabla f) = \partial^2 f/\partial x^2 + \partial^2 f/\partial y^2 + \partial^2 f/\partial z^2$

min : el mínimo valor de un conjunto de datos

max : el máximo valor de un conjunto de datos

clamp($x$) : $x$, cuando $x > 0$. 0 en otro caso

int($x$) : retorna la parte entera de $x$

# *Introducción*

## 1.1 Motivación

Este trabajo se inspira en la investigación previa desarrollada principalmente por Ralph Bittner. En [3], [4] y [5], Bittner justifica el interés en la tecnología de electro-obtención (EW) de cobre usando electrodos bipolares y desarrolla un modelo eléctrico para su simulación, el que se utiliza como base en este trabajo.

Debido a los convenientes requerimientos energéticos, el uso de electrodos bipolares en EW de cobre promete importantes ventajas económicas y operacionales respecto a la tecnología convencional que usa electrodos unipolares. Sin embargo, para hacer factible un proceso de EW de cobre usando electrodos bipolares, los prototipos existentes deben ser optimizados, ya que producen inaceptables depósitos de cobre, de espesor no homogéneo. La distribución del depósito de cobre esta relacionada directamente con la distribución del campo de densidad de corriente. Para apoyar el uso de electrodos bipolares en la EW de cobre, se desea estudiar diferentes estructuras geométricas desde el punto de vista eléctrico. Este objetivo motivó el desarrollo de un software que permita la simulación y el análisis de celdas de EW, pues se reducen los costos y el tiempo que implica el uso de prototipos.

## 1.2 Objetivos y Alcances

El objetivo general de este trabajo es desarrollar una aplicación gráfica para el estudio interactivo de un modelo base de celda electrolítica, apoyándose en técnicas de visualización de campos vectoriales.

**Alcances:**

- Crear una interfaz gráfica para el diseño geométrico de la celda, que implemente un modelo para el cálculo del campo vectorial densidad de corriente.
- Desarrollar algoritmos para la visualización interactiva de campos vectoriales distribuidos en el volumen de la celda.
- Los algoritmos de visualización deben poseer las siguientes características:
  - Permitir la exploración interactiva del volumen y realizar visualizaciones globales o locales del campo vectorial.
  - Permitir el mapeo visual de información escalar relativa al campo vectorial.
  - Visualización conjunta del campo vectorial y geometría de la celda electrolítica.

Es importante destacar que, aunque no estaba en los objetivos originales de este trabajo, durante su desarrollo, se concibió correcciones y mejoras al modelo base de la celda electro-obtención, lo que permitió hacer un aporte adicional en el campo de la modelación.

## 1.3 Electro-obtención de cobre

El proceso electroquímico en el cual se basa la electro-obtención de cobre se muestra en la Figura 1.1. Un rectificador suministra una corriente eléctrica continua que fluye desde el ánodo al cátodo a través de una solución ácida que contiene sulfato de cobre. Para mantener la composición del electrolito aproximadamente constante se requiere un circuito hidráulico que permita renovar continuamente el electrolito. Debido a la corriente impuesta, el cobre se deposita en el cátodo y el agua se descompone en el ánodo produciendo oxigeno gaseoso. La descomposición del ánodo implica la captación de electrones, lo que genera una corriente eléctrica que fluye desde el ánodo a la solución. El depósito de cobre esta relacionado con el flujo de electrones desde el cátodo a la solución. Cada ión cobre $Cu^{2+}$ capta dos electrones del cátodo transformándose en cobre metálico $Cu^0$, el que se adhiere al cátodo. En el interior del electrolito, la corriente eléctrica es producida por el movimiento de los iones. La potencia eléctrica proporcionada por el rectificador se consume parcialmente por pérdidas resistivas, pero es usada principalmente para producir las transformaciones químicas asociadas a las reacciones de reducción del ión cúprico en el cátodo y la oxidación del agua en el ánodo.

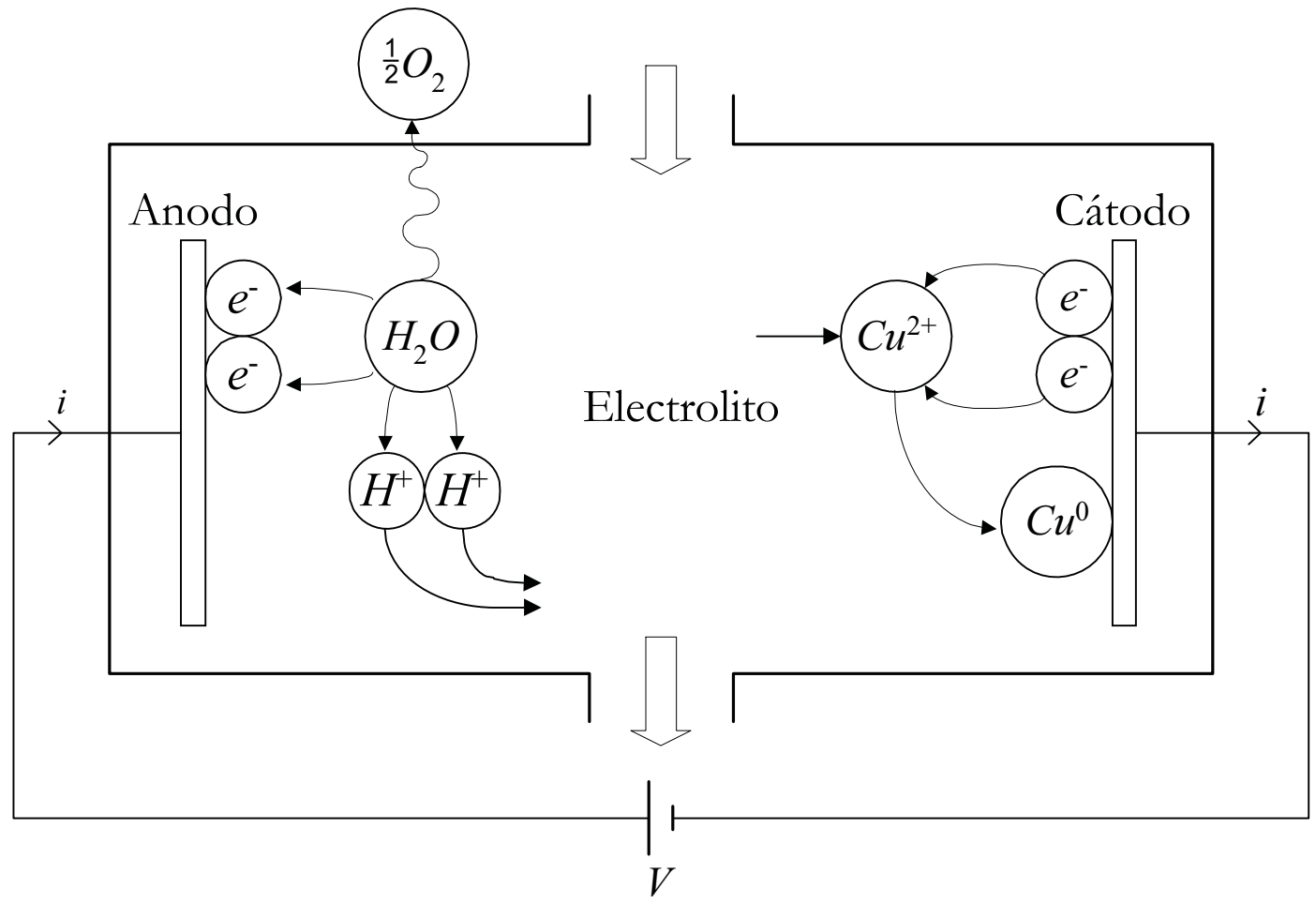

Figura 1.1: Proceso electroquímico básico de la electro-obtención de Cobre

Comercialmente, el proceso de electro-obtención de cobre se lleva a cabo en celdas rectangulares, en cuyo interior se colocan electrodos laminares, orientados todos paralelamente. El montaje convencional de electrodos hace uso de electrodos unipolares; cada electrodo es ánodo o cátodo, ánodos y cátodos se posicionan alternadamente, de modo que varios pares ánodo-cátodo se conectan en paralelo a una fuente de poder. Con esta configuración geométrica, las reacciones ocurren en ambos lados de cada electrodo y se obtiene cobre en las dos caras de cada cátodo.

En una celda usando tecnología de electrodos bipolares se usa un ánodo y un cátodo unipolar, y entre ellos se colocan electrodos bipolares. Los electrodos se llaman bipolares porque cada uno de ellos esta construido para actuar como ánodo por una de sus caras y como cátodo por la otra, de modo que todos los pares ánodo-cátodo están en conexión serie.

Los electrodos bipolares se dicen *flotantes*, porque están inmersos en la solución sin tener conexión eléctrica a una fuente de poder, la corriente fluye desde ánodo a cátodo atravesando cada electrodo bipolar. Para el mismo número de electrodos, la producción de cobre es la misma que en el caso unipolar. En este caso se obtiene cobre en cada electrodo bipolar, pero sólo en una de sus caras. La principal ventaja de esta tecnología es que en este caso se requiere una corriente mucho menor, pues el consumo de toda la celda es el mismo que el de un solo par de electrodos.

# 1.4 Modelo de la celda de electro-obtención

## 1.4.1 Modelo de base (Modelo de Bittner)

Existe poca experiencia en la modelación integral de este tipo de proceso. Reviste una gran dificultad debido a la interacción de variables eléctricas, químicas y mecánicas. En cualquier caso, el depósito de cobre esta condicionado principalmente por la densidad de corriente en cada punto del electrodo. Luego, para analizar la homogeneidad del depósito de cobre, se debe estudiar la distribución de la densidad de corriente en la superficie del electrodo.

Como una primera aproximación, el modelo desarrollado por Bittner considera un estado de operación ideal, en el que los iones inmersos en el electrolito son transportados sólo por el efecto del campo eléctrico. Considerando distribuciones de carga constantes, se recurre a la teoría electrostática para determinar el campo de corrientes estacionarias. El interés se enfoca en la geometría de la celda como variable para manipular la distribución del deposito de cobre en los cátodos. Es decir, la posición y dimensiones de los electrodos, los cuales se restringen a ser laminares. Debido a que las geometrías usadas convencionalmente son rectangulares, el modelo se implementa convenientemente basándose en el método numérico de diferencias finitas. Este genera un sistema de ecuaciones, donde las incógnitas son los potenciales en cada punto del dominio. La solución del problema electrostático requiere establecer condiciones de contorno en las interfaces donde se produce discontinuidad de parámetros. La condición de contorno en la interfase electrodo-electrolito presentaba un desafío, debido a la inexistencia de modelos para electrodos bipolares.

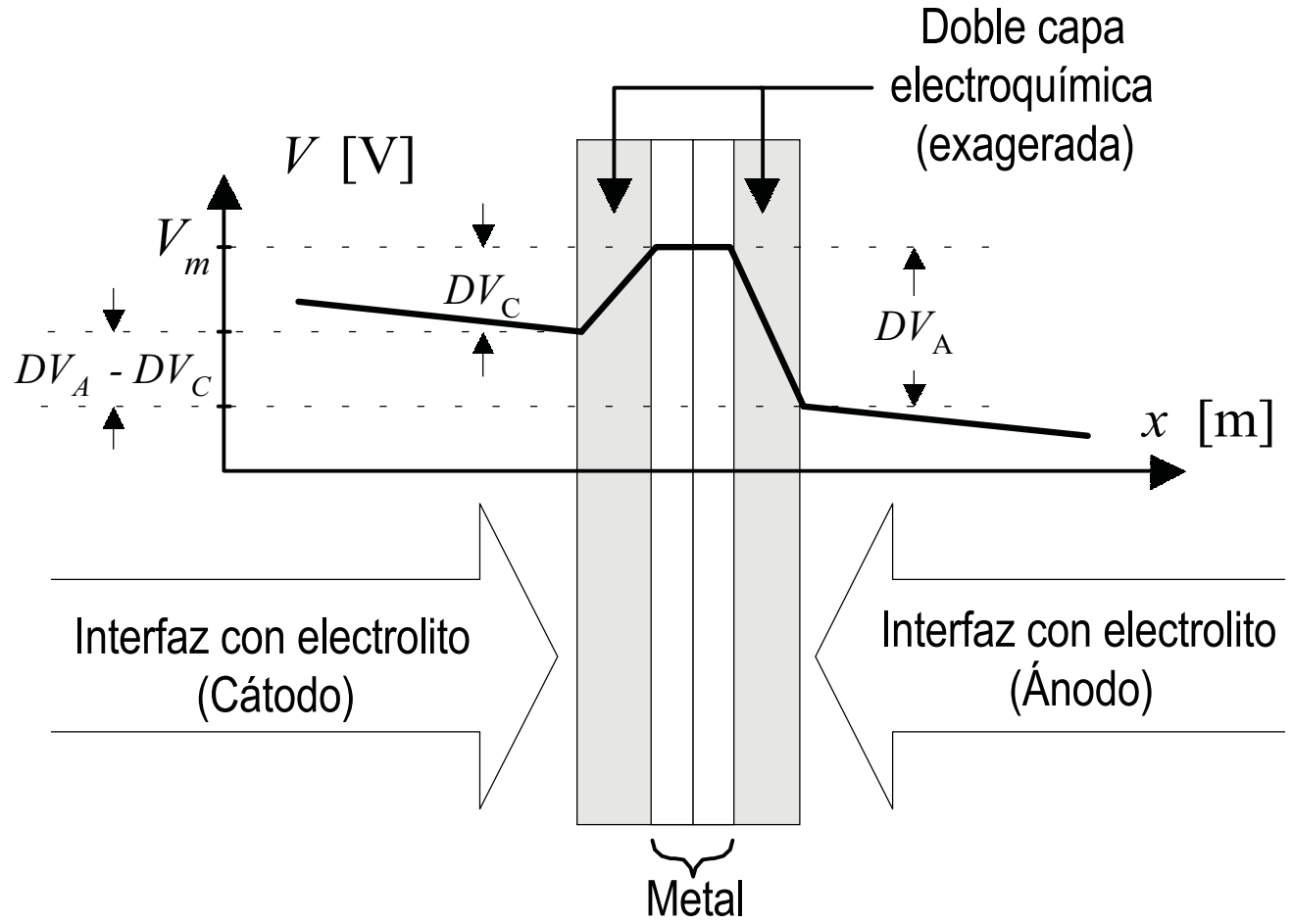

Figura 1.2: Sección transversal de electrodo bipolar laminar. $D V_{\mathrm{C}}$ y $D V_{\mathrm{A}}$ son los potenciales de electrodo.

En la interfase electrodo-electrolito existe una región microscópica denominada *doble capa electroquímica*, en la que se produce una variación del potencial eléctrico. Bittner redujo el modelo del proceso de electrodo a una discontinuidad de potencial como condición de contorno en la interfase electrodo-electrolito (Figura 1.2). El potencial en cualquier punto de la interfase, por el lado del electrolito, queda determinado a partir del potencial en la superficie del electrodo, *potencial metálico* ($V_m$), y un “salto de potencial”, *potencial de electrodo* ($D V$). El salto de potencial es una variable dependiente de la corriente que atraviesa la interfase, con función conocida.

En los electrodos unipolares el potencial metálico se conoce de antemano, pues estos electrodos se conectan físicamente a una fuente de voltaje (variable independiente), por ello la condición de contorno se aplica sin dificultad. Debido a que los electrodos bipolares son flotantes, no se conoce su potencial metálico (variable dependiente), por ello se busca una solución alternativa que no hace uso del potencial metálico, al relacionar directamente los potenciales del electrolito en las dos interfaces del electrodo bipolar. Tal solución utiliza la discontinuidad de potencial neta entre las dos caras del electrodo bipolar $(DV_A\text{-}DV_C)$, y se consigue manipulando directamente las ecuaciones de diferencia finita.

En su investigación, Bittner implementó el modelo para analizar básicamente tres tipos de geometrías de celdas, cuyos resultados se resumen en la Figura 1.3. Para las dos primeras celdas, los resultados de simulaciones hechas con el modelo base pudieron contrastarse con resultados obtenidos con prototipos experimentales. Se comprobó que al menos cualitativamente, el modelo predice adecuadamente la distribución de los depósitos de cobre, lo que valida los principios utilizados para desarrollar el modelo.

Desde el punto de vista hidráulico, la celda con electrodos centrados, Figura 1.3(a), constituye un único compartimiento que permite el flujo de electrolito. Sin embargo los depósitos de cobre resultan poco homogéneos, resultan condensados en el centro de los electrodos bipolares, mientras que en el cátodo unipolar ocurre lo contrario. Esto se debe a que la corriente tiende a rodear los electrodos bipolares para circular directamente entre el ánodo y cátodo unipolares

La celda con electrodos extendidos, Figura 1.3(b), presenta un comportamiento óptimo desde el punto de vista electrostático. Genera densidades de corriente uniformes, las que producen depósitos de cobre homogéneos. Sin embargo, esta alternativa presenta una mayor complejidad en el sistema de alimentación del electrolito, porque los electrodos bipolares dividen la celda en sub-celdas, las que deben ser alimentadas independientemente.

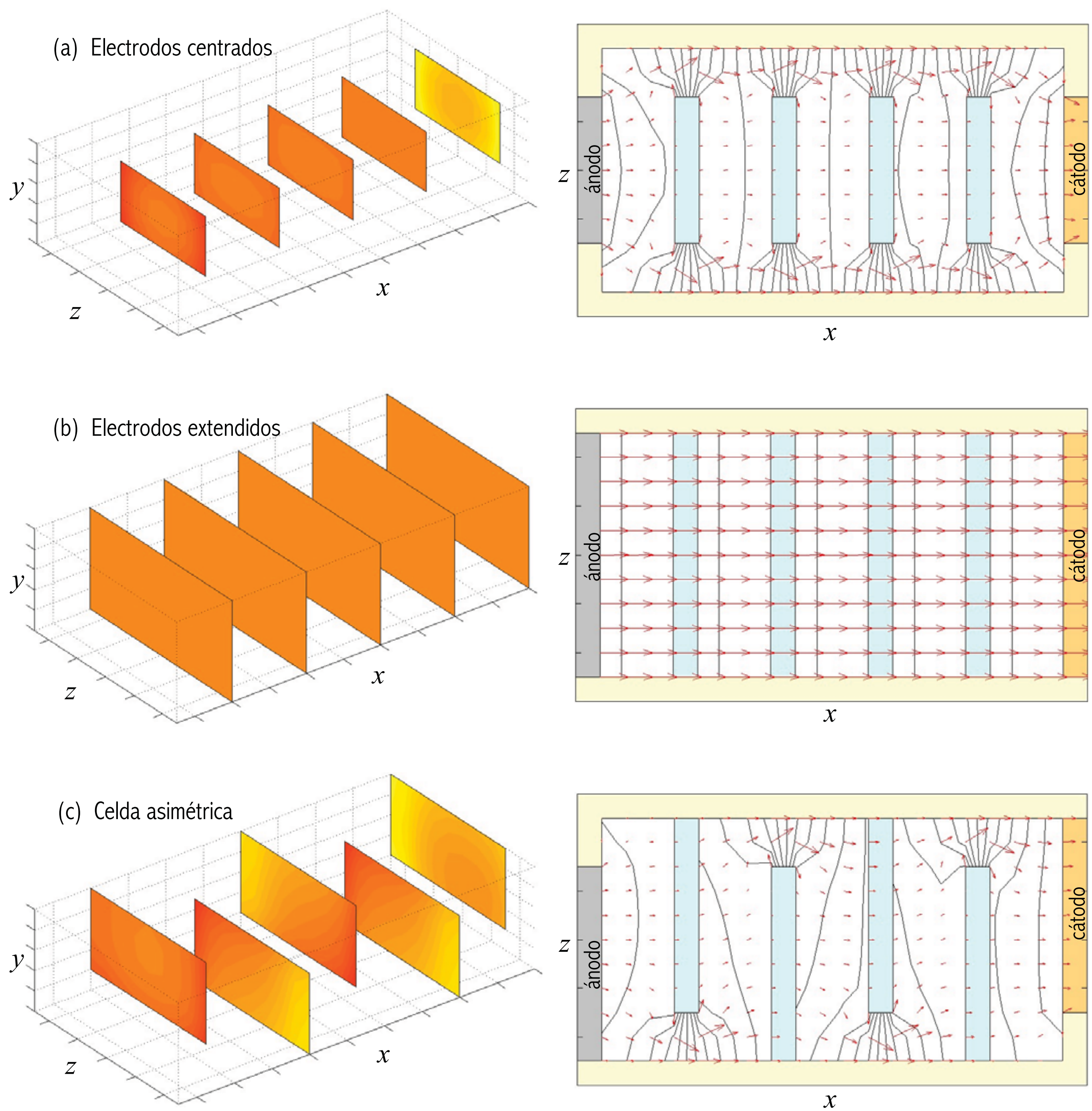

Figura 1.3: Resultados de la investigación de Bittner. Simulación, usando el modelo base, de 3 celdas de EW con 4 electrodos bipolares (electrodos centrales). (Izquierda) La distribución de densidad de corriente normal en las superficies de los cátodos codifica en color el grosor del depósito de Cu – color rojo indica un depósito delgado y color amarillo indica un depósito grueso. (Derecha) Campo eléctrico y líneas equipotenciales sobre un corte bidimensional de las celdas. La longitud de las flechas codifica la intensidad del campo eléctrico.

Con objeto de obtener una solución de compromiso, se simuló la celda asimétrica de la Figura 1.3(c). La idea era tener una única cámara desde el punto de vista hidráulico y obtener distribuciones de corriente más uniformes, al utilizar electrodos parcialmente extendidos. Sin embargo, en los bordes no extendidos de los electrodos se producen claramente los fenómenos indeseables vistos en la celda con electrodos centrados.

La existencia de corriente eléctrica a través de los electrodos bipolares implica un depósito de cobre en la superficie catódica de cada electrodo, esto muestra que el usar electrodos bipolares en celdas de EW de cobre tiene fundamento y se justifica su investigación, debido a las grandes ventajas eléctricas y metalúrgicas que se pueden obtener. Aunque ninguno de los prototipos de celda estudiados hasta el momento proporciona un desempeño favorable en todos los aspectos, se comprueba que el modelo resulta muy útil para comprender y predecir el funcionamiento de la celda de EW con electrodos bipolares, por lo que podría ser muy útil para obtener una configuración geométrica óptima.

### 1.4.2 Críticas al modelo de base y origen de un modelo mejorado

Los logros del trabajo de Bittner son indiscutibles, en particular, la modelación del proceso de electrodo como una discontinuidad de potencial parece un acierto importante. Sin embargo, la formulación del modelo de electrodo bipolar implementado en ecuaciones de diferencias finitas merece ciertas críticas.

Las ecuaciones de diferencia finita que modelan el electrodo bipolar fuerzan potencial constante en toda la superficie de la interfase cátodo-electrolito, [3], [4], [5], por el lado del electrolito. La anterior condición de equipotencialidad sólo es válida cuando los potenciales de electrodo son constantes. Esta fue una consideración en las primeras etapas de la modelación [3]. En la etapa final de modelación [4], [5], se considera que los potenciales de electrodo varían en toda la superficie del electrodo. Como se utilizan las mismas ecuaciones de diferencia, se produce una contradicción, pues el potencial en la superficie del metal no resulta constante. Tal contradicción no se hace evidente en las simulaciones, porque en los gráficos de potencial mostrados [5], no-se grafican los resultados reales obtenidos, por simplicidad sólo se toma una muestra del potencial metálico en un punto del electrodo y se repite en toda su superficie, por lo que el potencial metálico siempre se aprecia constante. Lo anterior se comprobó al estudiar el código que implementa el modelo base y despliega los resultados [6]. Al graficar los potenciales metálicos correctamente, se comprobó que es muy difícil notar el error gráficamente, porque los potenciales de electrodo ($D V$) varían relativamente muy poco sobre cada interfase. Sin embargo, como se ilustra en la Figura 1.4, siempre se cuantifican numéricamente y se comprueba que el potencial metálico ($V_m$) no resulta constante.

El hecho de forzar artificialmente constante el potencial en la interfase catódica, por el lado del electrolito, implica también forzar innecesariamente que el vector de campo eléctrico sea normal a la interfase en toda su superficie, como lo sería en un metal. Por último, las ecuaciones de diferencia relacionan los potenciales en el electrolito a ambos lados del electrodo bipolar, de tal modo que la orientación normal del campo, forzada en la interfase catódica, se propaga en cierto grado a la interfase anódica.

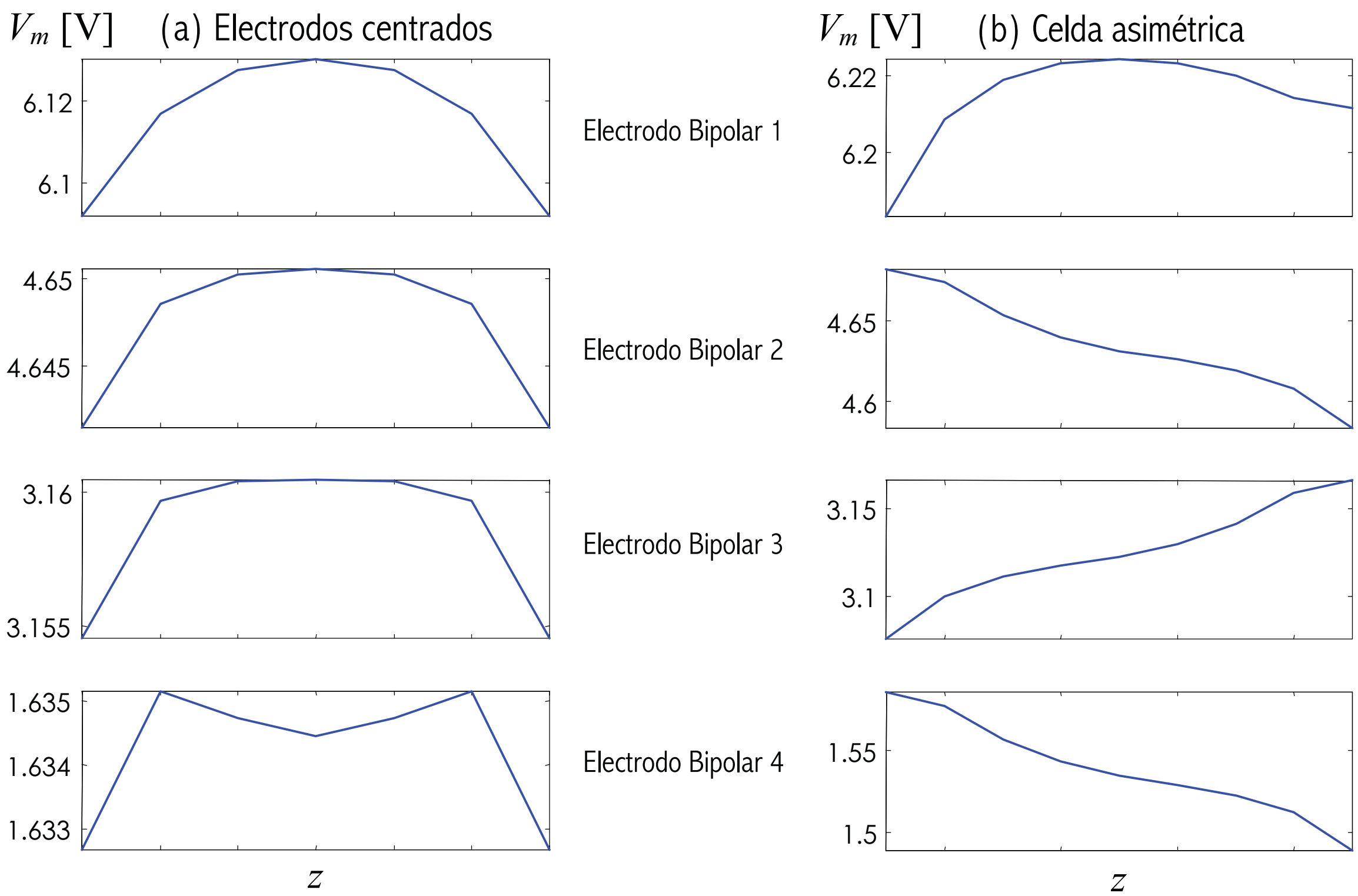

Figura 1.4: Potenciales metálicos de electrodos bipolares en modelo base. Perfiles de potencial correspondientes a celdas de la Figura 1.3 para $y$ =½ *altura de celda*.

A pesar de las imprecisiones en el modelo base, los resultados de las simulaciones realizadas concuerdan cualitativamente con resultados prácticos. No obstante, es posible que el modelo haya sido válido circunstancialmente, porque las configuraciones de prueba que se usó de referencia presentan un comportamiento que se ajusta a la suposición de potenciales de electrodo constantes, hecha implícitamente por el modelo. En efecto, en todas las configuraciones analizadas, los electrodos son láminas paralelas relativamente juntas, cuya área cubre una gran parte de la sección de la celda. Luego, salvo en los bordes de los electrodos, es lógico esperar que el campo eléctrico tienda a ser normal a la superficie de los electrodos y como las isosuperficies de potencial son paralelas a ellos, los potenciales de electrodo tenderían a ser constantes.

Aunque las condiciones forzadas por el modelo de Bittner pueden ser aproximadamente válidas para las configuraciones y condiciones de operación tomadas como referencia. Es posible que el modelo resulte inapropiado para configuraciones y/o condiciones de operación muy diferentes. Particularmente, porque como se explicará, la función que determina los potenciales de electrodo ($DV$) en cada punto del electrodo, en términos de la densidad corriente, tiene una característica fuertemente no lineal. Esto permite presumir que los potenciales de electrodo podrían tener una fuerte variabilidad en ciertas condiciones de operación.

Con objeto de mejorar la precisión de los resultados y validar el modelo electroestático para el análisis de geometrías arbitrarias, como las que se pretende soportar con el software que se desarrolla en el presente trabajo, se intentó desarrollar un mejor modelo de electrodo bipolar. Este objetivo se logró al desarrollar un método para determinar los potenciales metálicos en electrodos flotantes, el que tiene una base completamente analítica y constituye una importante contribución a la modelación desarrollada por Bittner. El método desarrollado se puede aplicar tanto a electrodos unipolares como bipolares y a cualquier tipo de geometría, no necesariamente rectangular. En particular, el conocimiento del potencial metálico en electrodos bipolares permite utilizar la misma condición de contorno empleada en electrodos unipolares, resolviendo así el problema electrostático en forma consistente.

## 1.5 Visualización

La corriente eléctrica puede ser imaginada macroscópicamente como un flujo de algún "fluido eléctrico". Históricamente, el modelo de fluido fue el primer modelo de la corriente eléctrica [15]. La *Visualización de Flujos* juega un importante rol en la ciencia e ingeniería [9]. El principal desafío de la Visualización de Flujos es encontrar maneras de representar y visualizar grandes conjuntos de datos (multidimensionales y multivariables), y hacerlo en una forma que sea tanto matemáticamente rigurosa como perceptiblemente tratable, de modo de proporcionar herramientas apropiadas para analizar e interpretar tales volúmenes de información.

Debido a la dimensionalidad de los datos generados por el modelo electroquímico en estudio, resulta indispensable disponer de herramientas de visualización adecuadas para su análisis. La herramienta más importante a desarrollar es una que permita la visualización tridimensional de un campo vectorial estacionario. En la investigación que realizó Bittner, la plataforma que utilizó para implementar el modelo proporcionaba una técnica elemental, que consistía en visualizar vectores bidimensionales como flechas orientadas, los que codifican la intensidad del campo con su longitud (Figura 1.3 - derecha). Como las flechas proporcionan una visualización local, es difícil comprender la estructura del campo vectorial, pues es necesario hacer una interpolación mental de las flechas adyacentes. Además, cuando las variaciones de la intensidad del campo son muy abruptas, las flechas se desvanecen y aportan poca información. En el caso tridimensional, las flechas resultan poco útiles, porque la dificultad para determinar su ubicación espacial produce una confusión visual.

La representación visual de campos vectoriales es un tema de continua investigación en visualización científica, pero existen técnicas adecuadas para muchos problemas. Para esta aplicación, se adoptó la técnica de visualizar líneas de flujo. Es una técnica popular por ser directa y poderosa. Además, es la única técnica que permite un rendimiento adecuado para aplicaciones interactivas y que, con una adecuada implementación, puede proporcionar una excelente representación visual. Es una técnica particularmente útil en problemas electrostáticos, donde la generación de líneas de flujo sobre superficies equipotenciales puede proporcionar una adecuada descripción de la estructura de un campo vectorial.

En este informe, se dará énfasis en la descripción de tópicos relativos a visualización científica. Se omiten detalles respecto a técnicas convencionales de computación gráfica [8], [19].

# *Modelación Tridimensional de una Celda de Electro-obtención*

## 2.1 Introducción

Como se anticipó en la sección 1.4.2, en este trabajo se logró mejorar el modelo base, principalmente debido al desarrollo de un método para la determinación del potencial de electrodos flotantes, el que se describe en la sección 2.3.2. Esta mejora permite modelar el electrodo bipolar sobre la base de condiciones de borde, en una forma completamente analítica, lo que elimina la restricción de electrodo laminar impuesta por la implementación numérica del modelo base.

Con objeto de implementar un modelo más preciso y una aplicación computacional que proporcione el mayor grado de libertad, en cuanto a las geometrías que se pueden analizar. En este trabajo, se decide implementar electrodos con espesor variable, en los cuales la reacción electroquímica ocurre en toda la superficie del electrodo, como debe ocurrir en un electrodo real. De este modo, el modelo debería proporcionar una representación más precisa del fenómeno real en los bordes de los electrodos, regiones de particular interés, y permitiría analizar geometrías mas variadas. Para implementar tal modelo, se reestructura la modelación electrostática presentada en [3], [4] y [5], para considerar todas las posibles orientaciones de las interfases.

## 2.2 Fundamentos de la teoría de electro-obtención de cobre

El mecanismo de conducción en el electrolito se basa en el movimiento de las partículas cargadas, los iones, los cuales son transportados por tres efectos: efecto del gradiente de potencial eléctrico (migración), efecto del gradiente de concentración (difusión) y el efecto del gradiente de densidad del electrolito (convección).

En una reacción de EW de cobre, el depósito de cobre puede ser afectado por el régimen de flujo, el que depende de la velocidad de circulación de la solución, de sus propiedades físicas y de la geometría de la celda. Dicho régimen modifica la velocidad de transferencia de masa y, si esta no es homogénea, la deposición de cobre será heterogénea. Además, es posible la retensión del electrolito e impurezas, con lo que la calidad química del producto puede ser degradada.

Considerando el proceso electroquímico en el cual se basa la electro-obtención de cobre descrito en la sección 1.2 (Figura 1.1, página 2), las transformaciones químicas asociadas a las reacciones de reducción del ión cúprico en el cátodo y la oxidación del agua en el ánodo, son respectivamente:

(2.1) $$Cu^{2+} + 2e \rightarrow Cu^0 \quad , e^0 = 0{,}34[V], \text{ y}$$

(2.2) $$2H_2O \rightarrow 4H^+ + 2O_2 + 4e \quad , e^0 = -1{,}23[V]$$

La reacción completa del proceso puede ser representada como:

(2.3) $$CuSO_4 + H_2O \xrightarrow{\text{Energía Eléctrica}} \downarrow Cu^0 + \tfrac{1}{2}O^2 \uparrow + H^2SO^4$$

Para que se produzca esta reacción, se tiene que aplicar una diferencia de potencial entre los electrodos, la que debe ser mayor que la diferencia entre los potenciales de equilibrio (potenciales de electrodo) de las semireacciones involucradas. En condiciones de temperatura y concentración estándar, esta diferencia es $E_0$ = 0,34[V] + (-1,23[V]) = -0,89[V]. Esto significa que para mantener la reacción en equilibrio dinámico en la dirección especificada en (2.3), se debe proporcionar energía con un voltaje mayor a 0,89[V]. Si se aplica un voltaje inferior, la reacción ocurre en el sentido opuesto, causando una corriente inversa.

Cuando las condiciones de operación son tales que la transferencia de masa (difusión) no limita la velocidad de la reacción, esta es controlada por la transferencia de carga (activación). En este caso, la forma aproximada de la característica corriente-voltaje de la reacción corresponde a la mostrada en la Figura 2.1(a). En esta, $j$ [$A/m^2$] es la densidad de corriente y $E_R$ es el potencial de equilibrio en condiciones no estándar, el que se calcula según la ecuación de Nernst [5].

El origen del potencial de electrodo $e_0$, asociado al cambio de fase, esta directamente relacionado con la formación de la denominada "doble capa electroquímica" en la interfase electrodo-electrolito. Cuando se pone en contacto dos fases con conductividades eléctricas significativas se produce una redistribución de carga. La región que contiene esta distribución se conoce como doble capa electroquímica, ya que el exceso total de carga a un lado de la interfase es de la misma magnitud y de signo opuesto al del otro lado. Esta distribución de carga causa un aumento gradual de potencial en el interior de la doble capa, en la dirección solución-electrodo, como se muestra en la Figura 2.1(b). En el metal se origina una densidad superficial de carga correspondiente a un exceso o déficit de electrones en la capa atómica superficial. En la solución existe una distribución de densidad volumétrica de carga, pero que esta confinada a una zona estrecha, cuyo orden corresponde al radio de los iones.

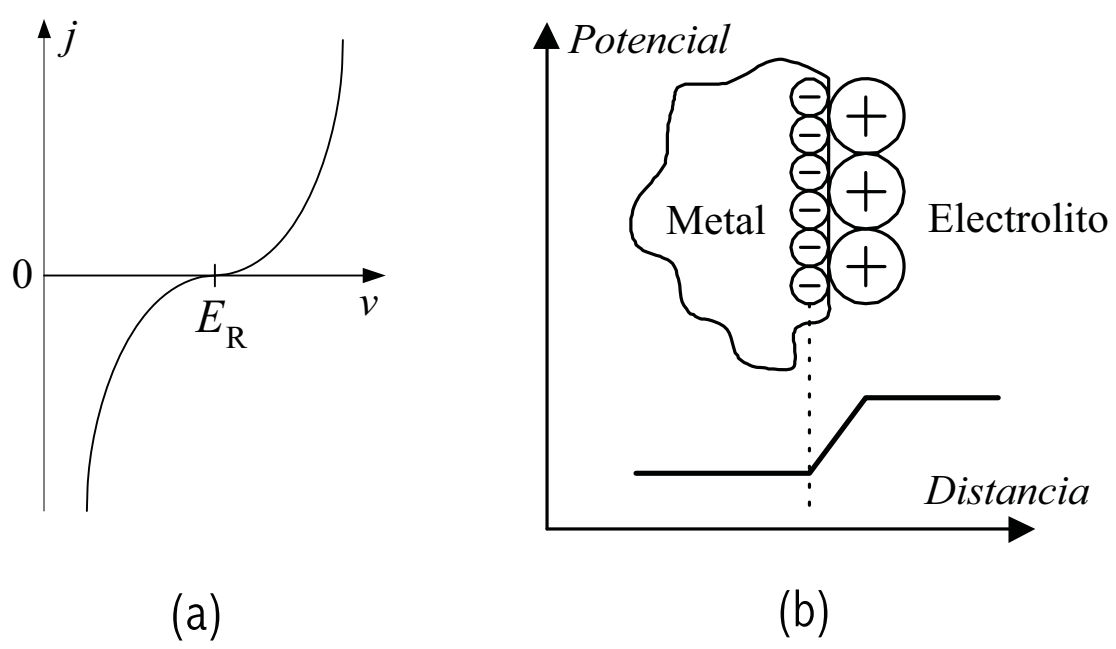

Figura 2.1: (a) Característica corriente-voltaje de un proceso de electrodo. (b) Doble capa electroquímica y potencial de electrodo asociado.

## 2.3 Formulación de la modelación física

### 2.3.1 Modelación electrostática

El objetivo de la modelación es obtener la distribución de la densidad de corriente eléctrica, en particular en la superficie de los electrodos catódicos, ya que dicha distribución se relaciona directamente con la del depósito de cobre. Se resuelve el problema en forma simplificada partiendo de una base de estado estacionario, es decir, considerando distribuciones de carga constantes que generan un campo de corrientes estacionarias. Sólo se considera la migración como mecanismo de conducción, es decir, la conducción debido al campo eléctrico. Lo anterior supone que la difusión es un proceso rápido (control por transferencia de carga) y que la solución es homogénea (no hay diferencias de concentración); no se considera el efecto del régimen de flujo.

Se debe determinar la distribución de potencial eléctrico ($V$) en todo el volumen de la celda. Basado en $V$, el vector de campo eléctrico ($\mathbf{E}$) y el vector densidad de corriente ($\mathbf{J}$) se obtienen respectivamente de:

(2.4) $$\mathbf{E} = -\nabla V$$

(2.5) $$\mathbf{J} = \sigma \mathbf{E}$$

donde $\sigma$ [$\Omega^{-1}m^{-1}$] es la conductividad eléctrica y $\nabla$ es el operador Nabla, el que en coordenadas rectangulares $\mathbf{x} = (x, y, z)$ se define como $\nabla = (\partial/\partial x)\,\hat{\mathbf{x}} + (\partial/\partial y)\,\hat{\mathbf{y}} + (\partial/\partial z)\,\hat{\mathbf{z}}$.

La homogeneidad del electrolito implica una conductividad eléctrica constante, en cuyo caso, los campos $\mathbf{E}$ y $\mathbf{J}$ son proporcionales. Luego, para un análisis cualitativo, el que es suficiente para verificar la homogeneidad de la distribución de la densidad de corriente, bastaría con obtener el campo eléctrico.

La solución general del problema electrostático de determinar $V$ involucra resolver un problema de condiciones de contorno. Se plantean condiciones de contorno en las interfaces donde ocurren discontinuidades de distribución de parámetros por cambio de material. En los materiales continuos, $V$ se determina resolviendo la ecuación de Poisson:

(2.6) $$\nabla^2 V = -\frac{\rho}{\varepsilon}$$

donde $\rho$[$C/m^3$] es la densidad de carga eléctrica volumétrica, $\varepsilon$[F/m] es la permitividad eléctrica del material. $\nabla^2 V$ es el laplaciano de $V$, el que en coordenadas rectangulares se define como $\nabla^2 V = \nabla \cdot (\nabla V) = \partial^2 V/\partial x^2 + \partial^2 V/\partial y^2 + \partial^2 V/\partial z^2$.

En medios homogéneos no existe densidad de carga eléctrica y la ecuación de Poisson toma la forma de la ecuación de Laplace:

(2.7) $$\nabla^2 V = 0$$

En el caso de la celda en estudio, se considera el electrolito como una solución conductora imperfecta homogénea, los electrodos como conductores perfectos y las paredes de la celda como dieléctrico. Luego, la distribución de potencial en el interior de la solución electrolítica queda determinada por la solución de la ecuación diferencial de segundo orden (2.7). La distribución de potencial en el interior de los electrodos queda determinada por la condición de equipotencialidad de un conductor perfecto. La homogeneidad de los materiales implica una conductividad $\sigma$ constante, por lo que la densidad de corriente puede ser evaluada cualitativamente a través del campo eléctrico.

Se requieren condiciones de contorno para las interfaces celda-electrolito y electrodo-electrolito. Para el caso de contacto entre celda y electrodo no se requiere condición de contorno adicional, pues en la interfase se considera el potencial por el lado del electrodo, el que se obtiene por la condición de equipotencialidad. La primera condición de contorno surge del hecho que en una interfaz dieléctrico-conductor, la componente del campo eléctrico normal a la interfase es nula en el lado del conductor, es decir:

(2.8) $$\nabla V \cdot \hat{\mathbf{n}} = 0$$

En (2.8) y en todas las expresiones futuras, $\hat{\mathbf{n}}$ es un vector unitario normal a la superficie de una interfase, orientado hacia el electrolito.

La región correspondiente a la doble capa electroquímica es no homogénea debido la existencia de densidad de carga. Como el ancho efectivo de esta zona es microscópico, la variación de potencial producida dentro de la zona se aprecia macroscópicamente como una discontinuidad de potencial en la interfase electrodo-electrolito. Al modelar la doble capa electroquímica con ancho nulo, la discontinuidad de potencial permite establecer la siguiente condición de contorno para todas las interfaces electrodo-electrolito:

$$V = V_{\mathrm{m}} - DV \tag{2.9}$$

donde $V$ es el potencial en el lado del electrolito y $V_{\mathrm{m}}$ es el *potencial metálico*, el potencial constante en la superficie del electrodo. La magnitud de la discontinuidad, $DV$, es el potencial de electrodo, medido en el sentido electrodo respecto a la solución electrolítica.

La curva de polarización de la Figura 2.1(a) establece implícitamente la relación entre los potenciales de electrodo y la corriente en cada punto de la interfaz electrodo-electrolito. Como en [4], [5], se considera una aproximación lineal de tal relación, la que para ánodo y cátodo respectivamente se expresa como:

$$DV_{\mathrm{A}} = -e_{\mathrm{R}} + K_{\mathrm{A}} \mathbf{J} \cdot \hat{\mathbf{n}} \tag{2.10}$$

$$DV_{\mathrm{C}} = +e_{\mathrm{R}} - K_{\mathrm{C}} \mathbf{J} \cdot \hat{\mathbf{n}} \tag{2.11}$$

donde $K_{\mathrm{A}}$ y $K_{\mathrm{C}}$ son constantes positivas adecuadas y $e_{\mathrm{R}}$ es el potencial de equilibrio de la reacción que ocurre en cada electrodo. Las relaciones (2.10) y (2.11) se expresan vectorialmente para que, independientemente de la orientación espacial del electrodo, siempre se considere la componente normal de la corriente, saliendo en el caso del ánodo y entrando en el caso del cátodo. Los signos diferentes en (2.10) y (2.11) surgen debido a que la curva de la Figura 2.1(a) considera la situación donde el voltaje aplicado se utiliza para

"polarizar" directamente la interfaz ánodo-electrolito, e inversamente la interfaz cátodo-electrolito.

Considerando las relaciones dadas en (2.4) y (2.5), las ecuaciones (2.10) y (2.11) se pueden rescribir convenientemente en función de potenciales:

(2.12) $$DV_A = -e_R - K_A \sigma \nabla V \cdot \hat{\mathbf{n}}$$

(2.13) $$DV_C = +e_R + K_C \sigma \nabla V \cdot \hat{\mathbf{n}}$$

donde $\sigma$ es la conductividad del electrolito.

Si los potenciales metálicos fueran variables independientes, las ecuaciones (2.7), (2.8), (2.9), (2.12), (2.13), junto a la condición de equipotencialidad, formarían un sistema de ecuaciones que permitirían encontrar la distribución de potencial en todo el dominio de la celda. En el caso de electrodos bipolares flotantes, el potencial metálico es una variable dependiente, luego para completar la solución del problema es necesario encontrar la forma de determinar el potencial metálico en electrodos flotantes, lo que se realiza en la siguiente sección.

### 2.3.2 Desarrollo de un método para la determinación del potencial de electrodos flotantes

Manteniendo la coherencia con la modelación del fenómeno de la doble capa electroquímica como una discontinuidad de potencial, se desea encontrar una relación entre el potencial del electrodo flotante y los potenciales de electrodo en toda la superficie de la interfase electrodo-electrolito. Como se describe en [5], la formación de los potenciales de electrodo esta determinada fundamentalmente por los mecanismos de acumulación de carga, debido al flujo de iones o electrones en la interfase electrodo-electrolito. De la teoría electromagnética [15], se obtiene una relación integral del flujo de carga (corriente) que atraviesa una superficie que encierra carga eléctrica, mediante la Ecuación de continuidad:

$$\oint_S \mathbf{J} \cdot d\mathbf{s} = -\frac{\partial}{\partial t} \int_{Vol(S)} \rho \, dVol \tag{2.14}$$

Esta ecuación relaciona el flujo de densidad de corriente que atraviesa una superficie cerrada con la variación instantánea de carga en el volumen encerrado por la superficie. El caso a considerar es una superficie cerrada, que encierre un electrodo flotante y que coincida con la interfase electrodo-electrolito por el lado del electrolito, de modo de encerrar completamente la doble capa electroquímica.

Debido que se usa la misma referencia para medir los potenciales de electrodo en interfaces anódicas y catódicas, el siguiente análisis se abstrae de la naturaleza de la reacción que ocurre en la superficie. Si se trata de un electrodo bipolar, los potenciales de electrodo deben ser calculados debidamente en cada región de su interfase con el electrolito. De modo que se supone que el resultado debería ser válido para electrodos unipolares y bipolares.

Como los potenciales de electrodo se definen para el estado de equilibrio dinámico, en el que cesa la acumulación de carga, la relación buscada entre el potencial del electrodo flotante y los potenciales de electrodo se obtiene en estado estacionario, lo que corresponde al caso cuando el término de la derecha en la ecuación (2.14) es nulo, es decir para

$$\oint_S \mathbf{J} \cdot d\mathbf{s} = 0 \tag{2.15}$$

Los potenciales de electrodo retienen la información relativa al proceso de acumulación de carga. De este modo, al incorporar de alguna forma la condición de contorno establecida en (2.9) en la ecuación (2.15), se obtiene la solución del problema.

Considerando las relaciones (2.4) y (2.5), la ecuación (2.15) es equivalente a:

$$\oint_S \sigma \nabla V \cdot d\mathbf{s} = 0 \qquad (2.16)$$

donde $d\mathbf{s}$ es un elemento diferencial orientable de superficie, el que tiene la forma $\hat{\mathbf{n}}\,dS$, siendo $dS$ la magnitud del diferencial de superficie y $\hat{\mathbf{n}}$ el vector normal unitario ya definido.

De acuerdo con la teoría del cálculo vectorial [16], la cantidad $\nabla V \cdot d\mathbf{s}$ corresponde a una diferencial total. Esta apreciación permite obtener una aproximación numérica de la solución, basándose en el Lema fundamental de derivación. Para la forma particular de $d\mathbf{s}$, el lema establece lo siguiente:

$$\frac{V(\mathbf{x}+d\mathbf{s}) - V(\mathbf{x}) - \nabla V \cdot d\mathbf{s}}{dS} \rightarrow 0, \text{ cuando } dS \rightarrow 0 \qquad (2.17)$$

De este modo, para $dS$ suficientemente pequeño, se puede considerar la siguiente aproximación

$$\nabla V \cdot d\mathbf{s} \approx V(\mathbf{x}+d\mathbf{s}) - V(\mathbf{x}), \text{ cuando } dS \rightarrow 0 \qquad (2.18)$$

Al combinar este resultado con la condición dada en (2.9) resulta:

$$\nabla V \cdot d\mathbf{s} \approx V(\mathbf{x}+d\mathbf{s}) + DV(\mathbf{x}) - V_{\mathrm{m}}, \text{ cuando } dS \rightarrow 0 \qquad (2.19)$$

Si la superficie $S$ se discretiza en $N$ elementos de superficie $\Delta S$, para $N$ suficientemente grande, el integrando de (2.16) se aproxima como en (2.19), entonces de (2.16) resulta:

$$\oint_S \sigma \nabla V \cdot d\mathbf{s} \approx \sum_{i=1}^{N} \sigma_i \left[ V(\mathbf{x}_i + d\mathbf{s}_i) + DV(\mathbf{x}_i) - V_m \right] = 0, \tag{2.20}$$

y en consecuencia:

$$V_m \approx \frac{\sum_{i=1}^{N} \sigma_i \left[ V(\mathbf{x}_i + d\mathbf{s}_i) + DV(\mathbf{x}_i) \right]}{\sum_{i=1}^{N} \sigma_i} \tag{2.21}$$

El resultado de (2.21) es válido en materiales no homogéneos y puede usarse con cualquier tipo de geometría. Las coordenadas $\mathbf{x}+d\mathbf{s}$ corresponden a puntos separados $\Delta S$ unidades de la superficie del electrodo en la dirección $\hat{\mathbf{n}}$ (Figura 4.1c, página 44). En general, los potenciales $V(\mathbf{x}+d\mathbf{s})$ deben estimarse mediante interpolación.

# *Visualización Interactiva de Campos Vectoriales mediante Líneas de Flujo*

## 3.1 Introducción

La característica más atractiva de las líneas de flujo, es la capacidad para entregar una representación continua de la estructura del campo vectorial. Existe la limitación de no poder obtener información del sentido de la línea de flujo. Para superar esta dificultad, suelen emplearse técnicas que segmentan las líneas, que pueden ser combinadas con animación. Como se indica más adelante, en el caso del problema electrostático, es posible determinar fácilmente el sentido de las líneas de flujo basándose en el uso de color. Por este motivo, no se realizarán esfuerzos adicionales para visualizar el sentido del campo, porque se considera más relevante conservar la característica de continuidad de las líneas.

Comúnmente se enfrentan dos problemas cuando se usan líneas de flujo. Un problema es que para obtener una adecuada representación del campo vectorial, no es muy obvio como distribuir las líneas de flujo en el espacio, usualmente se requiere participación del usuario en tal proceso. En esta aplicación, este problema no existe, pues en problemas electroestáticos se conoce una estrategia para lograr una adecuada visualización de la estructura del campo; la generación de líneas a partir de superficies equipotenciales. En esta aplicación, los electrodos constituyen adecuadas superficies equipotenciales, en particular porque interesa analizar el campo normal en sus superficies. En consecuencia, se puede concebir un método automático de generación de líneas de flujo. El segundo problema se refiere a la limitación del hardware gráfico convencional a la iluminación de primitivas de superficie. La iluminación de las líneas es indispensable para lograr una adecuada percepción de la orientación espacial, pero sin el soporte de hardware gráfico es difícil lograr un rendimiento para una aplicación interactiva. Una opción es visualizar las líneas de flujo como tubos construidos con polígonos, pero esto reduce considerablemente la cantidad de líneas de flujo visualizables. En esta aplicación, se utiliza una reciente técnica de iluminación de líneas que supera este inconveniente, al basar la implementación de la iluminación en el mapeo de textura, característica que puede beneficiarse del uso de hardware gráfico.

En suma a una eficiente iluminación que facilita la percepción de la orientación espacial, mediante un método que permite la percepción de profundidad, se puede proporcionar una adecuada indicación de la ubicación espacial de la línea de flujo. Adicionalmente, su puede colorear la línea de flujo de acuerdo a una cantidad escalar, en nuestro caso, es relevante visualizar el potencial eléctrico y la intensidad del campo vectorial $\mathbf{J}$. Como las líneas de flujo de corriente siempre se orientan del mayor a menor potencial, el mapeo del potencial en color es suficiente para permitir determinar el sentido de las líneas, cuando sea ambiguo.

Para la implementación de las características gráficas de la aplicación, se considera el uso de la librería gráfica OpenGL. Esta librería constituye una conveniente interfase software al hardware gráfico que soporta técnicas convencionales de computación gráfica para el rendering de puntos, líneas y polígonos.

## 3.2 Generación de líneas de flujo

### 3.2.1 Conceptos básicos

Las líneas de flujo, también llamadas curvas integrales, permiten apreciar gráficamente la estructura direccional de un campo vectorial estacionario. Son trayectorias, cuyos vectores tangentes coinciden con el campo vectorial.

Considerando un campo vectorial estacionario $\mathbf{J}$, si la línea de flujo se parametriza en función de la longitud de arco $\alpha$, su trayectoria $\mathbf{t}(\alpha)$ queda determinada por la siguiente ecuación diferencial:

$$\frac{d}{d\alpha}\mathbf{t}(\alpha) = \frac{\mathbf{J}}{|\mathbf{J}|} \equiv \hat{\mathbf{J}}(\mathbf{t}(\alpha)) \tag{3.1}$$

En (3.1) se define $\hat{\mathbf{J}}$ como el campo vectorial normalizado. Para campos vectoriales no estacionarios se pueden definir otros tipos de curvas integrales que consideran el tiempo en la integración [9], pero todas las curvas coinciden para campos estacionarios.

Para encontrar la línea de flujo que atraviesa un punto $\mathbf{x}$, (3.1) debe resolverse con la condición inicial $\mathbf{t}(0)=\mathbf{x}$. A las condicionales iniciales utilizadas para generar las curvas integrales se les llama *puntos semilla*.

### 3.2.2 Integración de líneas de flujo en un dominio descretizado

En (3.1) se indica como calcular las líneas de flujo en un dominio continuo. El caso de interés es cuando $\mathbf{J}$ esta definido en un dominio discretizado, en donde se debe usar interpolación para calcular el campo en coordenadas intermedias. En la práctica, a menos que la grilla tenga una pobre resolución, los errores debido a la interpolación son mucho más pequeños que los errores causados por un pobre método numérico de integración [9], [17]. Para obtener una adecuada precisión, se recomienda usar métodos de paso adaptivos con un adecuado control del error. Esto es particularmente recomendado en problemas electrostáticos, debido a la presencia de singularidades. En esta aplicación se usará un método basado en el descrito en [17], el cual considera interpolación trilineal y un método de integración Runge-Kutta de cuarto orden.

A partir de un punto $\mathbf{x}_1$, el método de Runge-Kutta de cuarto orden determina el siguiente punto $\mathbf{x}_2$ sobre la línea de flujo, ubicado a un paso $d$ delante de $\mathbf{x}_1$. El punto se estima en cuatro pasos, mediante la siguiente secuencia:

(3.2)
$$\mathbf{k}_1 = d\hat{\mathbf{J}}(\mathbf{x}_1), \qquad \mathbf{k}_2 = d\hat{\mathbf{J}}\left(\mathbf{x}_1 + \tfrac{1}{2}\mathbf{k}_1\right)$$
$$\mathbf{k}_3 = d\hat{\mathbf{J}}\left(\mathbf{x}_1 + \tfrac{1}{2}\mathbf{k}_2\right), \qquad \mathbf{k}_4 = d\hat{\mathbf{J}}\left(\mathbf{x}_1 + \tfrac{1}{2}\mathbf{k}_3\right)$$

(3.3)
$$\Delta\mathbf{x} = \frac{\mathbf{k}_1}{6} + \frac{\mathbf{k}_2}{3} + \frac{\mathbf{k}_3}{3} + \frac{\mathbf{k}_4}{6}$$

(3.4)
$$\mathbf{x}_2 = \mathbf{x}_1 + \Delta\mathbf{x}$$

Como estimación de error, se usa:

(3.5)
$$\Delta = \frac{1}{6}\left(\mathbf{k}_4 - d\mathbf{f}(\mathbf{x}_2)\right)$$

La expresión anterior corresponde a la diferencia entre las estimaciones entregadas por un método de cuarto orden y uno de tercer orden. En rigor esta fórmula estima el error de la fórmula menos precisa, pero como se indica en [17], en la mayoría de los casos se puede usar esta estimación con seguridad para controlar el tamaño del paso en el método de cuarto orden.

El objetivo del control adaptivo del tamaño del paso, es elegir $d$ tan grande como sea posible, mientras se satisface una tolerancia de error $TOL$ especificada por el usuario. Para un método de integración de cuarto orden, el término de error es proporcional a $d^5$. Por lo tanto, si un tamaño de paso $d$ genera un error $\Delta$, se puede obtener un tamaño de paso optimizado $d^*$ con:

(3.6) $$d^* = \left( \sqrt[5]{\rho_d \frac{TOL}{|\Delta|}} \right) d$$

con un factor de seguridad $\rho_d<1$.

El mecanismo de control del paso es el siguiente. Se calcula el siguiente punto, con un paso $d$ y se calcula $\Delta$ con ecuación (3.5). Si $|\Delta|$ es mayor que $TOL$, el punto se recalcula usando el paso $d=d^*$. De lo contrario, se continua con el siguiente punto usando $d=\min(d^*,d_{max})$, donde $d_{max}$ es el tamaño máximo de paso permitido. Como en [17], la integración de la línea se termina cuando $d$ se hace más pequeño que un valor $d_{min}$, el que debiera ser menor al tamaño de la grilla ($h$). Este mecanismo de control de longitud, básicamente trunca la línea cuando el campo tiende a anularse.

### 3.2.3 Selección de puntos semilla

Un problema común en la visualización de campos vectoriales usando líneas de flujo es la adecuada elección de los puntos semilla [18]. Idealmente, debiera controlarse la distribución de las líneas de flujo en lugar de la distribución de los puntos semilla, pero es mucho más simple controlar la distribución de los puntos semilla. Cuando es posible generar una gran cantidad de líneas de flujo, el posicionamiento de las líneas de flujo resulta menos importante.

Si la divergencia del campo vectorial es no nula, la densidad de las líneas de flujo no permanecerá constante. En algunas áreas las líneas de flujo pueden tender a juntarse, produciendo un aumento de la densidad local, mientras que en otras áreas pueden tender a expandirse, produciendo una disminución de la densidad local. En estos casos, para obtener resultados satisfactorios controlando sólo la densidad de los puntos semilla, suele limitarse la longitud de las líneas de flujo y, a partir de los puntos semilla, las líneas se integran una misma longitud hacia delante y hacia atrás.

Para visualizar la estructura de un campo electrostático resulta muy conveniente utilizar superficies isopotenciales como *superficies semilla*, es decir, superficies donde se distribuyen puntos semilla homogéneamente. Este método es muy útil en el caso electrostático, porque el campo electrostático siempre se orienta perpendicular a las superficies equipotenciales. Este método es particularmente útil para nuestro problema, porque la divergencia del campo eléctrico, y del campo de corrientes estacionarias, es nula en el electrolito, de modo que la densidad de líneas de flujo será aproximadamente constante en el interior de la celda.

## 3.3 Visualización de líneas de flujo

### 3.3.1 Iluminación de líneas en el espacio tridimensional

Para un objeto de dimensión $k>0$ en el espacio Euclidiano de dimensión $n>k$, la diferencia $n-k$ se define como la códimension del objeto. Esta definición es relevante porque la códimension determina la dimensionalidad del espacio normal del objeto, el que juega un papel importante en la descripción de la interacción de la luz con el objeto. Cuando un objeto tiene códimension 1, se le puede asignar un vector normal en forma natural y se pueden aplicar los modelos de iluminación usuales, se debe elegir el vector normal entre dos posibilidades. Sin embargo, cuando la códimension es mayor a 1, el espacio normal esta compuesto por un conjunto infinito de vectores y el modelo de iluminación debe ser extendido.

El caso tratado en libros populares sobre computación gráfica es el de superficies en un espacio tridimensional, correspondiente a objetos de códimension 1 ($k=2$, $n=3$). Las superficies se pueden caracterizar localmente por un vector normal unitario $\mathbf{N}$ en cada punto. En una superficie, el popular modelo de iluminación de Phong determina la intensidad de la luz en un punto mediante:

$$
\begin{aligned}
I &= I_{ambiente} + I_{difusa} + I_{especular} \\
&= k_a + k_d \mathbf{L} \cdot \mathbf{N} + k_s (\mathbf{V} \cdot \mathbf{R})^s
\end{aligned}
\tag{3.7}
$$

siendo $\mathbf{L}$ la dirección de la luz, $\mathbf{V}$ la dirección de vista y $\mathbf{R}$ el vector de reflexión unitario. $\mathbf{R}$ es el vector en el plano $\mathbf{L}$-$\mathbf{N}$, con el mismo ángulo a la superficie normal que el que describe la luz incidente (Figura 3.1).

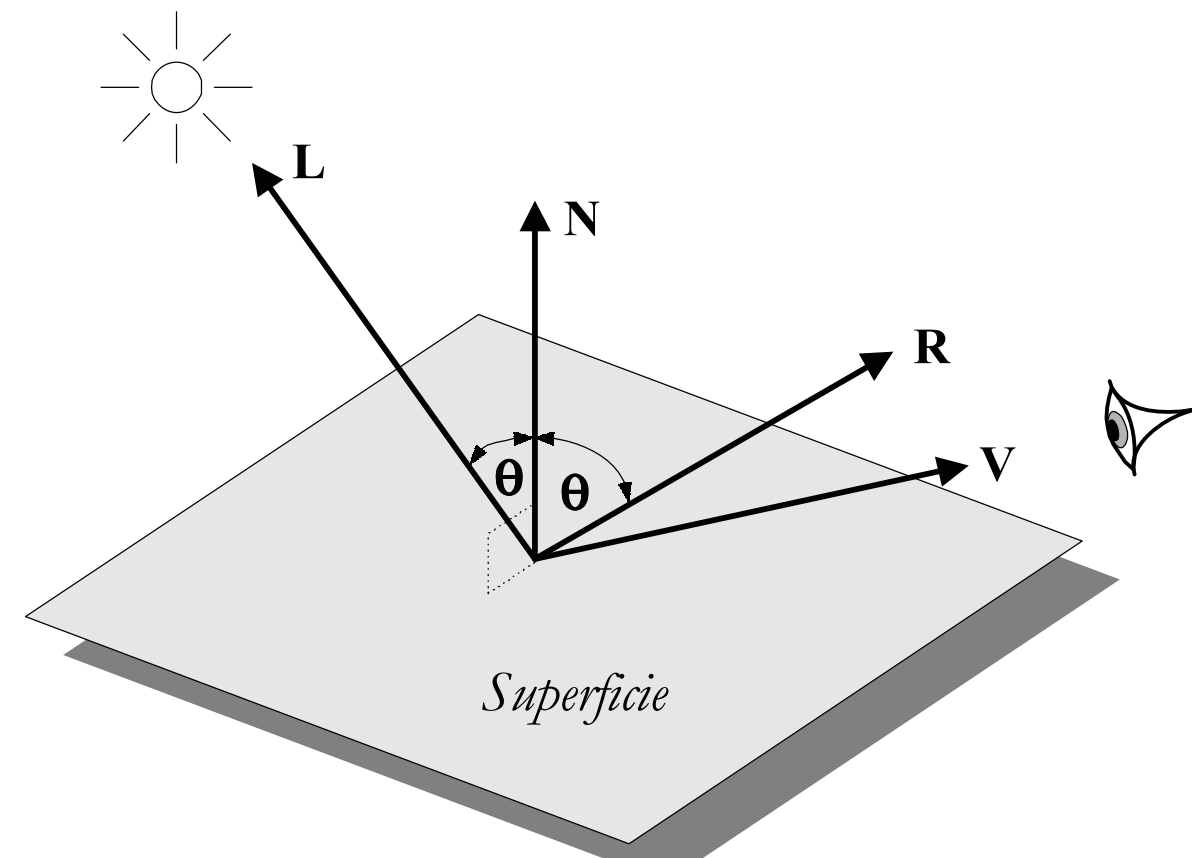

Figura 3.1: Modelo de reflexión Phong para primitivas de superficie.

En (3.7), el primer término representa la intensidad de la luz ambiente debido a múltiples reflexiones en el entorno. El segundo término describe la reflexión difusa debido a la Ley de Lambert. Los objetos con reflexión difusa pura se observan igualmente brillantes desde todas las direcciones, debido a la independencia del vector de vista. Las reflexiones especulares se centran en el vector de reflexión $\mathbf{R}$. El exponente $s$ controla el ancho del brillo especular.

En el caso de primitivas líneas, la códimension es 2, el vector normal y el vector de reflexión no están definidos en forma única, como se muestra en la Figura 3.2(a), sino que existe un conjunto infinito para cada uno de ellos. Los modelos clásicos de reflexión de superficies se pueden generalizar para códimensiones mayores a 1, mediante un método para la selección de un vector particular en cada conjunto. En [1], Banks muestra una formulación física de tales generalizaciones.

Para líneas en $\Re^3$, los resultados de la generalización del modelo de iluminación se explican a continuación. Para la descripción de los vectores seleccionados es conveniente notar que, como se muestra en la Figura 3.2(b), cualquier vector **X** puede descomponerse en dos componentes ortogonales $\mathbf{X}_T$ y $\mathbf{X}_N$, correspondientes a las proyecciones sobre **T** y el espacio normal, tal que $\mathbf{X} = \mathbf{X}_N + \mathbf{X}_T$.

De todos los posibles vectores normales, se elige el que es coplanar al vector de luz **L** y al vector unitario tangente a la línea **T**. De todos los posibles vectores de reflexión, **R** se elige tal que su componente tangente es opuesta a la componente tangente de **L**, y de modo que se minimiza la distancia entre **R** y **V**. Tal vector satisface las siguientes relaciones:

(3.8) $$\mathbf{R}_\mathbf{T} = -\mathbf{L}_\mathbf{T}, \text{ y}$$

(3.9) $$\mathbf{R}_\mathbf{N} = \left|\mathbf{L}_\mathbf{N}\right| \frac{\mathbf{V}_\mathbf{N}}{\left|\mathbf{V}_\mathbf{N}\right|}$$

Se podrían obtener fórmulas explicitas para **N** y **V**, de modo de calcular la intensidad mediante (3.7). Pero como los vectores seleccionados dependen de los vectores **L** y **V**, es conveniente expresar los productos puntos **L·N** y **V·R** directamente en términos de **L**, **V** y **T**.

Debido a la elección de **N**, se obtiene:

(3.10) $$\mathbf{L} \cdot \mathbf{N} = \left|\mathbf{L}_N\right| = \sqrt{\left|\mathbf{L}\right|^2 - \left|\mathbf{L}_T\right|^2} = \sqrt{1 - (\mathbf{L} \cdot \mathbf{T})^2}$$

Como se indica en [1], el producto punto **L·N** es siempre positivo, porque para códimensiones mayores a 1, en un objeto no se pueden distinguir lados frontal y posterior, de modo que el ángulo entre un vector y el espacio normal no puede ser mayor a 90º.

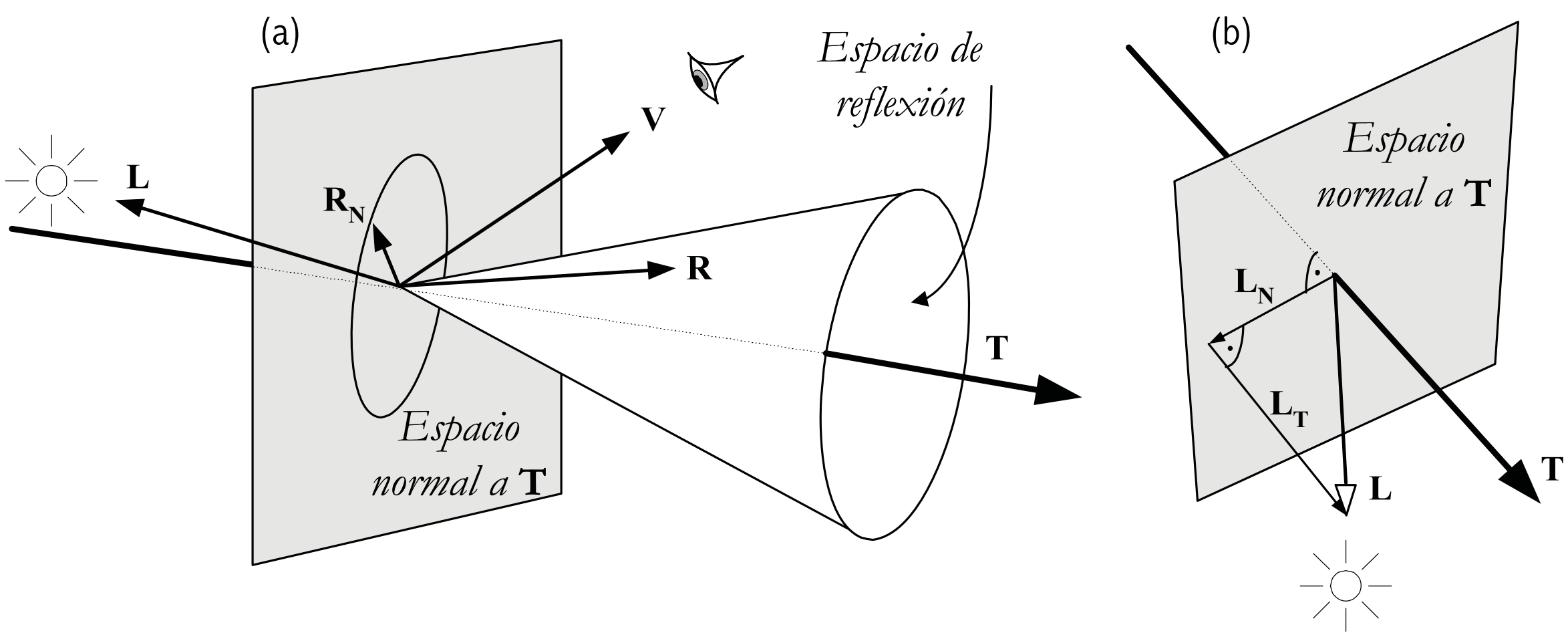

Figura 3.2: Modelo de reflexión Phong adaptado para primitivas líneas. (a) Para una primitiva de línea **T,** hay infinitos vectores normales pertenecientes a un plano normal e infinitos vectores de reflexión, pertenecientes a un cono centrado en **T**. (b) Cualquier vector (**L**) puede descomponerse en dos componentes ortogonales ($\mathbf{L}_\mathrm{T}$ y $\mathbf{L}_\mathrm{N}$), correspondientes a las proyecciones sobre **T** y el espacio normal.

El producto punto **V·R** se puede descomponer de la siguiente forma

$$\mathbf{V}\cdot\mathbf{R} = (\mathbf{V_T} + \mathbf{V_N})\cdot(\mathbf{R_T} + \mathbf{R_N})$$

Al considerar las restricciones de (3.8) y (3.9), se puede obtener

$$\mathbf{V}\cdot\mathbf{R} = \sqrt{1-(\mathbf{L}\cdot\mathbf{T})^2}\sqrt{1-(\mathbf{V}\cdot\mathbf{T})^2} - (\mathbf{L}\cdot\mathbf{T})(\mathbf{V}\cdot\mathbf{T}) \tag{3.11}$$

En [1] se indica que para códimensiones mayores a 1, en el modelo difuso ocurre un curioso fenómeno, el que incrementa la brillantez total de una imagen, haciéndola más uniforme y dificultando la percepción espacial. En [18] se da una explicación al fenómeno para el caso de líneas en $\Re^3$. Para compensar el efecto de exceso de brillo cualitativamente, Banks sugiere exponenciar el término de intensidad difusa usando una potencia $p(n,k)$ y propone un valor de $p$=4.7635 para el caso de líneas en $\Re^3$. En consecuencia, para líneas en $\Re^3$ el término difuso se calculará como:

$$I_{difusa} = k_d(\mathbf{L}\cdot\mathbf{N})^p \tag{3.12}$$

En [18] se indica la obtención de mejores resultados con un valor $p=2$, por lo que se considerará $p$ como un parámetro ajustable por el usuario.

Cuando el ángulo entre $\mathbf{V}$ y el espacio de reflexión es mayor a 90º, $\mathbf{V}\cdot\mathbf{R}$ es negativo y es razonable recortarlo a cero, pues se puede considerar que en tal caso el observador esta demasiado alejado del espacio de reflexión, de modo que no recibe luz reflejada. En tal caso, la intensidad especular esta dada por:

$$I_{especular} = k_s\left(\text{clamp}(\mathbf{V}\cdot\mathbf{R})\right)^s \quad (3.13)$$

donde $\text{clamp}(x)=0$, cuando $x<0$.

## 3.3.2 Rendering optimizado de líneas iluminadas

Debido a la dependencia de los vectores **N** y **R** en la posición de la fuente de luz, para cada nueva posición de la fuente de luz se requiere el recálculo apropiado de los vectores normal y de reflexión, cálculo que no realiza el hardware gráfico. Luego, a pesar que el modelo de iluminación de Phong es implementado en el hardware gráfico convencional, no se puede utilizar en forma directa. En [18] se desarrolla un método que evita el recálculo explícito de la normal, al evaluar los productos puntos definidos en (3.10) y (3.11) explotando la característica de mapeo de textura, disponible en aceleradores gráficos. De modo que se puede realizar el cálculo de la iluminación de líneas completamente en el hardware gráfico y así obtener una alta eficiencia.

El método supone la disponibilidad de una API gráfica similar a OpenGL, la que proporciona una interfase software al hardware gráfico. En esta librería, para cada vértice se puede especificar un vector homogéneo de coordenadas de textura. Convencionalmente, las primeras componentes de este vector se usan como índices en un mapa de textura uni-, bi-, o tri-dimensional. El mapa de textura puede contener colores y/o transparencias, las cuales se pueden usar para modificar, en varias formas, el color original del fragmento en el pipeline gráfico. Además, es posible cambiar las coordenadas de textura usando una matriz de transformación de textura de dimensión 4x4.

La matriz de transformación de textura resulta la característica clave que permite realizar cálculos de sombreado, usando el hardware de mapeo de textura. La técnica consiste en usar el vector unitario tangente a la línea, $\mathbf{T}$, como coordenada de textura $\mathbf{t_0}$ en cada vértice. Los productos punto de interés se calculan en el hardware gráfico al usar la siguiente matriz de transformación de textura:

$$\mathbf{M} = \frac{1}{2}\begin{bmatrix} L_x & L_y & L_z & 1 \\ V_x & V_y & V_z & 1 \\ 0 & 0 & 0 & 0 \\ 0 & 0 & 0 & 2 \end{bmatrix} \qquad (3.14)$$

El resultado del producto con la matriz, es el vector de textura homogéneo transformado $\mathbf{Mt_0} = [t_1, t_2, t_3, t_4]^{\mathrm{T}}$. Son de interés las dos primeras componentes del vector, $t_1$ y $t_2$, dadas por:

$$t_1 = \tfrac{1}{2}(\mathbf{L} \cdot \mathbf{T} + 1)$$
$$t_2 = \tfrac{1}{2}(\mathbf{V} \cdot \mathbf{T} + 1)$$

Tanto $t_1$ como $t_2$ pertenecen al rango [0,1] por lo que se pueden usar como índices en un mapa de textura bidimensional $P(t_1, t_2)$, en donde los cálculos de iluminación pueden estar precalculados basándose en las relaciones:

(3.15) $$\mathbf{L}\cdot\mathbf{T} = 2t_1 - 1$$
$$\mathbf{V}\cdot\mathbf{T} = 2t_2 - 1$$

Al usar un modo de textura que establezca como color del fragmento, el color de su textura $P(t_1, t_2)$, se obtiene una imagen equivalente a la obtenida mediante la iluminación con una fuente de luz puntual. El color es establecido sólo en los vértices, entre los vértices se realiza interpolación lineal de las coordenadas de textura, de modo que entre los vértices la iluminación no resulta exacta, pues se usa vectores tangente no normalizados. Sin embargo, el efecto no es notable en la práctica.

Cuando se usa una fuente de luz puntual, localizada en la mismo posición del observador, el vector **L** y el vector **V** son idénticos, por lo que la ecuación (3.11) se simplifica a:

(3.16) $$\mathbf{V}\cdot\mathbf{R} = 1 - 2(\mathbf{L}\cdot\mathbf{T})^2 = 1 - 2(2t_1 - 1)^2$$

En este caso, el único producto punto a evaluar es **L·T**, por lo que basta utilizar un mapa de textura unidimensional. De acuerdo a (3.12), (2.13), (3.15) y (3.16), su inicialización debe hacerse mediante la siguiente formula

(3.17) $$P(t_1) = k_a + k_d\left[\sqrt{1-(2t_1-1)^2}\right]^p + k_s\left[\text{clamp}\left(1-2(2t_1-1)^2\right)\right]^s$$

Para la aplicación de esta investigación, se opta por el último caso descrito, no solo porque es más simple y eficiente, sino porque garantiza una adecuada iluminación independientemente de la posición de vista, y porque se libera al usuario de la necesidad de ubicar apropiadamente la fuente de luz. Los vectores tangentes **T** en cada vértice, se obtienen en el proceso integrativo que genera las líneas de flujo, que corresponden al campo **J** normalizado evaluado en los vértices ($\hat{\mathbf{J}}$).

### 3.3.3 Antialiasing de líneas

Debido a la naturaleza discreta de los dispositivos de visualización, las líneas visualizadas suelen presentar formas dentadas, especialmente las líneas casi horizontales o verticales. El efecto se denomina alias y puede ser muy molesto, especialmente en aplicaciones interactivas. Existen varias técnicas efectivas para suprimir el alias, las que pueden ser soportadas por el hardware gráfico. El método utilizado en la aplicación es el soportado por OpenGL [13]. Para cada uno de los píxeles que intercepta la línea, se calcula un valor de cobertura, basado en la fracción del área del píxel que es cubierta por la línea. A continuación, la opacidad de los píxeles interceptados se multiplica por el valor de cobertura calculado. Al habilitar un modo de mezcla de opacidades adecuado, el color de cada píxel resulta ser una mezcla del color de la línea y el color del objeto de fondo, con lo que se logra una visualización mucho más continua de la línea.

### 3.3.4 Coloración de las líneas de flujo

La coloración de las líneas de flujo para codificar una cantidad escalar resulta una herramienta de visualización muy poderosa. En nuestro caso, es de interés visualizar la intensidad del campo vectorial y el potencial eléctrico. Lo ideal, sería poder obtener los colores para las componentes ambiente y difusa de un mapa de color, usando una textura indexada por la cantidad escalar a mapear. Sin embargo, para implementar el mecanismo de iluminación optimizado se utilizan las tres componentes disponibles del vector de textura homogéneo, por lo que no se puede considerar un índice adicional para indexar el color.

Sin embargo, la librería gráfica OpenGL proporciona un modo de mapeo de textura que resuelve el problema, pues permite modular (multiplicar) el color de la textura con el color base del objeto. El color base se puede definir para cada vértice separadamente.

Un inconveniente de esta implementación es que el brillo especular resulta coloreado, en lugar de ser de color constante, como establece el modelo de iluminación de Phong. Como se señala en [18], esta es una limitación menor si se usan colores brillantes de base. A pesar de resultar coloreado, el brillo especular se puede identificar claramente, siendo muy importante para la comprensión de la estructura espacial del campo.

OpenGL utiliza el método de sombreado incremental denominado Gouraud [8], [19], el cual calcula el color a lo largo de un segmento de línea mediante interpolación del color de los vértices. Para que el mapeo en color de la cantidad escalar sea suficientemente preciso, se debe limitar la separación máxima de los vértices adyacentes sobre la línea de flujo. Esto se puede hacer simplemente utilizando adecuadamente el parámetro $d_{\mathrm{max}}$, definido en el proceso de integración de las líneas de flujo (sección 3.2.2).

### 3.3.5 Percepción de profundidad

Para mejorar la percepción espacial de escenas tridimensionales complejas, resulta muy importante la sugestión de profundidad. Esta se logra ajustando el color de los objetos de acuerdo a su distancia a la cámara. El objetivo es que los objetos más cercanos se observen brillantes y que los objetos lejanos se observen obscuros.

En el caso de las líneas de flujo, hay situaciones donde las líneas de flujo se ordenan, tales que constituyen estructuras similares a superficies. Debido a que el modelo de iluminación no distingue entre interior y exterior, en ciertos casos, la estructura espacial de tales seudo superficies no es clara a primera vista. La sugestión de profundidad puede facilitar la identificación de estructuras lejanas y cercanas, mejorando la percepción espacial. El mejoramiento es aun más importante cuando los objetos son rotados interactivamente.

# *Desarrollo de la Aplicación*

## 4.1 Introducción

Presentados los fundamentos teóricos en los que se basa este proyecto, en este capítulo se describen los aspectos más relevantes del diseño y la implementación del software desarrollado.

Debido a que el modelo base de la celda de electro-obtención fue modificado, consecuentemente resultó necesario desarrollar una nueva implementación numérica del modelo. Debido a que los tiempos de simulación promedio del modelo base eran considerables, se hizo un esfuerzo por optimizar la implementación numérica del nuevo modelo.

Basándose en lo mencionado en la sección 3.2.3 (página 31), al generar puntos semilla sólo en la superficie de los electrodos, distribuidos homogéneamente, y sin limitar la longitud de las líneas, se espera que las líneas de flujo recorran el volumen completo de la celda y revelen la estructura global del campo vectorial ($\mathbf{E}$ o $\mathbf{J}$). Esta estrategia constituye la base de un método automático de generación de líneas de flujo implementado en esta aplicación.

Con todas sus características, la implementación de las líneas de flujo que se contempló constituye una herramienta muy poderosa. No obstante, se consideraron algunas herramientas de visualización adicionales para mejorar la funcionalidad de la aplicación.

## 4.2 Modelación numérica de la celda de electro-obtención

### 4.2.1 Consideraciones y restricciones del modelo

El modelo considera una celda cuyo dominio se reduce a un arreglo tridimensional de puntos en coordenadas rectangulares, usando una unidad de discretización $h$[m]. Para proporcionar la máxima flexibilidad, los electrodos se consideran como paralelepípedos en los cuales la reacción electroquímica ocurre en toda su superficie. Para los electrodos bipolares, como en el modelo desarrollado por Bittner, se los restringe a estar orientados paralelamente en una determinada dirección, pero el espesor de las secciones anódica y catódica se puede variar independientemente.

La formulación del modelo básicamente diferencia el electrodo bipolar del unipolar por el uso de distintas fórmulas de cálculo de los potenciales de electrodo. Salvo la diferenciación que debe hacerse entre electrodos unipolares y bipolares, para determinar si se debe calcular el potencial metálico, las ecuaciones de borde para ambos electrodos son las mismas. Por lo anterior, la implementación del modelo utiliza convenientemente un electrodo generalizado, el que puede funcionar como electrodo unipolar o bipolar, su geometría se muestra en la Figura 4.1(a). La distinción de sección izquierda y derecha es para diferenciar las ecuaciones que se usan para calcular los potenciales de electrodo, ecuación (2.12) o (2.13). En electrodos unipolares se usa la misma ecuación en ambas secciones, (2.12) en ánodos y (2.13) en cátodos. En electrodos bipolares se usa ecuación (2.13) en la sección izquierda y ecuación (2.12) en la sección derecha.

Como se aprecia en la Figura 4.1(b), se evita que la división entre secciones ocurra en los puntos de la grilla de muestreo, porque en la división la reacción es indefinida. Se define la coordenada $x_0$ para determinar implícitamente la posición de la división de sección. Si impone la restricción $x_1 < x_0 \leq x_2$, para que cada sección tenga un espesor de por lo menos $h/2$. Si se usa un espesor de $h/2$ en ambas secciones, el electrodo opera como el electrodo bipolar laminar modelado por Bittner.

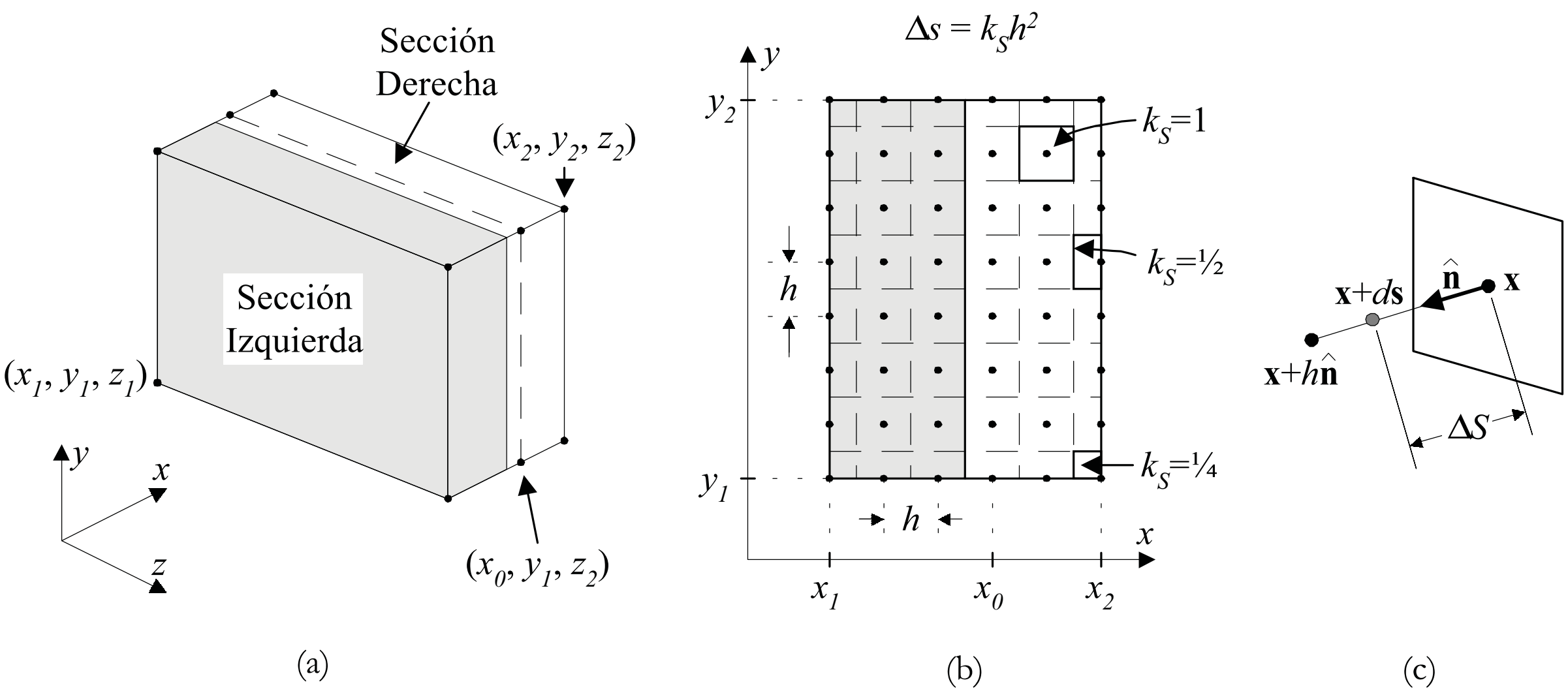

Figura 4.1: Geometría del electrodo modelado. (a) Estructura general. (b) Detalle en separación de secciones izquierda y derecha, y discretización de la superficie. (c) Relaciones relevantes en un elemento de superficie

## 4.2.2 Algoritmo de cálculo

### 4.2.2.1 Algoritmo básico

#### a) *Cálculo de la distribución de potencial*

La solución del problema del problema electrostático se realiza mediante el método de diferencias finitas [5], [15], el que aproxima la ecuación diferencial (2.7) y las condiciones diferenciales de borde, por ecuaciones de diferencia, las que resultan al truncar la expansión en serie de Taylor de una función. Las ecuaciones de diferencia determinan el potencial en cada punto del electrolito en función de los potenciales de los puntos adyacentes. Las ecuaciones para el cálculo de los potenciales de electrodo y los potenciales metálicos de electrodos flotantes generan ecuaciones de diferencia adicionales. En conjunto, las ecuaciones de diferencias forman un único sistema de ecuaciones lineales, cuyas incógnitas son los potenciales en el electrolito, los potenciales de electrodo y los potenciales de electrodos flotantes.

Para modelar la reacción electroquímica en toda la superficie del electrodo, se aumentó considerablemente el conjunto de ecuaciones de diferencia respecto al que se utilizaba en el modelo base, las que ya eran un número apreciable. Por esto se omite el detalle de las ecuaciones de diferencia.

A diferencia del modelo desarrollado por Bittner, en el que sistema de ecuaciones se resuelve mediante un método directo, en este trabajo se decide utilizar un método de solución iterativo (Gauss-Seidel), lo que se justifica en la sección 4.2.2.2. El término del proceso iterativo se realiza mediante el control del máximo error absoluto en el potencial y con la especificación de un número máximo de iteraciones para control de posible divergencia.

La distribución de potencial en el volumen rectangular de la celda se almacena convenientemente en un arreglo tridimensional escalar. En las interfaces donde se produce discontinuidad de potencial, en el arreglo se almacena el potencial correspondiente al electrolito, salvo en la interfase metal-dieléctrico, donde se considera el potencial del metal. En las regiones correspondientes al interior de electrodos, incluida la interfase metal-dieléctrico, el potencial metálico se almacena una vez finalizado el proceso iterativo.

Las ecuaciones para calcular los potenciales de electrodo, (2.12) y (2.13), utilizan sólo la componente normal del gradiente de potencial a la superficie del electrodo. En las aristas y vértices del electrodo, la superficie normal puede ser indefinida, por lo que se deben calcular de forma especial. En un punto perteneciente a una *arista interior* de un electrodo (arista sin incluir vértices), la superficie normal al gradiente sólo esta definida cuando la arista es tangente a la celda. En un vértice de un electrodo, la superficie normal al gradiente sólo esta definida cuando el vértice es tangente a una arista de la celda. Cuando la superficie normal al gradiente esta definida, el potencial de electrodo se calcula con (2.12) o (2.13), según corresponda. El criterio que se aplica cuando la superficie normal es indefinida, es considerar como potencial de electrodo un promedio de los potenciales de electrodo vecinos. Sólo se consideran los puntos vecinos en las caras adyacentes no tangentes a la celda.

El cálculo de los potenciales de electrodos flotantes requiere definir elementos que discreticen la superficie del electrodo. La discretización elegida se muestra en la Figura 4.1(b). Los elementos de superficie se centran en la grilla de muestreo, resultando un elemento con área no constante $\Delta S = k_S h^2$. El cálculo se efectúa a partir de una versión optimizada de la ecuación (2.21), la que se muestra en la sección 4.2.3.1.

### b) *Cálculo del campo densidad de corriente*

Una vez terminado el proceso iterativo en el que se calcula la distribución de potencial, se debe determinar el campo vectorial densidad de corriente en base al cálculo del gradiente de potencial, como establecen ecuaciones (2.4) y (2.5). Como en el caso de la distribución de potencial, el campo densidad de corriente se almacena convenientemente en un arreglo tridimensional que representa el volumen de la celda, pero en este caso el arreglo obviamente es vectorial, en cada punto de la grilla se deben almacenar las tres componentes escalares del campo vectorial.

A continuación se describe en detalle la forma en que se realiza el cálculo del gradiente de potencial en el volumen de la celda.

Dado un punto $P$ con potencial $V$, cada componente del campo eléctrico **E** ($E_x$, $E_y$, o $E_z$) en $P$, se calcula usando los potenciales de los puntos adyacentes, $V_{\text{Back}}$ y $V_{\text{Forward}}$, colineales en la dirección de la componente ($x$, $y$ o $z$). $V_{\text{Back}}$ es el potencial del punto adyacente más cercano al origen y $V_{\text{Forward}}$ el del punto más lejano al origen. Para cada componente se distinguen tres alternativas de calculo que se aplican dependiendo de la interfase a la que pertenece $P$, por ejemplo para determinar $E_x$ las tres posibilidades son:

*Diferencia central*: $$E_x(P) = \frac{V_{x\,Back} - V_{x\,Forward}}{2h}$$

*Diferencia directa hacia atrás*: $$E_x(P) = \frac{V_{x\,Back} - V}{h}$$

*Diferencia directa hacia adelante*: $$E_x(P) = \frac{V - V_{x\,Forward}}{h}$$

En las interfaces, se considera siempre el campo eléctrico en el lado del electrolito, salvo cuando $P$ no tiene un punto vecino en el electrolito (interfase electrodo-celda = metal-dieléctrico), donde se considera el potencial del metal y el campo eléctrico es nulo.

El caso de diferencia central, se aplica cuando la componente del campo eléctrico es continua en el punto $P$, es decir, los puntos adyacentes pertenecen al electrolito. Cuando hay discontinuidad de potencial, se usa sólo el punto vecino que pertenece al electrolito ($V_{\text{Back}}$ o $V_{\text{Forward}}$).

Las aristas y vértices de los electrodos deben cumplir la condiciona de derivada normal nula en la interfaz con los limites de la celda. El campo eléctrico se inicializa con componentes normales nulas en tales zonas, de modo que sólo se deben modificar las componentes tangenciales.

#### 4.2.2.2 Conveniencia y necesidad de un método iterativo para solucionar el sistema de ecuaciones

**a) *Consideraciones sobre convergencia***

En una primera etapa de desarrollo del modelo base (modelo de Bittner), se consideró los potenciales de electrodo constantes y conocidos, de modo que se tenía como únicas incógnitas los potenciales en el electrolito, los que se determinaban mediante un sistema de ecuaciones lineales. Para la solución del sistema de ecuaciones se consideró la alternativa de un método de solución directo como más eficiente. Para poder resolver el sistema de ecuaciones por un método directo, las incógnitas deben estar organizadas como un arreglo unidimensional. Debido a que las incógnitas se almacenan un arreglo tridimensional, tubo que aplicarse una costosa transformación lineal que mapea el dominio tridimensional en un arreglo unidimensional.

Posteriormente, Bittner consideró los potenciales de electrodo como incógnitas adicionales que debían determinarse a partir de (2.10) y (2.11). En lugar de resolver un único sistema de ecuaciones que incluyera todas las incógnitas del sistema (potenciales en el electrolito y potenciales de electrodo), Bittner separo "los espacios de solución" y incorporó un ciclo iterativo adicional para actualizar los potenciales de electrodo, una vez obtenido los potenciales de electrolito. Es decir, resolvía primero el sistema de ecuaciones original para determinar los potenciales del electrolito, considerando constantes los potenciales de electrodo, luego evaluaba los potenciales de electrodo, considerando constantes los potenciales de electrolito, y posteriormente repetía el proceso en forma iterativa hasta lograr convergencia. Este algoritmo resultó divergente y para lograr convergencia tubo que incorporar al proceso iterativo lo que denomino "factor de actualización". El método que utilizó para obtener convergencia es un método convencional para acelerar y mejorar la convergencia de sistemas de ecuaciones resueltos por un método iterativo, se denomina *Método de Sobrerelajación Sucesiva* (SOR).

Intuitivamente, en este trabajo se pensó que la "separación de espacios de solución" podía ser una técnica numéricamente inapropiada. No estaba en los objetivos del proyecto hacer un análisis numérico del algoritmo base, pero conociendo algunos conceptos básicos respecto a la convergencia de métodos iterativos para solucionar sistemas de ecuaciones [2], [7], se pudo hacer un análisis somero. Este análisis indica que, por lo menos para la estructura de las matrices utilizadas en este problema, la "separación de espacios de solución" es inapropiada. La "separación de espacios de solución" siempre introduce problemas adicionales de divergencia al algoritmo, problemas que no necesariamente se presentan cuando no se separan los espacios de solución.

Si el sistema de ecuaciones original se hubiera resuelto con un método iterativo, a pesar de haber hecho la separación de espacios de solución, aparentemente no habría diferencia con haber resuelto un único sistema de ecuaciones por un método iterativo, por lo que quizás no se hubieran presentado problemas de divergencia. Al resolver el problema utilizando un único sistema de ecuaciones que incluya todas las incógnitas, al parecer es posible que no se presente el problema de divergencia y que no sea necesaria utilizar la técnica SOR. Efectivamente, el algoritmo de cálculo utilizado en este proyecto emplea un único sistema de ecuaciones y no fue necesario utilizar la técnica SOR para lograr convergencia.

### b) *Conveniencia de un método iterativo*

El método iterativo es la mejor opción para el tipo de problemas donde se generan sistemas de ecuaciones con matrices ralas y de gran dimensión [2], [7], como sucede en este problema. Debido a que las matrices involucradas en la solución del sistema de ecuaciones lineales a resolver tiene un muy pequeño porcentaje de elementos no nulos, un algoritmo iterativo resulta más eficiente, pues con un método directo se realiza una gran cantidad de operaciones innecesarias. El método iterativo reduce el espacio de almacenamiento y logra mayor precisión que un método directo en el mismo tiempo de procesamiento. Por ejemplo, se puede cuantificar a priori el impacto de usar un método iterativo para solucionar nuestro problema. Se supone que el conjunto de todas las ecuaciones involucradas se resuelve con un sistema de ecuaciones de orden $n$. La ecuación de diferencia finita promedio para cada punto es de la forma $V_0=k_1V_1+ k_2V_2+ k_3V_3+ k_4V_4+ k_5V_5+ k_6V_6$. De modo que para cada iteración, un método iterativo realiza $6n$ multiplicaciones. La *eliminación Gaussiana* es el método directo más eficiente y se ha calculado que el número de multiplicaciones y divisiones que realiza es $n^3/3+n^2+n/3$. De este modo, en el tiempo que se resuelve directamente el sistema usando eliminación Gaussiana, se podrían realizar aproximadamente $n^2/18+n/6$ iteraciones con un método iterativo. Para los órdenes involucrados este número de iteraciones es muy grande, con este número se podría resolver muchas veces el sistema con una adecuada precisión.

Un método iterativo corrige los errores de redondeo e incluso equivocaciones de cálculo a medida que el proceso avanza, además permite controlar la precisión de los resultados. Por el contrario, un método directo (eliminación Gaussiana) propaga los errores de redondeo en forma proporcional a su orden y su índice de condición. Una ventaja adicional del método iterativo es que como condiciones iniciales se pueden utilizar los resultados obtenidos en una solución previa del sistema, con objeto de refinar los resultados.

Esta claro que en teoría la elección indicada para resolver el sistema de ecuaciones lineales es un método iterativo. Esto es seguro para una aplicación escrita en un lenguaje de programación general, como el que se utiliza en esta aplicación (C++). Cabe destacar que en la investigación de Bittner se utilizó el paquete matemático MATLAB para implementar los algoritmos. Esta plataforma esta optimizada para algoritmos vectorizados, de modo que penaliza importantemente algoritmos iterativos. Se pudo comprobar que bajo MATLAB, un método directo resulta más eficiente que un método iterativo. Pero esto es una excepción que sólo puede darse en ambientes de programación optimizados para algoritmos vectorizados, como lo es MATLAB.

### c) *Necesidad de un método iterativo*

Finalmente, el método iterativo no sólo es más eficiente y preciso que un método directo, simplifica notablemente la implementación del algoritmo al permitir el uso de una estrategia de programación orientada a objetos, la que se describe en la siguiente sección.

Considerando la aparente necesidad de resolver un único sistema de ecuaciones que incluya todas las incógnitas, para ser resuelto por un método directo, todas las incógnitas deben organizarse en un arreglo unidimensional. Al considerar los potenciales de electrodo como incógnitas, la necesaria y costosa transformación de las incógnitas ordenadas en un volumen tridimensional a un vector unidimensional sería muy difícil de implementar. La causa principal de la dificultad es que, debido a la discontinuidad de potencial, las incógnitas correspondientes a los potenciales de electrodo coinciden espacialmente con incógnitas de potencial en el electrolito. La transformación de las incógnitas que se usó en la implementación anterior ahora no sería biyectiva. La transformación se complica porque los potenciales de electrodo no ocupan completamente el volumen. Para una aplicación general como la que se implementa en es proyecto, la que contempla el uso de múltiples electrodos, no necesariamente ubicados simétricamente en el espacio y no necesariamente de las mismas dimensiones, la transformación sería extremadamente difícil de implementar.

La complicación anterior prácticamente descarta el uso del método directo. El método iterativo no se ve afectado por esta complicación porque no requiere presentarse en la forma matricial convencional para ser implementado.

#### 4.2.2.3 Estrategia de programación orientada a objetos y selección binaria de las ecuaciones de diferencia

A pesar de la conveniente formulación vectorial que se realizó en esta investigación, por eficiencia, la implementación debe ser escalar como en el modelo desarrollado por Bittner. Debido al gran numero de ecuaciones para determinar el potencial que se deben considerar en cada punto de una celda, es importante realizar eficientemente la selección de la ecuación. La selección de la ecuación a utilizar en cada punto de la grilla de muestreo se efectúa en cada iteración, pero se hace muy eficientemente basándose en una abstracción de objetos electrodos y un objeto electrolito.

El objeto electrolito define sólo una ecuación de diferencia en cada punto de su dominio, para el caso que no existen electrodos: una ecuación para el interior, una ecuación para cada cara, una ecuación para cada arista y una ecuación en cada vértice.

Los objetos electrodos determinan el potencial en su volumen, y deben seleccionar la ecuación a utilizar en cada punto, de acuerdo a las combinaciones que resultan al ser o no tangentes sus 8 caras a la celda. Como en el objeto electrolito, los puntos de la superficie se agrupan por caras, aristas y vértices, pero en este caso, en cada grupo hay más de una ecuación a considerar. Se aplica un esquema de selección binaria, basado en la condición de tangencia a la celda de cada una de sus 8 caras, condición que se determina sólo una vez, al inicializar el proceso iterativo.

El resultado neto de la clasificación binaria es que con sólo 8 consultas binarias (el estado de tangencia de las 8 caras), se determina el conjunto completo de ecuaciones en la superficie de un electrodo. El algoritmo funciona siempre que el objeto electrolito no modifique el potencial en el dominio de los electrodos, para ello utiliza una máscara que indica la presencia o no de electrodo en cada punto del dominio. Dicha máscara se inicializa antes de comenzar el proceso iterativo.

La optimización del cálculo del campo eléctrico es menos relevante, porque sólo se efectúa una vez finalizado el proceso iterativo, no obstante, también se realiza. Se aprovecha la abstracción de objetos electrodos y electrolito, y se utiliza un esquema similar de selección binaria de las ecuaciones, de acuerdo al estado de tangencia de las interfaces de los electrodos.

### 4.2.3 Optimizaciones

#### 4.2.3.1 Fórmula recursiva para el cálculo de potenciales de electrodos flotantes

En el contexto de una solución numérica global del problema electrostático, se puede obtener una importante simplificación de (2.21), la que elimina la dependencia en los potenciales de electrodos, simplificando y haciendo más eficiente el algoritmo de cálculo. Si los potenciales se calculan en forma iterativa, en el siguiente orden:

1. Potenciales de electrodo
2. Potenciales del electrolito en la interfase electrodo-electrolito
3. Potenciales del electrolito en el resto del dominio
4. Potenciales de electrodos flotantes,

entonces, al reemplazar apropiadamente (2.9) en (2.21) se obtiene:

(4.1) $$V_m^{(n)} = V_m^{(n-1)} + \frac{\sum_{i=1}^{N} \sigma_i \left[ V^{(n)}(\mathbf{x}_i + d\mathbf{s}_i) - V^{(n)}(\mathbf{x}_i) \right]}{\sum_{i=1}^{N} \sigma_i}$$

En particular, para el caso de electrolito homogéneo ($\sigma$ constante), se tiene:

(4.2) $$V_m^{(n)} = V_m^{(n-1)} + \frac{\sum_{i=1}^{N} \left[ V^{(n)}(\mathbf{x}_i + d\mathbf{s}_i) - V^{(n)}(\mathbf{x}_i) \right]}{N}$$

Notando que la serie en las fórmulas recursivas (4.1) y (4.2) corresponde a la versión discreta de la integral de flujo de la ecuación (2.15), se aprecia que $V_\mathrm{m}$ converge a un valor determinado cuando la integral de flujo es nula, lo que concuerda con la condición física de equilibrio dinámico.

Cuando se calculan los potenciales en el orden señalado, al usar (4.1) o (4.2), se tiene que no es necesario almacenar los potenciales de electrodos, lo que simplifica importantemente las estructuras de datos necesarias para almacenar los resultados. Si se desea conocer los potenciales de electrodo una vez terminado el proceso iterativo, su valor se puede conocer en términos del potencial metálico del electrodo y el potencial del electrolito en la interfase.

#### 4.2.3.2 Eliminación de la necesidad de interpolación tridimensional en coordenadas rectangulares

Para evaluar los potenciales $V(\mathbf{x}+d\mathbf{s})$ que aparecen en (4.2), por su eficiencia y simplicidad, se decide usar simplemente interpolación trilineal. Debido a que $\Delta S<h$ y al uso de electrodos rectangulares, como se muestra en la Figura 4.1(c), $\mathbf{x}+d\mathbf{s}$ siempre se ubica entre dos puntos adyacentes de la grilla de muestreo. De modo que la interpolación trilineal se reduce efectivamente a una interpolación lineal. En consecuencia, la fórmula (4.2) puede optimizarse para electrodos rectangulares en coordenadas rectangulares, incluyendo la interpolación lineal, con lo cual resulta:

(4.3) $$V_m^{(n)} = V_m^{(n-1)} + \frac{h}{N}\sum_{i=1}^{N} k_S(\mathbf{x}_i)\left[V^{(n)}(\mathbf{x}_i + h\hat{\mathbf{n}}_i) - V^{(n)}(\mathbf{x}_i)\right]$$

Considerando que no hay flujo de corriente a través una cara del electrodo tangente a la celda (2.8), en la implementación de la formula anterior resulta conveniente separar el aporte de cada cara en la serie. De modo que para cada electrodo se consideran sólo las caras no tangentes a la celda.

#### 4.2.3.3 Estimación de condiciones iniciales para potenciales de electrodos flotantes

Para acelerar la convergencia se usa un algoritmo de predicción, orientado a celdas con electrodos laminares, el que inicializa el potencial metálico de los electrodos flotantes. Todos los electrodos se ordenan en la dirección $x$, de acuerdo a su posición central $x_C=(x_1+x_2)/2$. El potencial metálico en electrodos flotantes se interpola linealmente, de acuerdo a la coordenada $x_C$, sobre la base del potencial metálico de los dos electrodos no flotantes que lo rodean. En la interpolación se considera los saltos de potencial de todos los electrodos ubicados entre los dos electrodos unipolares, debido a los potenciales de electrodo para corriente nula.

### 4.2.4 Resumen del Algoritmo de cálculo

De acuerdo al orden de cálculo preestablecido, una vez que se ha definido la geometría de la celda y se ha especificado los potenciales de electrodos unipolares, el algoritmo global se aplica como se resume a continuación:

**Inicialización:**

{

- Inicializar potencial en el volumen de la celda.
- Inicializar potencial metálico en electrodos flotantes (bipolares).
- Para cada electrodo:
  {
  - Inicializar variables que indican si sus caras son o no tangentes a la celda.
  - Inicializar puntero a función de cálculo de $DV$ en secciones izquierda y derecha, de acuerdo a tipo de electrodo.

  }
- Inicializar máscara binaria tridimensional que indica presencia de electrodos en la celda.
- *NúmeroDeIteraciones* = 0; *ErrorAbsoluto* = ∞;

}

**Iteración:**

Mientras (*ErrorAbsoluto* > *ErrorPermitido*) y (*NúmeroDeIteraciones* < *MáxIteracionesPermitido*)

{

- *ErrorAbsoluto* = 0;
- Para cada electrodo, actualizar potencial en interfases con electrolito*.
  Para cada punto $\mathbf{x}$ en la interfase electrodo-electrolito:
  $V(\mathbf{x}) = V_{\mathrm{m}} - DV(\mathbf{x});$
- Actualizar potencial en el dominio no ocupado por electrodos*.
- Actualizar potencial metálico en electrodos flotantes* (4.3).
- *NúmeroDeIteraciones* = *NúmeroDeIteraciones* + 1;

}

**Termino:**

{

- Para cada electrodo, almacenar el potencial metálico en su volumen interior.
- Calcular el campo de densidad de corriente: $\mathbf{J} = -\sigma\nabla V$
- Visualizar los resultados.

}

---

* Para cada potencial actualizado, se debe calcular la variación absoluta de potencial en la iteración, y si es superior a *ErrorAbsoluto*, se debe asignar a *ErrorAbsoluto*.

## 4.3 Generación automática de las líneas de flujo

### 4.3.1 Integración de las líneas de flujo

Para generar líneas de flujo desde la superficie de los electrodos hacia su exterior, se debe elegir adecuadamente el sentido de la integración de la línea, pues el método de integración previamente formulado siempre integra la línea en el sentido del campo vectorial. En la Figura 4.2 se muestran los casos a considerar. Resulta natural pensar en la evaluación del signo del producto punto $\hat{\mathbf{J}} \cdot \hat{\mathbf{n}}$ como criterio para determinar el sentido de integración. Sin embargo, tal criterio no se puede aplicar en las aristas de los electrodos.

Se desea determinar el sentido de integración de la misma forma en toda la superficie del electrodo, incluyendo las aristas, para ello se concibió el siguiente algoritmo. A partir de un punto semilla $\mathbf{x}_0$, posicionado sobre la superficie del electrodo, se estima un punto adyacente $\mathbf{x}^*$ en el sentido del campo vectorial. $\mathbf{x}^*$ debe ser suficientemente cercano a la superficie del electrodo como para estar contenido en su interior, si es que el campo se orienta hacia el electrodo. Luego, si el vértice $\mathbf{x}^*$ esta contenido en el interior del electrodo, la integración de la línea se realiza en contra del campo, de lo contrario se efectúa a favor del campo. Considerando la Figura 4.1(a), se tiene que un punto $(x,y,z)$ esta contenido en el interior de un electrodo cuando se cumplen las tres condiciones siguientes: $x_1<x<x_2$, $y_1<y<y_2$, y $z_1<z<z_2$.

Cuando la integración se realiza en contra del campo, básicamente se aplica el mismo algoritmo de integración descrito en la sección 3.2.2, sólo se debe reemplazar *d* por *–d* en la ecuación (3.2).

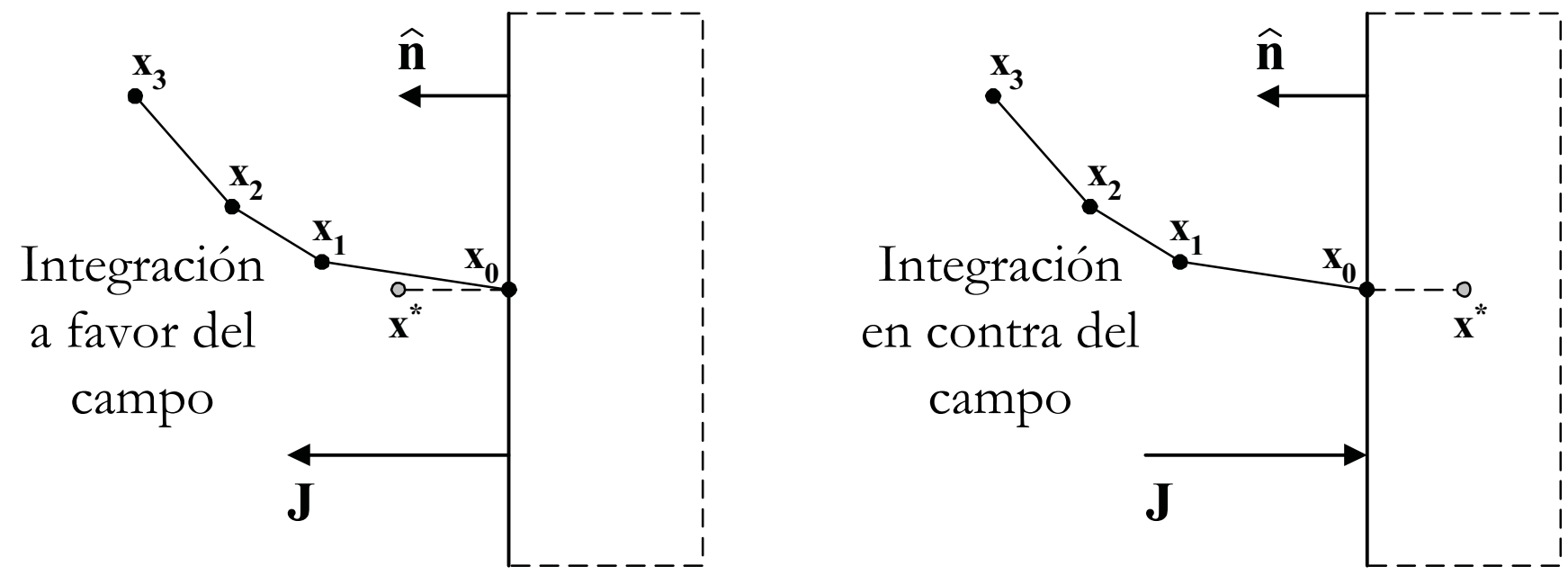

Figura 4.2: Determinación del sentido de integración de líneas de flujo generadas en la superficie de electrodos. Vista de la sección transversal de un electrodo.

El vértice $\mathbf{x}^*$ se calcula mediante la siguiente ecuación:

(4.4) $$\mathbf{x}^* = \mathbf{x_0} + \delta\, h\hat{\mathbf{J}}(\mathbf{x_0})$$

donde $h$ es la unidad de discretización del dominio y $\delta$ es un factor suficientemente pequeño (en la práctica resultó adecuado $\delta$=0,2).

### 4.3.2 Generación automática de puntos semilla y control de la cantidad de líneas de flujo visualizadas

Es fundamental que el usuario pueda especificar de alguna manera la cantidad de líneas de flujo visualizadas, para ello se considera la especificación de una densidad de líneas de flujo por unidad de superficie $\rho_S$ [líneas/$h^2$]. Esta densidad generaría una determinaría cantidad de puntos semilla para cada electrodo, los que se deben distribuir homogéneamente en su superficie. Por conveniencia, la densidad se define adimensional, se consideran las longitudes en términos del número de intervalos de discretización $h$.

Se debe resolver el problema de la distribución homogénea de una determinada cantidad de puntos sobre la superficie del electrodo, la que debería generar una grilla rectangular de puntos semilla en la superficie de cada cara del electrodo. Si dicha grilla se determinara separadamente para cada cara, el problema sería trivial. Sin embargo, en las aristas del electrodo las grillas de las caras adyacentes no necesariamente serían coincidentes, por lo que no se lograría una sensación visual de distribución homogénea de puntos semilla en la superficie del electrodo. Por lo anterior, se plantea el problema de obtener una grilla para la superficie total del electrodo. El problema requiere la determinación del número de puntos semilla que se ubicaran sobre las aristas, los que determinarán grillas de muestreo uniformes en el interior de las caras.

Considerando que en unidades de discretización $h$, las dimensiones del electrodo son: $I$ en la dirección $x$, $J$ en la dirección $y$, $K$ en la dirección $z$. Se desea determinar el número de puntos semilla sobre las aristas, siendo estos $n_{\mathrm{i}}$, $n_{\mathrm{j}}$, y $n_{\mathrm{k}}$, en las dirección $x, y, z$ respectivamente (Figura 4.3).

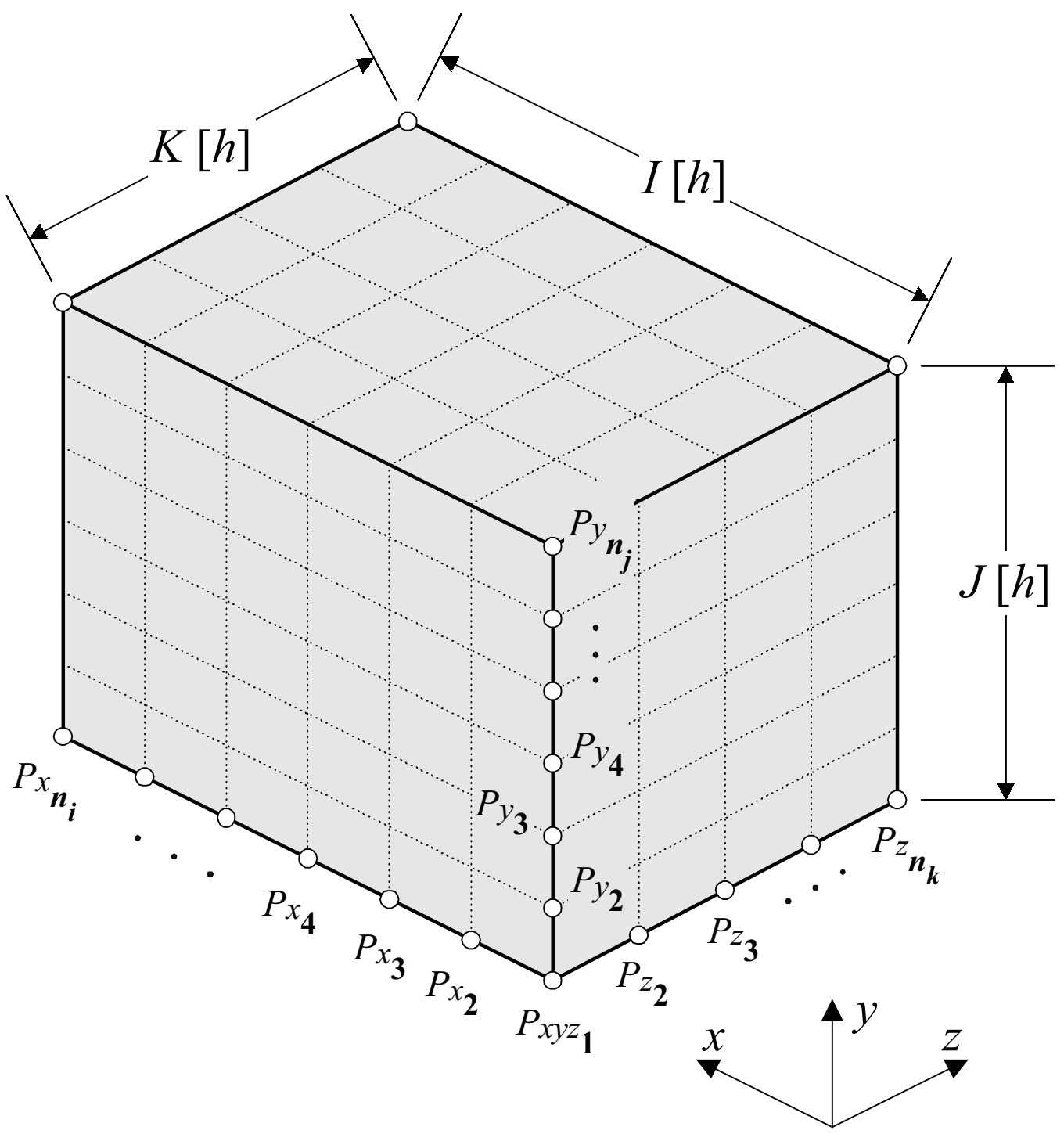

Figura 4.3: Distribución homogénea de puntos sobre la superficie de un electrodo.

El procedimiento detallado para obtener la solución de este problema se incluye en el Anexo A (página 102), la solución adaptada se resume a continuación.

$$
\begin{aligned}
n_i &= \max\left\{3, \operatorname{int}(n_i^* + 0.5)\right\}, \\
n_j &= \max\left\{3, \operatorname{int}(n_j^* + 0.5)\right\} \text{ y} \\
n_k &= \max\left\{3, \operatorname{int}(n_k^* + 0.5)\right\}
\end{aligned}
\tag{4.5}
$$

donde int es la función que retorna la parte entera de un numero real,

$$
\left.
\begin{aligned}
n_i^* &= \left( (I+J+K)\frac{\rho_8}{4} + \sqrt{\rho_S - \rho_{\min}} \right) I, \\
n_j^* &= \left( (I+J+K)\frac{\rho_8}{4} + \sqrt{\rho_S - \rho_{\min}} \right) J, \\
n_k^* &= \left( (I+J+K)\frac{\rho_8}{4} + \sqrt{\rho_S - \rho_{\min}} \right) K
\end{aligned}
\right\} \rho_S \geq \rho_8,
$$

$$
\rho_8 = \frac{4}{IJ + JK + IK}, \text{ y}
$$

$$
\rho_{\min} = \frac{4(IJ + JK + IK) - (I+J+K)^2}{(IJ + JK + IK)^2} = \rho_8 - \left( \frac{I+J+K}{IJ+JK+IK} \right)^2
$$

La solución garantiza que tanto $n_i$, $n_j$, y $n_k$ sean mayores a 2, independientemente de la densidad que especifique el usuario, lo que asegura un mínimo de 3 puntos semilla en cada arista del electrodo. Esta solución difiere de la obtenida en el Anexo A, porque en (4.5) se utiliza una cota mínima de 3, a diferencia de la cota de 2 usada en (A.9). Este cambio no afecta la validez de la solución, se hizo para asegurar que en el interior de las caras se coloque al menos un punto semilla.

En el proceso de generación de líneas de flujo, se utiliza un valor común de ρ para todos los electrodos y los parámetros $n_i$, $n_j$, y $n_k$ se obtienen para cada electrodo. Para evitar generar inútilmente dos líneas de flujo en cada punto de la grilla que coincide con alguna arista del electrodo, en el proceso de generación de líneas de flujo, la grilla se recorre sólo por el interior de cada cara, luego se recorren los interiores de cada arista sola una vez (sin incluir los vértices) y finalmente los 8 vértices.

### 4.3.3 Control de la longitud de las líneas de flujo

Después de obtener resultados preliminares, se consideró conveniente el uso de mecanismos de control de longitud adicionales al establecido por el esquema básico de integración (sección 3.2.2, página 29).

#### 4.3.3.1 Eliminación del efecto zigzag

Cuando las líneas de flujo atraviesan regiones donde el campo mantiene una dirección homogénea, pero se producen discontinuidades debido a que invierte su sentido en forma abrupta, suelen producirse efectos oscilatorios indeseados. La integración de la línea se concentra en la región de discontinuidad, porque en cada paso de integración se invierte el sentido del campo y con ello el sentido de la integración de la línea. A este fenómeno indeseado se le denominó *efecto zigzag*. Como el campo mantiene una dirección homogénea, el paso de integración tiende a concentrarse en $d_{max}$, por lo que la integración de la línea continua indefinidamente. Para el caso de las celdas en estudio, en condiciones normales de operación, el efecto zigzag se produce principalmente en las proximidades de electrodos unipolares, como se ilustra en la Figura 4.4(a).

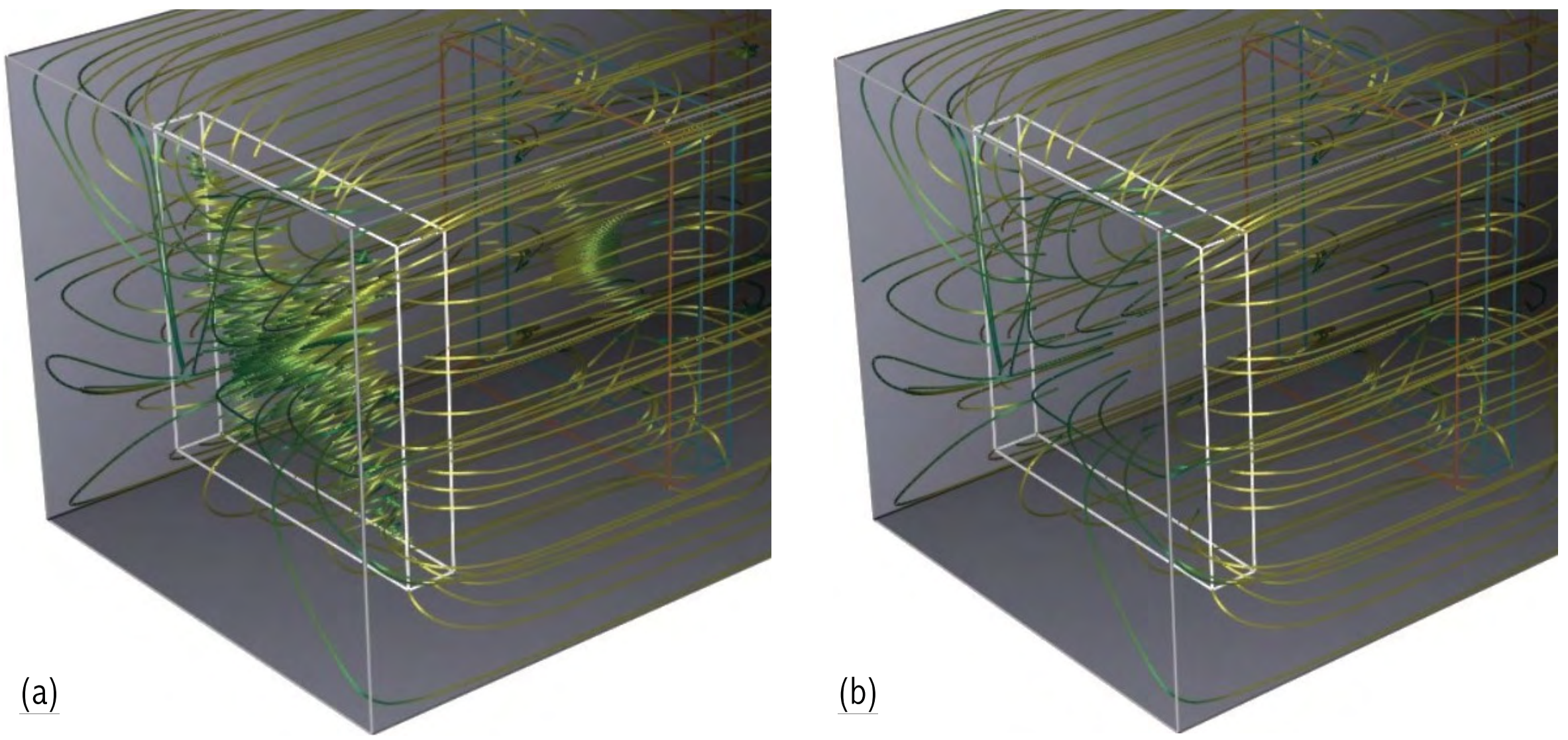

Figura 4.4: Efecto "zigzag"; en regiones donde el campo vectorial se invierte abruptamente o es nulo, se pueden producir oscilaciones indeseadas de las líneas de flujo. Esto ocurre en (a), en el interior de electrodos rectangulares. En (b), se aplica estrategia de truncado de líneas de flujo cuando su curvatura cambia abruptamente sobre 90º, lo que elimina el efecto zigzag.

Dependiendo del espesor del electrodo y del tamaño máximo de paso permitido, las líneas podrían ser truncadas por el esquema básico de integración, debido a que en el interior de los electrodos se almacena campo nulo. Aunque no se trunquen, el efecto puede pasar inadvertido si las oscilaciones quedan restringidas al interior de los electrodos.

Como el efecto se podría producir en cualquier región de la celda que presente tal tipo de discontinuidad, y como los recursos desperdiciados para almacenar e intentar dibujar los segmentos de línea indeseados pueden ser importantes, es necesario impedir la formación de estas oscilaciones. Para ello se incorporó una condición adicional que trunca la integración de las líneas de flujo. Esta condición consiste en calcular el producto punto del campo normalizado, en cada par de vértices consecutivos generados por la integración, y truncar la línea si dicho producto es negativo. La condición tiene el efecto de restringir las variaciones de la curvatura de la línea de flujo en cada paso de integración a menos de 90º, lo que consigue el efecto deseado. Como en el proceso normal de integración se debe obtener el campo normalizado en cada vértice, el costo adicional de la condición de truncado es un producto punto. En la Figura 4.4, se muestra un caso práctico donde se produce el efecto y como es eliminado usando la condición de truncado incorporada.

#### 4.3.3.2 Truncado de las líneas de flujo para evitar que atraviesen electrodos bipolares

Debido a que se contempla la generación de líneas en la superficie de todos los electrodos y al hecho que las líneas de flujo tienden a atravesar los electrodos bipolares, cuando los electrodos bipolares se alinean paralelamente se puede producir una excesiva cantidad de líneas de flujo. Se desea poder controlar en cierto grado la densidad de líneas de flujo a partir de la densidad superficial de líneas generadas, para ello es necesario evitar que las líneas atraviesen los electrodos. Lo anterior requiere que para cada vértice generado en el proceso integrativo, se determine si esta contenido en el interior de cualquier electrodo, cuando se detecta un vértice en el interior de un electrodo se termina la integración.

#### 4.3.3.3 Proporcionando grado de libertad para que el usuario controle la longitud de las líneas de flujo

Adicionalmente, se determina el uso de cotas de longitud máxima para las líneas de flujo, se definen dos cotas; una para las líneas generadas desde electrodos unipolares y la otra para las líneas generadas desde electrodos bipolares. Este mecanismo de control requiere que en el proceso integrativo se acumule la cantidad $|\mathbf{\Delta x}|$ definida en la sección 3.2.2. El objetivo de este control, es proporcionar cierto grado de libertad al usuario para que controle la longitud de las líneas de flujo, lo que puede resultar útil para realizar algún tipo de análisis.

## 4.4 Herramientas complementarias de visualización

### 4.4.1 Visualización del depósito de cobre

Esta herramienta se orienta a la visualización cualitativa del depósito de cobre en la superficie de los electrodos, pues es uno de los objetivos de análisis más relevante. Como se ha dicho, el depósito de cobre se relaciona directamente, con la componente de la densidad de corriente normal a la superficie catódica. Resulta entonces adecuado el mapeo en color de tal magnitud en la superficies catódicas. Para proporcionar una herramienta de análisis más completa, se considera también el mapeo de la componente de la densidad de corriente normal, en las superficies anódicas. La condición de operación de referencia que se estudia es; la corriente entrando a los cátodos y saliendo de los ánodos. Consecuentemente, la cantidad mapeada en la superficie de los cátodos es $-\mathbf{J}\cdot\hat{\mathbf{n}}$ y en los ánodos $\mathbf{J}\cdot\hat{\mathbf{n}}$, siendo $\hat{\mathbf{n}}$ el vector superficial unitario orientado hacia el electrolito.

Se definió un modelo de electrodo unipolar que se puede orientar en cualquier dirección ortogonal. Sin embargo, en la implementación del modelo de electrodos bipolares, se los restringe a una orientación predeterminada (ortogonales al eje $x$), lo que es suficiente para el análisis de celdas prácticas con electrodos laminares. Por tal motivo, en principio sólo se considera el mapeo en color de la densidad de corriente en las "caras principales", las caras ortogonales a la dirección preestablecida. Si el interés se concentra en electrodos laminares, es conveniente esta restricción, porque en los bordes de los electrodos se pueden usar colores sólidos predeterminados que identifiquen el tipo de electrodo.

La técnica de mapeo se implementa mediante las capacidades de aplicación de textura de la librería gráfica OpenGL. Para lograr el objetivo deseado, las caras principales se subdividen en un arreglo rectangular de polígonos, y para cada vértice generado, se debe especificar como coordenada de textura la densidad de corriente correspondiente, adecuadamente escalada. El escalamiento se hace de modo de utilizar completamente el mapa de color en cada cara principal.

La calidad y precisión de la visualización depende de la cantidad de polígonos en que se subdivide cada cara principal. Para que el usuario pueda especificar la resolución de la subdivisión de una manera uniforme para todos los electrodos, se considera el uso de una densidad de vértices por unidad de superficies $\rho_S$ [vértices/$h^2$]. Tal densidad genera una grilla de vértices, de la misma forma como se generan los puntos semilla que generan líneas de flujo. Planteando el problema de distribución homogénea de puntos como en el caso de los puntos semilla, se obtiene que el número de vértices por lado es $H\sqrt{\rho_S}$, donde $H$ es la longitud del lado respectivo en unidades de discretización $h$.

## 4.4.2 Extractor de rebanadas

Una herramienta simple y muy útil la constituye el extractor de rebanadas [9]. Esta herramienta consiste en un plano, el cual puede ser ubicado por el usuario en el espacio tridimensional. Sobre el plano se mapea una cantidad escalar relevante codificada en color, la que se obtiene en la posición que ocupa el plano en el interior de la celda. Los datos mapeados se interpolan sobre el plano para visualizar una imagen.

Para la aplicación se dispondrá de tres extractores de rebanadas, uno por cada eje coordenado. Cada extractor es ortogonal a su eje coordenado y el usuario lo puede desplazar a lo largo del eje, entre los límites de la celda. Como cantidades escalares a mapear en color se consideran el potencial eléctrico y la intensidad de la densidad de corriente. Cada uno de los extractores de rebanadas es un polígono texturizado y se implementa de la misma forma que la herramienta utilizada para visualizar el depósito de cobre. Como en la herramienta anterior, se utiliza una densidad de vértices por unidad de superficies $\rho_S$ [vértices/$h^2$], la que será aplicada a los tres extractores de rebanadas.

### 4.4.3 Cursor tridimensional

Como complemento a las herramientas visuales cualitativas, es necesaria una herramienta cuantitativa que permita conocer las magnitudes de los datos en el volumen de la celda. Una herramienta simple y útil consiste en un cursor tridimensional. El cursor actúa como una sonda que entrega el valor de los datos correspondientes a su posición en la celda. Se deben proporcionar controles adecuados para que el usuario mueva el cursor interactivamente en el volumen de la celda.

# 4.5 Consideraciones sobre interactividad

## 4.5.1 Sistema de visualización 3D

### 4.5.1.1 Modelo de Cámara

El rendering optimizado de líneas de flujo hace posible contemplar una aplicación interactiva. La interactividad se logra mediante un adecuado modelo de cámara para que el usuario se desplace en el espacio tridimensional. El modelo elegido para esta aplicación se ilustra en la Figura 4.5. La posición de la cámara define la posición del observador. Se define un objetivo de referencia (punto de enfoque), hacia donde apunta la cámara. La posición de la cámara respecto al punto de enfoque se define en términos de tres ángulos de rotación; elevación ($\phi$), azimut ($\theta$) y torsión ($\omega$). En el plano de proyección (pantalla), sólo se visualiza la porción del espacio contenida en el volumen visual, el que permanece estático respecto al sistema de coordenadas del observador (Figura 4.5b).

Como grados de libertad para el usuario se considera el control de: los ángulos de rotación de la cámara, la distancia al punto de enfoque (*zoom*), el movimiento del punto de enfoque en forma perpendicular al plano de vista (plano de proyección) y un modo de autoenfoque, el que se describe más adelante.

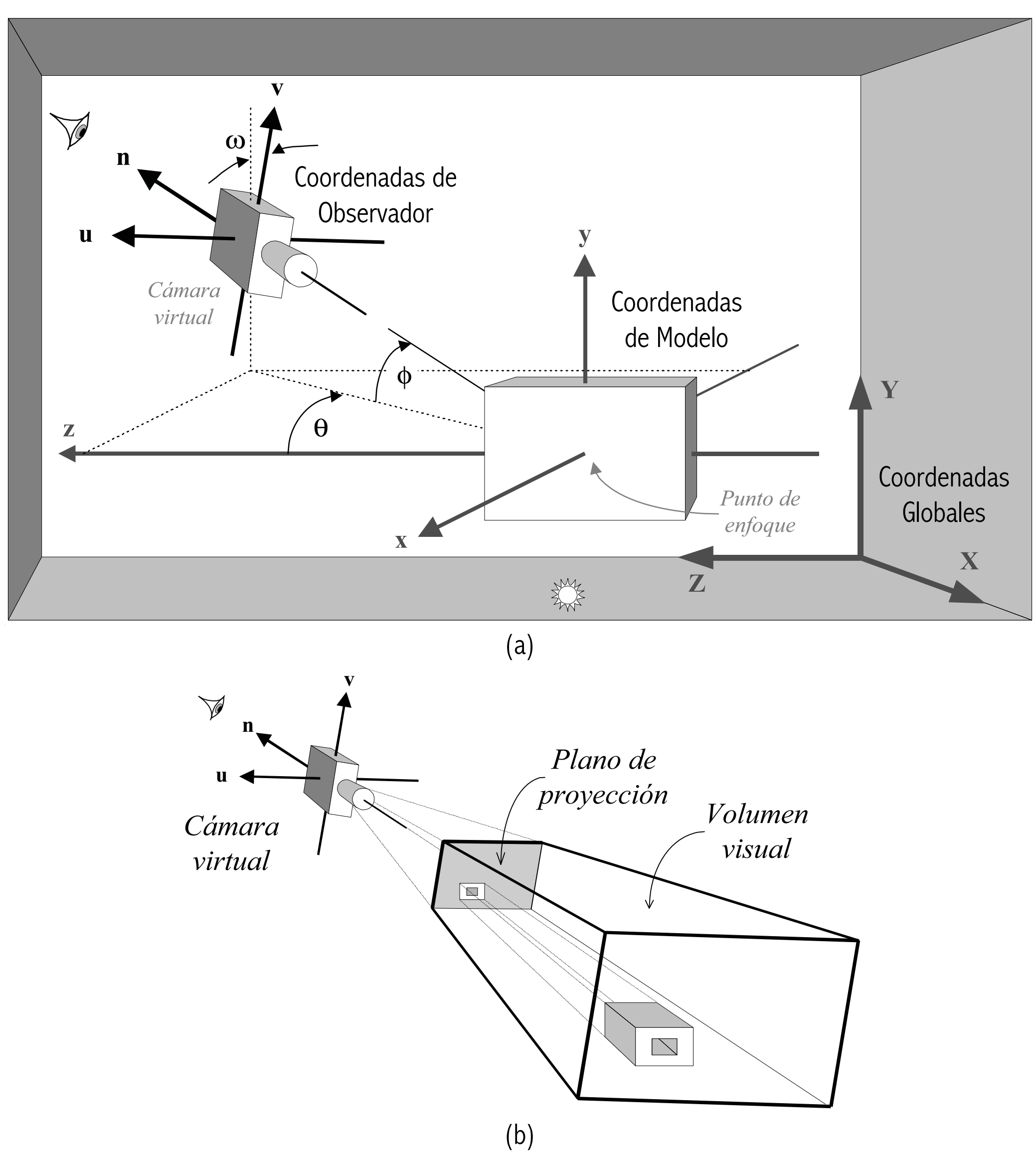

Figura 4.5: Modelo de cámara. (a) Grados de libertad de movimiento de la cámara. (b) Volumen visual.

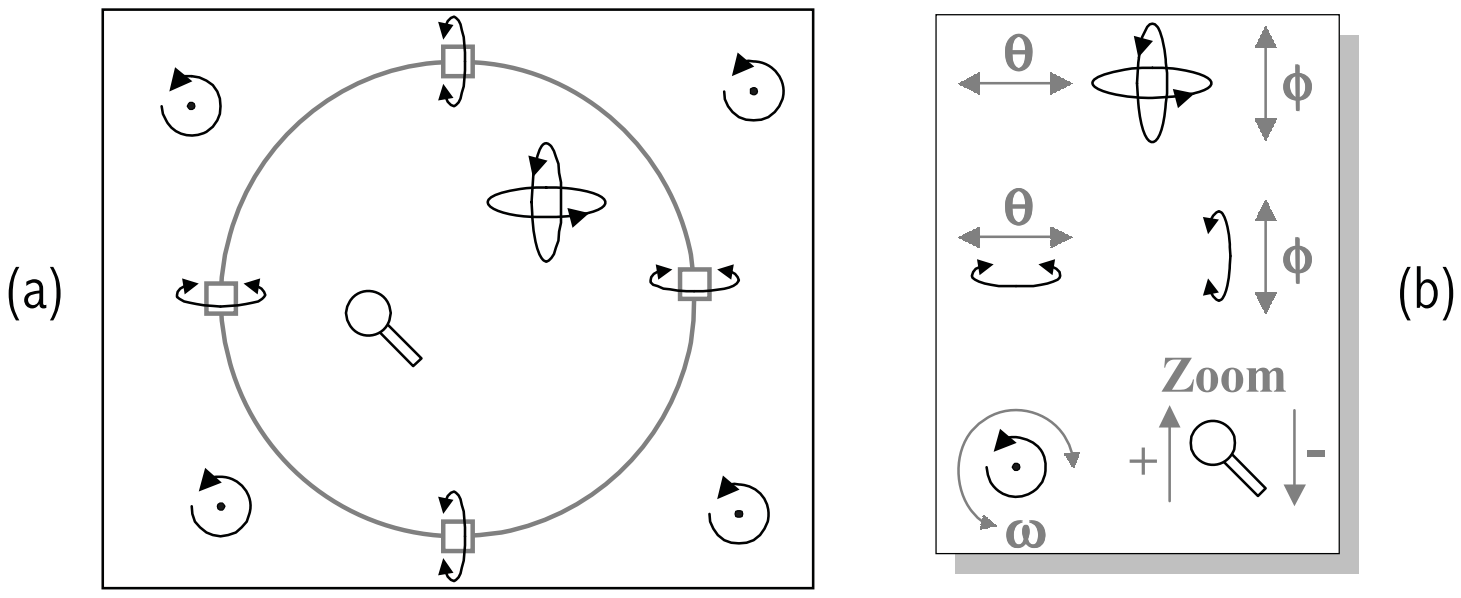

Figura 4.6: Abstracción de “trackball” para controlar desplazamientos de la cámara con el mouse. (a) Representación en pantalla. (b) Tipos de movimientos y cursores asociados.

#### 4.5.1.2 Control de los movimientos de la cámara mediante el mouse

Para una adecuada interactividad, resulta muy conveniente controlar los movimientos de la cámara con el mouse. Para este propósito se consideró la implementación de un “trackball virtual”, el que se muestra en pantalla básicamente como una circunferencia (Figura 4.6a). Para mover el trackball, se debe arrastrar el mouse con el botón principal presionado, previamente se debe haber habilitado un modo de control de rotaciones de la cámara, el que visualiza el trackball en pantalla.

El trackball se puede desplazar verticalmente, horizontalmente o ser rotado en forma perpendicular a la pantalla. Estos tres grados de libertad permiten controlar los tres ángulos de rotación de la cámara. Como sólo existen dos grados de libertad para mover el mouse, para el movimiento del trackball se utiliza un criterio conveniente basándose en la posición inicial del puntero del mouse, en el momento de comenzar a arrastrarlo. Cuando el puntero del mouse se encuentra inicialmente sobre el trackball (en el interior de la circunferencia), el mouse realiza desplazamientos verticales y horizontales del trackball, en forma simultanea. Como se aprecia en la Figura 4.6a, existe la posibilidad de colocar inicialmente el mouse sobre uno de cuatro cuadrados que forman parte del trackball. En estos casos, sólo se realiza un desplazamiento horizontal o vertical. En cambio, cuando el puntero del mouse se encuentra inicialmente en la periferia del trackball (en el exterior de la circunferencia), el mouse se usa para rotar el trackball en forma perpendicular a la pantalla.

Para que el usuario no se confunda, mientras esta activo el trackball, el cursor del mouse toma una forma representativa del tipo de movimiento que este efectuando el trackball. Si sólo se mueve el mouse, sin mantener presionado el botón principal, el cursor cambia de forma de acuerdo a su posición en la pantalla, para indicar el tipo de movimiento de trackball que se iniciará al presionar el mouse. En la Figura 4.6(b) se muestra la forma básica de los cursores elegidos y los tipos de movimientos de cámara asociados.

Como complemento al trackball, se contemplan otros modos que permiten manipular la cámara mediante el mouse. Por ejemplo, al activar el modo zoom, los desplazamientos verticales del mouse se utilizan para realizar acercamientos o alejamientos. Debido a la geometría rectangular, resultan muy útiles controles que ajusten la cámara, para conmutar en forma directa a cada una de las vistas ortogonales de la celda. etc.

Para facilitar el acceso a los distintos modos de operación de la cámara, resulta muy conveniente una barra de herramientas con botones representativos de los distintos modos de operación, los que pueden ser seleccionados rápidamente con el mouse. Entre otros botones, es útil por ejemplo uno que produzca cambios del ángulo de azimut en 180º. Si se esta enfocando un electrodo, este botón permite enfocar rápidamente la cara opuesta del electrodo, esto resulta particularmente útil con electrodos laminares.

#### 4.5.1.3 Selección de electrodos con el puntero del mouse

Para distintos propósitos, resulta muy útil poder seleccionar electrodos con el puntero del mouse. Con la librería gráfica OpenGL, esto puede hacerse en forma muy simple. Además del modo de operación por defecto (Render), OpenGL proporciona un modo de operación denominado "Selection" orientado a la selección de objetos en la pantalla. La capacidad de selección interactiva de electrodos se utiliza básicamente para obtener información individual y modificar parámetros individuales de electrodos, como su geometría.

#### 4.5.1.4 Autoenfoque

Por defecto, el punto de enfoque se inicializa en el centro de la celda. Considerando la capacidad para seleccionar un electrodo con el mouse, es útil proporcionar un modo de autoenfoque, el que seleccione en forma óptima el punto de enfoque y el volumen visual con sólo seleccionar un electrodo. Para implementar esta característica, se resolvió el problema de autoenfoque de una esfera de posición y radio conocido (Figura 4.7). Luego, para auto enfocar cualquier objeto, basta con determinar una esfera límite que contenga el objeto, lo que es muy simple de realizar para un paralelepípedo (electrodo o celda).

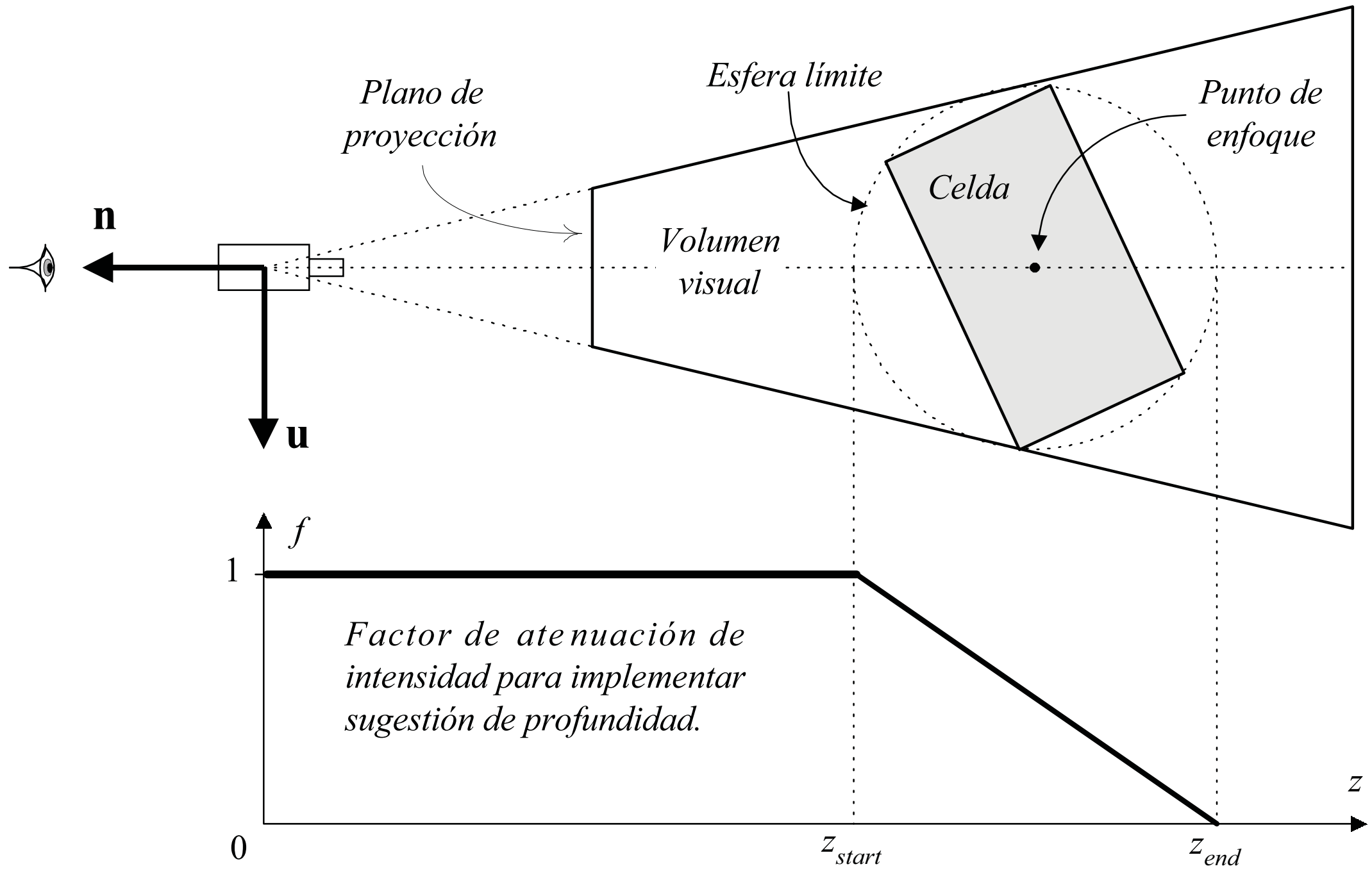

Figura 4.7: Geometría en la que se basa la implementación de autoenfoque y sugestión de profundidad.

## 4.5.2 Sugestión de profundidad

OpenGL proporciona varias alternativas para implementar la característica de sugestión de profundidad comentada en la sección 3.3.5. La forma más simple de hacerlo es utilizando niebla (fog) en modo RGBA. En modo RGBA, el color con niebla resulta de la siguiente expresión

$$C = fC_i + (1 - f)C_f$$

El color resultante es una mezcla del color RGBA de la niebla ($C_f$) con el color RGBA del fragmento (píxel) de entrada ($C_i$), usando un factor de mezcla de niebla $f$.

Al usar color RGBA de niebla negro, $C_f$ = (0,0,0,1), el color "con niebla" se calcula efectivamente como

$$C = fC_i$$

Existen tres formas de evaluar el factor de mezcla de niebla. La forma más conveniente para esta aplicación es usar un factor de atenuación lineal, porque este modo permite implementar un mecanismo automático de sugestión de profundidad, en el sentido que siempre entrega resultados satisfactorios sin requerir ajuste de parámetros, a diferencia de lo que ocurriría con los otros modos de niebla. El factor de mezcla de niebla lineal es:

$$f(z) = \frac{z_{end} - z}{z_{end} - z_{start}}$$

donde $z_{start}$ y $z_{end}$ son parámetros ajustables y $z$ es la distancia entre el centro de vista (cámara) y el centro del fragmento.

Para implementar la sugestión de profundidad se utilizó básicamente el método ilustrado en la Figura 4.7. Los valores de $z_{start}$ y $z_{end}$ se asignan a partir de la posición del punto de enfoque y el radio de la esfera límite. Adicionalmente, se proporciona una forma simple para ajustar el grado de percepción de profundidad en forma relativa. Esta consiste en permitir que el usuario modifique las posiciones de $z_{start}$ y $z_{end}$, al expresarlas en términos relativos del radio de la esfera límite. Se proporciona entonces al usuario un par de factores para ponderar el radio de la esfera límite. Por defecto estos factores valen 1 y se tiene el caso mostrado en la Figura 4.7. Si por ejemplo, el usuario especifica un factor de ponderación 2 para $z_{start}$, $z_{start}$ se ubica 2 radios de la esfera limite delante del centro de enfoque, por lo que el factor de atenuación de intensidad $f$ disminuye a la mitad.

### 4.5.3 Segmentación interactiva

Debido a que en una celda la cantidad de electrodos y líneas de flujo puede ser considerable, es muy importante disponer de herramientas para poder restringir la visualización a un conjunto reducido de datos, y así poder eliminar problemas de oclusión de objetos de interés. Como las líneas de flujo se agrupan por electrodos, se puede considerar un simple y poderoso mecanismo de segmentación de los datos visualizados. Este se basa en la capacidad para seleccionar un electrodo con el mouse. El mecanismo de segmentación consiste en tres modos de visualización, los que se aplican en forma independiente a electrodos y líneas de flujo.

Los modos son:

- Visualizar todos los objetos (todos los electrodos y/o todas las líneas de flujo);
- Visualizar sólo los objetos relativos al electrodo seleccionado con el mouse (el electrodo propiamente tal y/o todas las líneas de flujo generadas de tal electrodo); o
- No visualizar objetos (ningún electrodo y/o ninguna línea de flujo).

Por ejemplo, se podría visualizar todos los electrodos y sólo las líneas de flujo generadas desde el electrodo seleccionado.

Además, para solucionar problemas de oclusión, el usuario puede complementar la característica de segmentación con la capacidad para explorar interactivamente el volumen de la celda y la capacidad para controlar la longitud de las líneas de flujo desde los electrodos.

### 4.5.4 Interfaz Gráfica de Usuario

Para lograr una adecuada interactividad, la interfaz gráfica de usuario juega un rol fundamental. Debe proporcionar controles para acceso rápido a todas las características usadas con mayor frecuencia. Se debe proporcionar una adecuada interfaz para el diseño geométrico de la celda y para controlar el método numérico de cálculo de potenciales (simulador). Para que el usuario tenga un control de la calidad y velocidad de la visualización, todas las características de visualización y generación de líneas de flujo se implementan como estados de operación conmutables.

## 4.6 Implementación

La implementación de la aplicación computacional formulada en este proyecto dio origen al programa denominado "EWCellCAD", diseñado para ejecutarse en sistemas operativos Windows 9x/NT [14]. La aplicación se implemento en lenguaje C++, usando el compilador C++Builder 4.0 [12].

Como ya se señalo, el código de rendering se implementó utilizando la librería gráfica OpenGL 1.1 [10], [13]. Para optimizar el código de rendering, las estructuras de datos se organizaron para minimizar los cambios de estado, se utilizaron objetos textura y se hizo uso intensivo de arreglos de vértices para minimizar la sobrecarga de llamada a funciones por vértice.

# *Resultados*

La aplicación desarrollada se evaluó en un PC con procesador Pentium II de 350MHz, con 64MB RAM. El PC se equipó con una tarjeta aceleradora gráfica Matrox Millenium G200 (8MB), la cual incluye aceleración hardware para OpenGL.

En las siguientes secciones se presentan secuencias de imágenes de resultados, organizadas como casos de estudio. En todos los casos, los electrodos se pueden diferenciar por sus colores; los ánodos son de color gris, los cátodos de color cobre y la sección anódica de los electrodos bipolares es de un tono celeste.

## 5.1 Interfaz Gráfica de Usuario

La GUI de la aplicación desarrollada se muestra en la Figura 5.1. El diseño satisface las consideraciones sobre interactividad previamente discutidas. Como se observa, a la derecha de la ventana de visualización se proporciona un acceso rápido a todos los parámetros y modos de operación, organizándolos en un conjunto de etiquetas. En la Figura 5.2, se puede ver el detalle de las etiquetas no mostrado en la Figura 5.1. A continuación se describen algunas características generales del funcionamiento de la interfaz gráfica.

Se proporciona una barra de herramientas que permite conmutar entre distintos estados de operación del mouse (parte superior derecha del área de menús), usados para manipular la cámara interactivamente. La abstracción de trackball para controlar los movimientos de la cámara se muestra en la Figura 5.5(a) y en la Figura 5.7(e) (circunferencia de color verde).

El diseño geométrico de la celda se realiza empleando los controles de la etiqueta "Geometry" en conjunto con una barra de herramientas correspondiente, la que se muestra en la parte superior izquierda del área de menús. Los botones de la barra de herramientas permiten: conmutar el estado del mouse al de selección de electrodo, redimensionar la celda, agregar a la celda los distintos tipos de electrodos (ánodo, cátodo y bipolar), y eliminar el electrodo seleccionado con el mouse. Con la etiqueta "Geometry" se puede manipular la posición y las dimensiones del electrodo seleccionado, además su potencial metálico. La posición se especifica mediante las coordenadas $(x_1, y_1, z_1)$ definidas en la Figura 4.1(a). Los controles de la etiqueta impiden que el electrodo se ubique fuera de la celda y proporcionan una útil realimentación visual de la ubicación del electrodo relativo a la celda.

La etiqueta "Simulation" proporciona acceso al cálculo numérico de potenciales y densidad de corriente. El proceso iterativo se organizo en dos ciclos anidados. Cuando se presiona el botón "Step" se ejecuta el ciclo interior, un determinado número de iteraciones. Cuando se presiona el botón "Run" se ejecuta el ciclo exterior, el que en cada iteración ejecuta el ciclo "Step". El objetivo de esta jerarquía es permitir la visualización de resultados parciales en forma eficiente. La visualización de resultados supone la disponibilidad del campo densidad de corriente. Cuando no se habilita la visualización de resultados, el campo densidad de corriente sólo se calcula al terminar completamente el proceso iterativo. Cuando se habilita la visualización de resultados, esta se actualiza una vez en cada ciclo "Step", de modo que el cálculo del campo densidad de corriente se realiza sólo una vez por ciclo "Step".

En la etiqueta "Streamlines" se dispone de controles para manipular todos los parámetros relacionados a la generación de las líneas de flujo. La especificación de longitudes máximas desde electrodos unipolares o bipolares puede emplearse como una herramienta de segmentación.

Bajo el botón "Visualization" se muestran las etiquetas "Streamlines", "Electrodes" y "Probes". Las dos primeras permiten controlar la visualización y segmentación de los objetos asociados. En la etiqueta "Probes" se concentran las herramientas de visualización adicionales para explorar interactivamente los datos. En particular, en la Figura 5.1 se muestra los controles asociados al cursor 3D.

La aplicación permite almacenar y recuperar archivos que almacenan la geometría, potenciales eléctricos y parámetros de la celda. No se almacena las líneas de flujo ni el campo de densidad de corriente pues, a diferencia del cálculo de potenciales, estas cantidades se calculan muy rápidamente, de modo que se contribuye a reducir el tamaño del archivo.

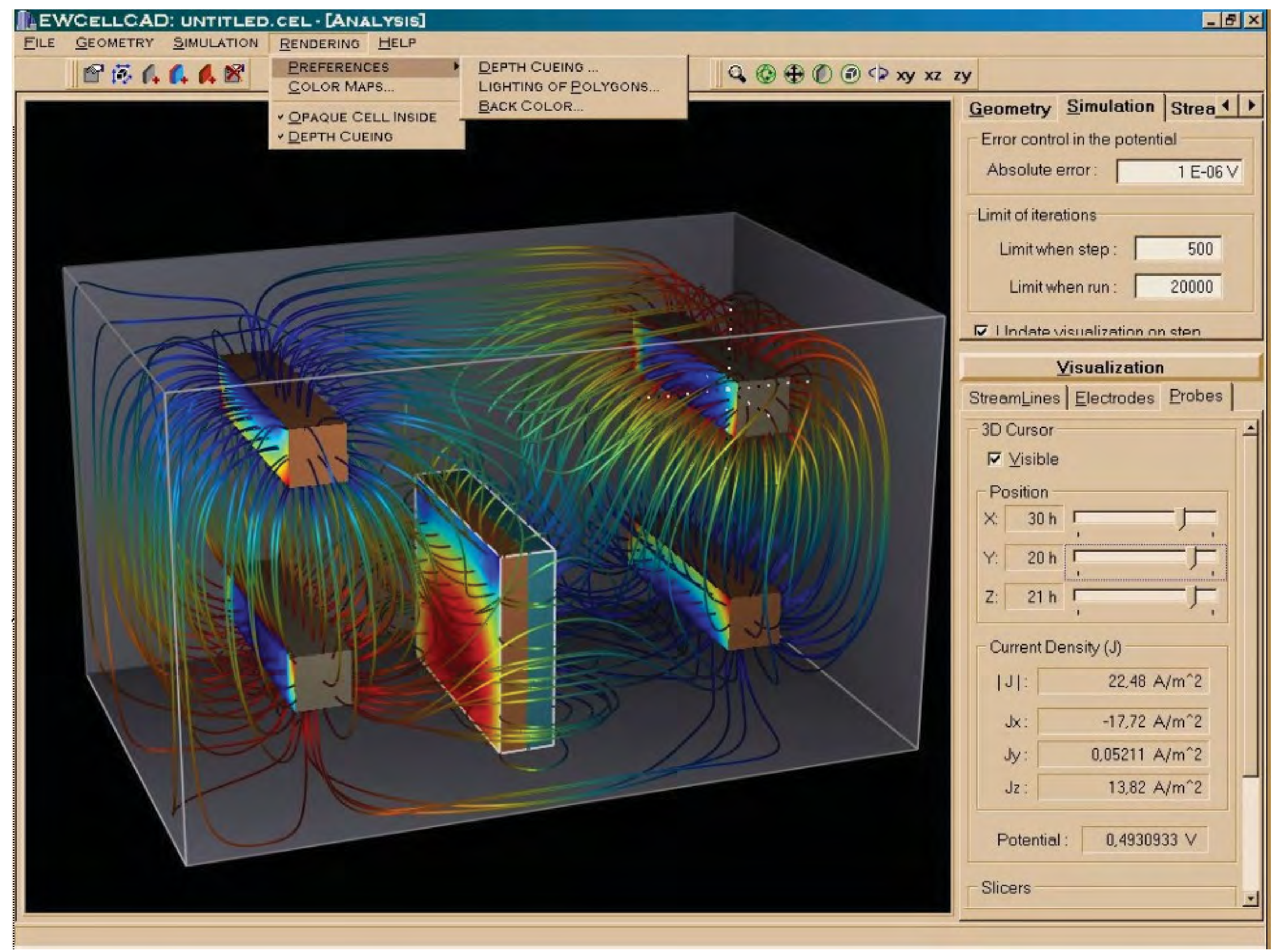

Figura 5.1: GUI. Sobre las líneas de flujo, el color mapea la magnitud relativa del potencial eléctrico. Considerando que las líneas se orientan de mayor a menor potencial y que en el mapa de color empleado (jet) el azul corresponde a las magnitudes mínimas y el rojo a las máximas, se puede determinar claramente el sentido de las líneas de flujo

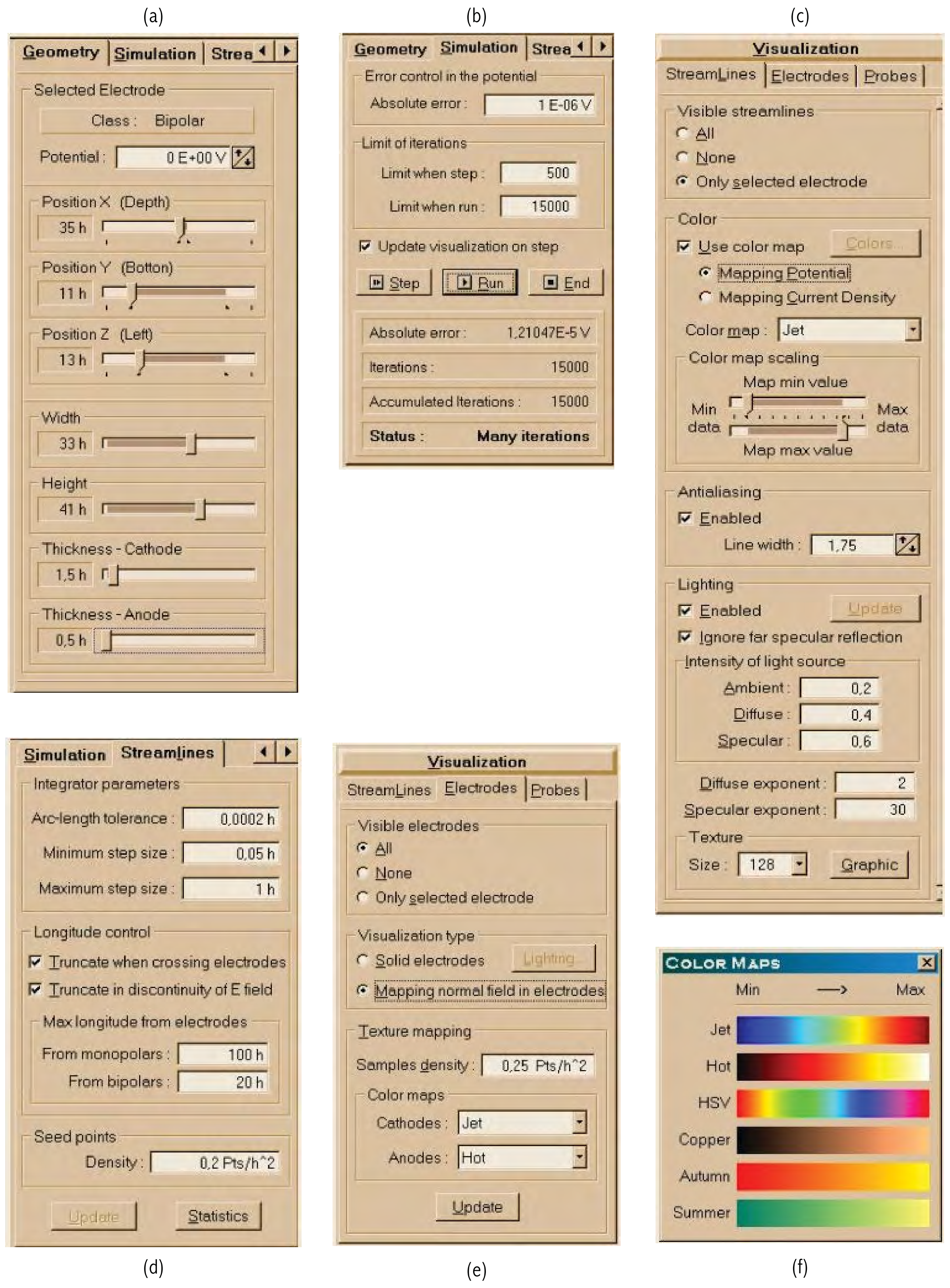

Figura 5.2: Detalle de los controles principales de la GUI y mapas de color disponibles.

## 5.2 Características visuales de las líneas de flujo

En primer lugar, en la Figura 5.3 se ilustra la efectividad y utilidad de la técnica antialiasing aplicada a las líneas. En la Figura 5.3(a), se aprecia que el efecto del alias es muy molesto y la situación empeora en el caso interactivo, debido a que al aliasing espacial se agrega el aliasing temporal. No obstante, en la Figura 5.3(b) se observa que para imágenes estáticas, la técnica antialiasing empleada resulta muy satisfactoria.

En la Figura 5.4, se muestra el mejoramiento progresivo en la visualización de las líneas de flujo al aplicar las características de iluminación, mapeo de color y percepción de profundidad. Con cada característica agregada, el mejoramiento de la visualización es evidente. El mejoramiento más dramático ocurre al iluminar las líneas, Figura 5.4(b), se hace visible la orientación espacial y mejora la percepción de la curvatura de las líneas.

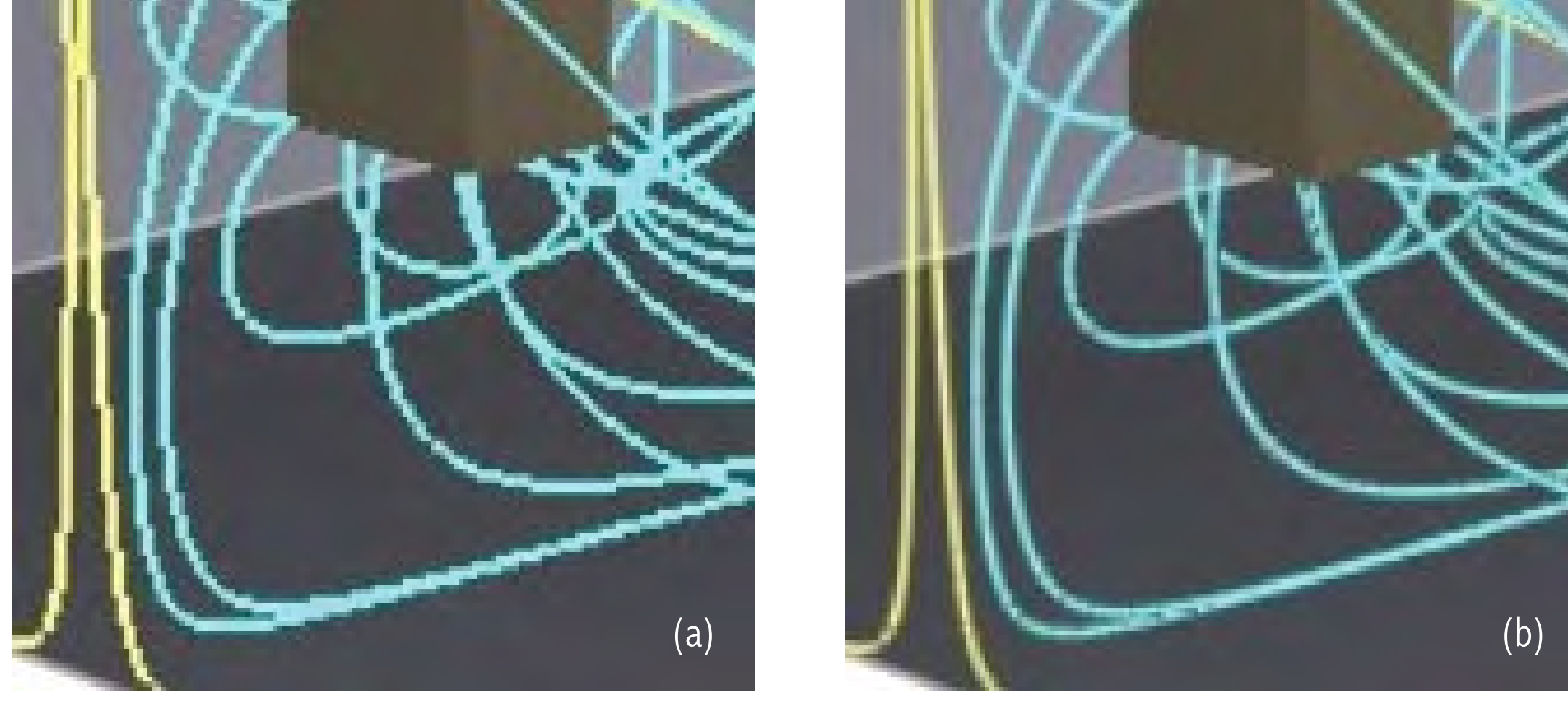

Figura 5.3: Antialiasing de líneas. En (a), las líneas de flujo exhiben alias. En (b), las líneas de flujo son suavizadas al aplicar una técnica antialiasing.

En las Figuras 5.4 (c) y (d), se utiliza color para codificar la intensidad del campo densidad de corriente sobre las líneas de flujo. Además del beneficio de la información visualizada, mejora la percepción de la ubicación espacial de las líneas de flujo, debido a la correlación espacial del color de la cantidad mapeada. Se confirma el requerimiento de colores brillantes para poder apreciar adecuadamente el brillo especular, en las regiones de color azul y rojo disminuye la percepción del brillo especular.

Una aplicación del uso de color particularmente útil es la codificación del sentido del campo vectorial, debido a que las líneas de flujo se orientan del mayor a menor potencial. En la Figura 5.1 se ilustra claramente esta aplicación.

En la Figura 5.4(d), se mejora la impresión espacial de la estructura del campo al aplicar percepción de profundidad. Los beneficios de la iluminación como de la percepción de profundidad se pueden aprovechar mucho más cuando la celda se rota interactivamente. En la interfaz de usuario se proporciona un control para ajustar la intensidad de la sugestión de profundidad. En la En la Figura 5.4(d), la atenuación de la intensidad con la profundidad fue relativamente leve para no ocultar completamente las estructuras más lejanas. Si el grado de percepción de profundidad se ajusta para ocultar objetos lejanos, estos se hacen visibles a medida que se rota la celda.

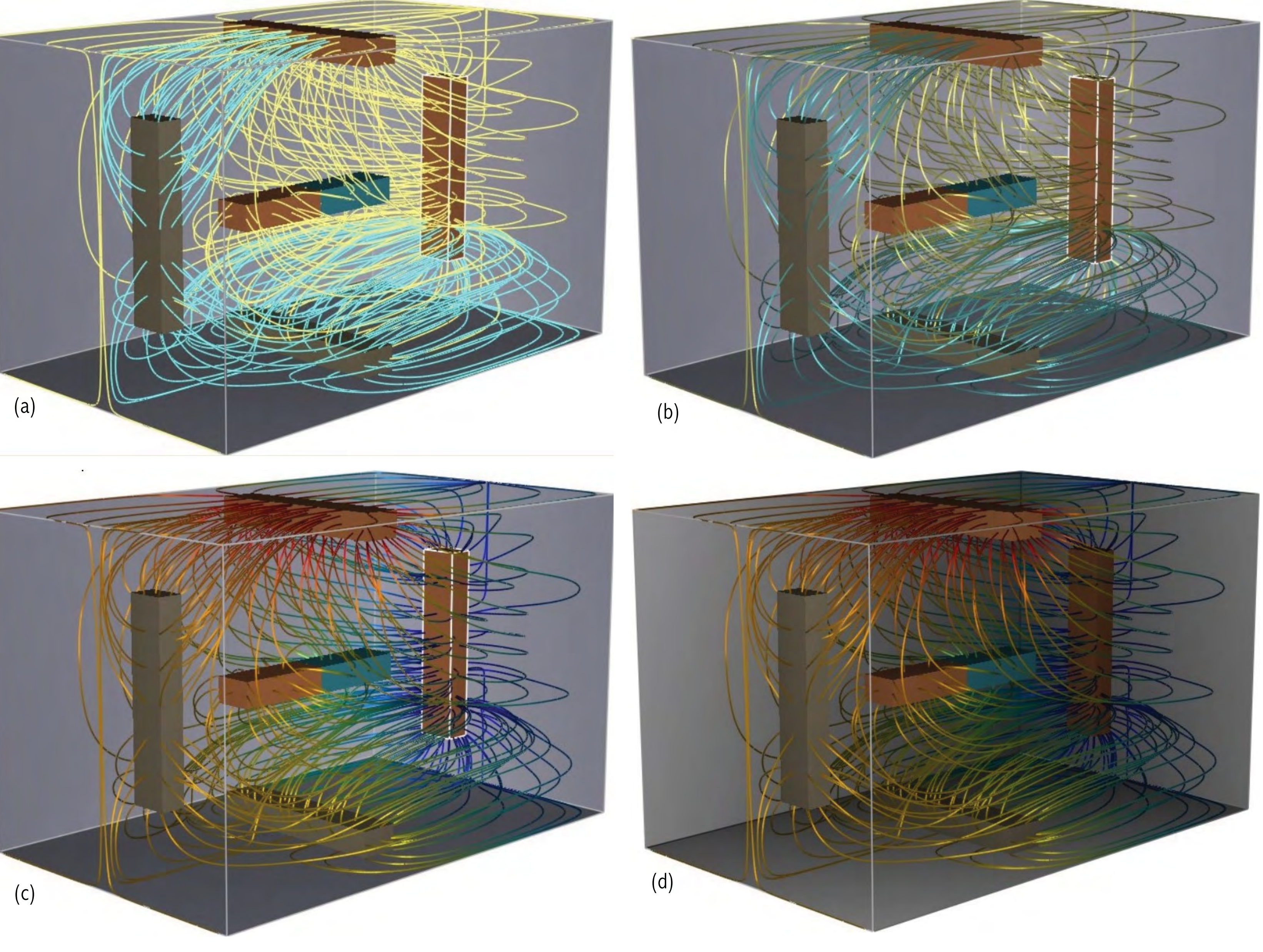

Figura 5.4: Características visuales de las líneas de flujo. En (a), las líneas de flujo no son iluminadas (se usa color constante en toda la longitud). (b) a (d) muestran el efecto de una adecuada iluminación. En (c) y (d), se usa el color para codificar la intensidad del campo vectorial. Finalmente, en (d) se incorpora sugestión de profundidad, mejorando la percepción de la ubicación espacial de las líneas de flujo

## 5.3 Celda de electro-obtención con un electrodo bipolar

Para ilustrar la utilidad de las herramientas de visualización y el correcto funcionamiento del modelo mejorado de electrodo bipolar, a continuación se realiza un simple análisis de una celda con un electrodo bipolar. En la Figura 5.5, se muestra el caso donde se aplica una diferencia de potencial de 1[V] entre ánodo y cátodo. En la en la Figura 5.6, se muestra el caso donde se aplica una diferencia de potencial de 3[V] entre ánodo y cátodo. En las líneas de flujo, el color (jet) codifica la intensidad del campo vectorial densidad de corriente. En los electrodos, se usa color (jet) en las caras principales para codificar sólo la componente normal de la densidad de corriente.

En la Figura 5.5(a), sobre las líneas de flujo el mapa de color se escala entre las magnitudes mínima y máxima de la intensidad de campo (escalamiento por defecto). Sin embargo, es difícil diferenciar las magnitudes relativas. Desde la Figura 5.5(b) en adelante, el mapa de color se escala en un rango de intensidades de campo de menor magnitud. Se aprecia mucho mejor la distribución espacial de la intensidad del campo, esta es muy fuerte en los bordes del electrodo bipolar. En la Figura 5.5(a), al hacer un acercamiento en los bordes del electrodo bipolar, se puede comprobar que existen líneas de color rojo, pero la intensidad disminuye abruptamente hacia el exterior del electrodo, por lo que el mapa de color se concentra en una región de cambio muy estrecha y no se aprecia el cambio gradual de color. En la práctica, se comprobó que en general, la capacidad para escalar arbitrariamente el mapa de color es fundamental.

Al observar las superficies catódicas en las Figuras 5.5 (a) y (b), se aprecia que la intensidad del campo se concentra en una región circular del electrodo bipolar y se podría suponer lo mismo respecto al depósito de cobre. Sin embargo, en el cátodo ocurre lo contrario. Por lo anterior, se podría deducir que las líneas de flujo que atraviesan el electrodo bipolar no llegan al cátodo, pero la alta concentración de líneas de flujo y su bajo contraste dificulta confirmar tal suposición.

En las Figuras 5.5 (c) y (d), se utiliza la capacidad de visualización selectiva de las líneas de flujo para confirmar claramente la suposición anterior. En la Figura 5.5(c), se visualizan sólo las líneas de flujo generadas desde el ánodo. Se comprueba que rodean el electrodo bipolar y al llegar al cátodo, se concentran en sus bordes. En la Figura 5.5 (d), se visualizan sólo las líneas de flujo generadas desde el electrodo bipolar y se corrobora la suposición. Se aprecia que la mayoría de las líneas de flujo se cierran en el mismo electrodo.

Si en la Figura 5.5(d), se mapea el potencial eléctrico sobre las líneas de flujo, se puede apreciar el sentido de las mismas. Con este método, se comprobó claramente que en el electrodo bipolar ocurre la reacción electroquímica contraria a la deseada, por lo que no se deposita cobre en el electrodo bipolar.

Al aplicar una diferencia de potencial de 3[V] entre ánodo y cátodo, en la Figura 5.6 (a) se aprecia claramente que esta vez la corriente atraviesa la mayor parte de la cara principal del electrodo bipolar, pero en los bordes las líneas de flujo tienden a rodearlo. Se observa que la intensidad de la densidad de corriente aumento globalmente en el volumen de la celda, pero aun se concentra fuertemente en los bordes del electrodo bipolar. Respecto al depósito de cobre, se puede apreciar una distribución relativamente más homogénea, respecto al caso de polarización débil. Sin embargo, en los bordes de los electrodos, el depósito sigue siendo mucho más débil que en el resto de la superficie.

En las Figuras 5.6 (b) a (d), se visualizan sólo las líneas de flujo generadas desde el electrodo bipolar, lo que confirma lo señalado anteriormente y permite apreciar detalles de la deformación del campo en la superficie del electrodo. Se puede apreciar que la corriente no es necesariamente normal en las caras principales del electrodo bipolar y que las líneas de flujo no se relacionan localmente entre las caras catódica y anódica.

El comportamiento cualitativo presentado por la celda concuerda con los resultados esperados cuando se usa un electrodo bipolar. Debido a la barrera energética que imponen los potenciales de electrodo del electrodo bipolar, la corriente tiende a rodearlo. El resultado es un depósito de cobre no homogéneo que se concentra en el centro del electrodo. Aun más, cuando el potencial aplicado a la celda no es suficiente para polarizar el electrodo bipolar, en el electrodo bipolar ocurre la reacción electroquímica inversa a la deseada y la corriente originada en el ánodo ni siquiera lo atraviesa, en consecuencia, no se deposita cobre en el electrodo bipolar.

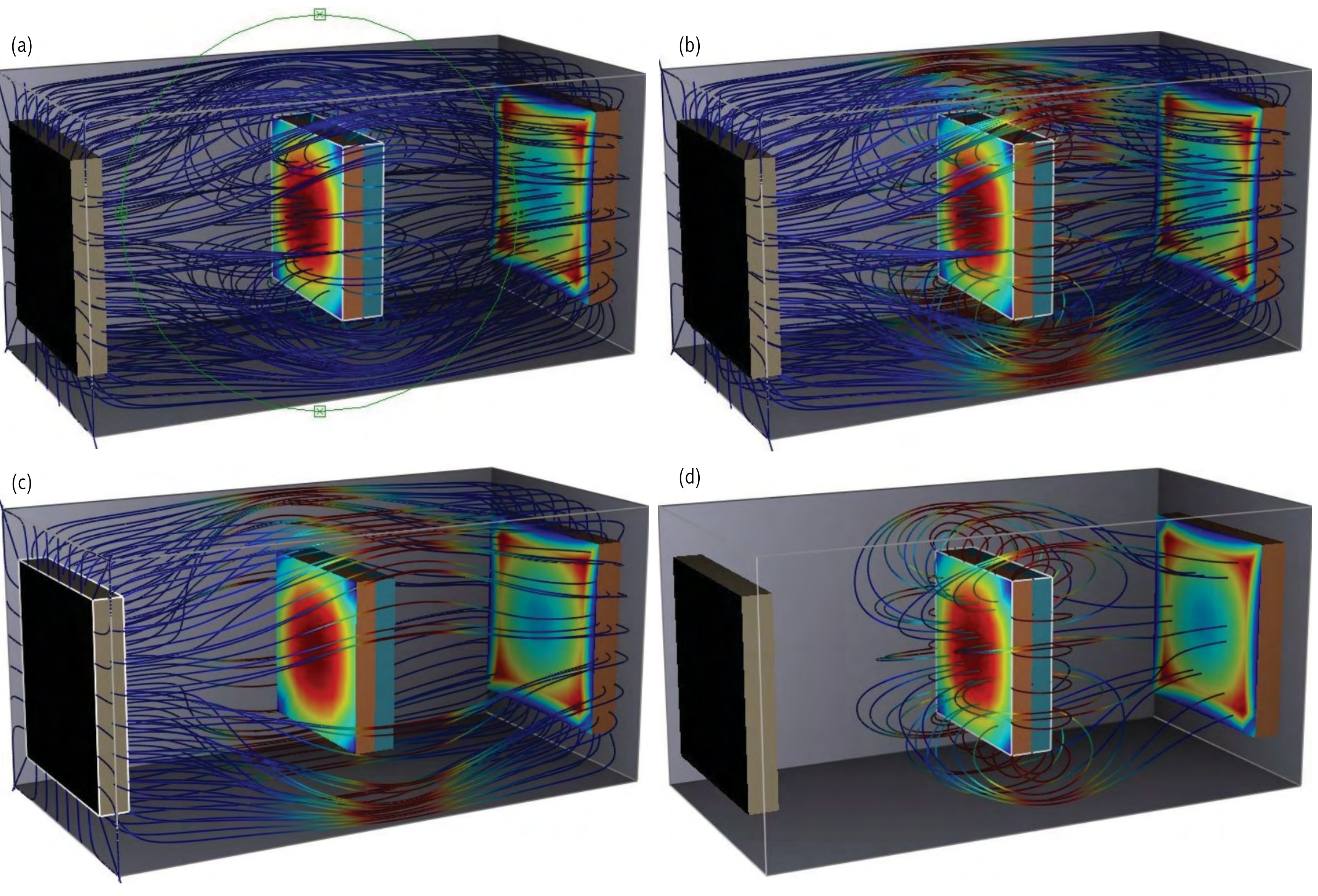

Figura 5.5: Celda de electro-obtención con un electrodo bipolar insuficientemente polarizado. Sobre la caras principales de los electrodos se usa un mapa de color que codifica la intensidad de campo normal (depósito de cobre). En (a), sobre las líneas de flujo, el mapa de color se escala entre las magnitudes mínima y máxima de la intensidad de campo. En (b), se aprovecha mejor el mapa de color, escalándolo a un rango de intensidades de campo de menor magnitud. La visualización segmentada de las líneas de flujo permite apreciar que la corriente originada desde el ánodo prácticamente no atraviesa el electrodo bipolar. En (c), se visualizan sólo las líneas de flujo integradas desde el ánodo (electrodo izquierdo). En (d), se visualizan sólo las líneas de flujo complementarias; las integradas desde el electrodo bipolar (electrodo central).

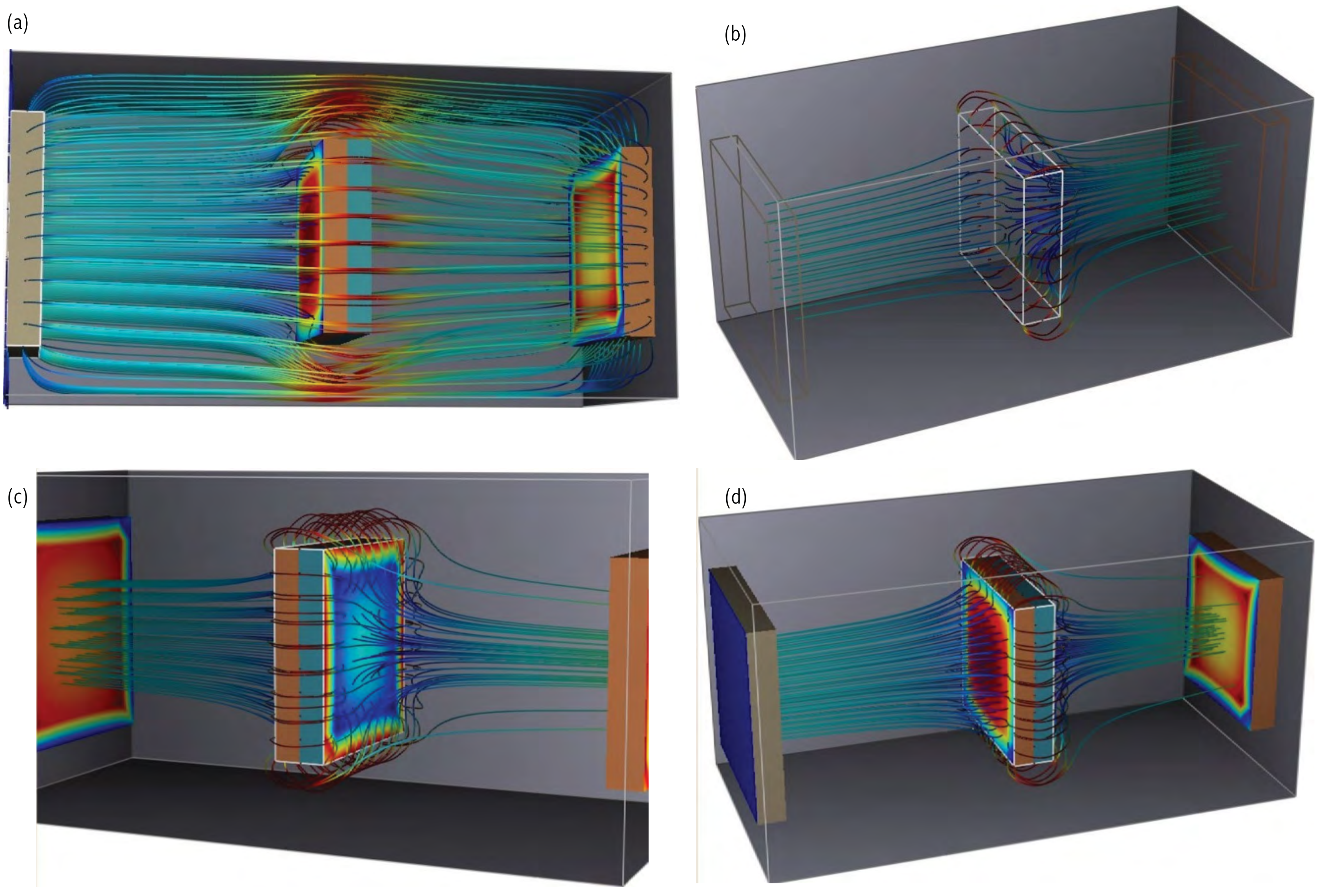

Figura 5.6: Celda de electro-obtención con electrodo bipolar polarizado. A diferencia de la celda con electrodo bipolar débilmente polarizado (Fig. 4.5), en (a) se aprecia que la corriente del ánodo atraviesa la mayor parte de la superficie del electrodo bipolar. En (b) a (d), se visualizan sólo las líneas generadas desde el electrodo bipolar, lo que ratifica lo anterior. En (c) y (d) se muestra detalles de la deformación del campo en la superficie del electrodo bipolar, en cara anódica y catódica respectivamente

## 5.4 Celda de electro-obtención asimétrica con 4 electrodos bipolares

A continuación se muestran algunos resultados obtenidos al simular una celda de electro-obtención asimétrica con varios electrodos bipolares laminares, los que ocupan parcialmente la sección de la celda.

En la Figura 5.7(a), se muestra una vista ortogonal en perspectiva desde la parte inferior de la celda, donde el color de las líneas de flujo codifica su intensidad. En la Figura 5.7(b), se muestra un plano de corte correspondiente a una celda muy parecida, analizada con la aplicación prototipo desarrollada por Bittner. Aunque con ambas visualizaciones, básicamente se puede obtener la misma información, las líneas de flujo continuas visualizan de forma más clara la estructura global del campo vectorial. En el caso tridimensional, las flechas son poco útiles porque no se puede determinar su ubicación espacial, y cuando las variaciones de la intensidad del campo son muy abruptas, las flechas se desvanecen y no aportan información.

Desde la Figura 5.7(c) a Figura 5.7(e), se ilustra la utilidad de algunas de las herramientas de visualización adicionales; segmentación y extractor de rebanadas. Finalmente, en la Figura 5.7(e) se muestra un acercamiento para ilustrar la capacidad para realizar observaciones locales más detalladas.

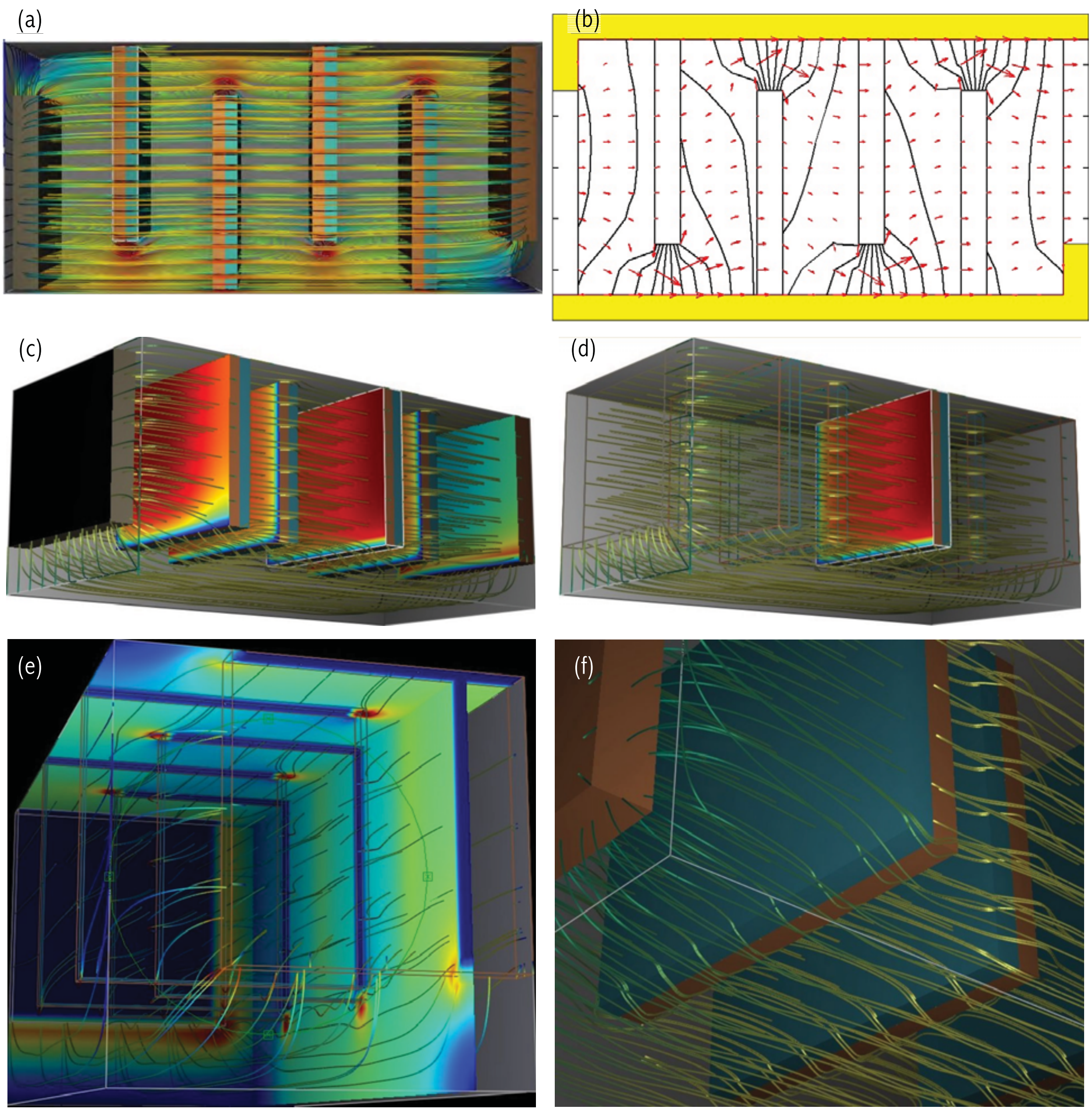

Figura 5.7: Celda de electro-obtención asimétrica con 4 electrodos bipolares. En (a), se muestra una vista ortogonal inferior en perspectiva, generada con la aplicación de esta investigación. El color (jet) de las líneas, mapea la intensidad del campo. (b) es un ejemplo comparativo de la limitada visualización de resultados que se disponía con el prototipo desarrollado por Bittner. En (c) y (d), se mapea en color (jet) la intensidad de campo normal en las caras catódicas, lo que permite predecir la distribución del depósito de cobre. (d) ejemplifica la utilidad de la capacidad de visualización selectiva de electrodos para eliminar problemas de oclusión. En (e), se utilizan los "extractores de rebanadas" para visualizar la intensidad del campo densidad de corriente. Finalmente, en (f) se muestra un acercamiento para ilustrar la calidad visual de las líneas de flujo lograda en vistas locales. Se usa un mapa de color brillante (summer) para enfatizar la iluminación.

# *Discusión y Conclusiones*

## 6.1 Discusión y Conclusiones

Para estudiar diferentes estructuras geométricas, de modo de optimizar el proceso de EO de cobre usando electrodos bipolares flotantes, es importante disponer de una herramienta de simulación y análisis, basada en un adecuado modelo del proceso. En este trabajo se implementó una aplicación computacional orientada al logro de dicha herramienta, que satisface adecuadamente los objetivos planteados en la sección 1.2. En suma, se hizo una contribución a la modelación de celdas de electrolíticas. En particular, se obtuvo una formulación analítica para calcular el potencial metálico en electrodos flotantes.

Con esta aplicación, el diseñador puede estudiar la influencia en el funcionamiento de la celda de: la geometría de la misma, los potenciales de alimentación y el efecto de diferentes tipos de electrodos. Estos últimos se especifican mediante los parámetros que aproximan la función de dependencia de los potenciales de electrodo en términos de la densidad de corriente (curva de polarización). Aunque el origen de la aplicación se motiva en el interés de estudiar celdas de EW de cobre con electrodos bipolares, el software se puede utilizar para estudiar celdas de EW convencionales, las que hacen uso exclusivo de electrodos unipolares.

La aplicación desarrollada proporciona una GUI que facilita el diseño geométrico de las celdas, permite simular la celda con control del error numérico y proporciona herramientas de visualización adecuadas para analizar los datos, multidimensionales y multivariables, generados por el modelo. En particular, se implementó una técnica muy eficaz para la visualización interactiva de campos vectoriales estacionarios. Como complemento a la herramienta de visualización de campo vectorial, se implementaron algunas herramientas de visualización adicionales, relativamente simples, pero muy efectivas para análisis cualitativo de las magnitudes de los datos. Estas herramientas se potencian con herramientas interactivas para visualización selectiva de datos, y en conjunto se logra una software bastante completo para la simulación y el análisis de celdas de electro-obtención.

La herramienta de visualización más poderosa desarrollada es la capacidad para visualizar el campo vectorial como líneas continuas de flujo. Esta herramienta permite obtener una visualización de la estructura global del campo vectorial y también logra visualizaciones locales que permiten apreciar detalles más finos. Mediante una adecuada iluminación de las líneas de flujo se logra apreciar su orientación en el espacio, y mediante un mecanismo de sugestión de profundidad, se logra mejorar la percepción de la ubicación espacial de las líneas de flujo. Adicionalmente, las líneas de flujo se pueden colorear para codificar la intensidad de la densidad de corriente o el potencial eléctrico. El último caso permite determinar el sentido de las líneas de flujo, cuando no esta claramente definido. Debido a que la técnica de iluminación de líneas utilizada permite efectuar los cálculos en el hardware gráfico, se pudo concebir una aplicación interactiva. Al rotar interactivamente la celda, las características de iluminación de líneas y percepción de profundidad son potenciadas, y se facilita la comprensión de la estructura del campo vectorial.

La aplicación proporciona una conveniente interfaz para controlar la generación de líneas de flujo, de modo de obtener concentraciones de líneas adecuadas para distintos objetivos de análisis. La generación de líneas de flujo es automática y logra una adecuada descripción de la estructura del campo vectorial sin mayor asistencia del usuario.

Las herramientas de visualización del software desarrollado resultan inútiles si no se tiene un adecuado modelo del proceso que se simula. Un importante logro de este trabajo fue el hacer mejoras al modelo base de la celda de electro-obtención. Estas mejoras permiten suponer una mayor precisión del modelo, se aumenta el rango de geometrías que se puede simular y se amplia las posibilidades de desarrollo del modelo.

Los resultados generales obtenidos con el nuevo modelo desarrollado en este trabajo concuerdan cualitativamente con el modelo desarrollado por Bittner, en lo que respecta a la estructura global del campo de densidad de corriente obtenida y la posible distribución del depósito de cobre. A diferencia del modelo desarrollado por Bittner, donde en el electrodo bipolar se forzó corriente normal en cara catódica y donde se relacionó localmente las corrientes en cara catódica y anódica, con el nuevo modelo no siempre ocurren tales situaciones y sin embargo se obtienen los mismos resultados generales. Lo anterior confirma los argumentos que motivaron la búsqueda de un mejor modelo y permite suponer que el nuevo modelo representa con mayor precisión el fenómeno real.

Además de las mejoras en la formulación teórica del modelo, se realizaron importantes mejoras en su implementación numérica. Se mejoró la convergencia, velocidad y potencial de desarrollo del modelo. No se hicieron mediciones comparativas precisas, pero se comprobó claramente que la nueva implementación resulta notablemente más eficiente.

Debido a que la solución del sistema de ecuaciones se realiza por un método iterativo, es simple incluir ecuaciones no lineales en el sistema. De este modo, ahora es posible considerar mejores aproximaciones en la curva de polarización, respecto a la pobre aproximación lineal a la que se estaba limitado con el modelo base, al usar un método directo de solución. Para lograr convergencia, en el modelo mejorado hubo que restringir la constante que define la aproximación lineal de la curva de polarización a pequeños valores. Se comprobó que tal restricción también existe en el modelo base, e incluso la restricción es más fuerte. Esto quizás impide apreciar la magnitud real del efecto de la variación de los potenciales de electrodo y evidencia la necesidad de mejorar la aproximación.

El uso de un método iterativo proporciona una ventaja adicional que se aprovecho en la implementación del software. El método iterativo permite controlar la precisión y rapidez de la simulación, de acuerdo al interés del usuario. Cada vez que se reinicia el proceso iterativo de cálculo, se tiene la opción de utilizar condiciones iniciales por defecto o mantener el estado actual de los potenciales, los que pueden ser inicializados a partir de datos previamente guardados en un archivo. Lo anterior permite hacer refinamientos de una solución a un bajo costo computacional. Además, se puede simular con rapidez cambios menores en las condiciones de operación, sin tener que reiniciar completamente el proceso iterativo. Por ejemplo, a partir de una celda previamente simulada, se puede analizar el efecto de mover, agregar o quitar electrodos, modificar el voltaje de alimentación, etc.

La aplicación recién descrita del método iterativo se puede complementar con la capacidad para visualizar con rapidez las líneas de flujo, a medida que son modificadas al progresar el proceso iterativo. Debido a que los potenciales calculados con el método iterativo se aproximan a la solución correcta, en una forma aparentemente asintótica. La convergencia numérica de la solución se visualiza como si fuera un proceso físico transiente. Esta claro que el modelo sólo considera una situación de estado estacionario, pero el efecto de visualización transiente quizás puede ser una herramienta útil para especular la evolución transiente del proceso físico, entre distintos estados estacionarios de operación.

## 6.2 Trabajos futuros

Como futuras mejoras en el área de visualización, resulta muy útil desarrollar una técnica para controlar la transparencia de las líneas de flujo, de modo de implementar una técnica de visualización selectiva de datos más completa. La idea es poder controlar la opacidad de las líneas de acuerdo a alguna función de interés, de modo de poder concentrar la visualización en ciertos datos de interés sin ocultar el resto de las líneas de flujo. La transparencia podría ser especificada en términos de una cantidad escalar como el potencial o la intensidad del campo eléctrico, o de acuerdo a una región espacial de interés, etc. Por ejemplo, se podría visualizar opacas las líneas generadas desde cierto electrodo y semitransparentes el resto de las líneas.

Para el análisis de los campos escalares, resulta útil incorporar algoritmos de rendering volumétrico, los que entregarían mucho más información que los extractores de rebanadas.

En lo que respecta al modelo, todavía hay mucho por desarrollar. Se debe incorporar las variables químicas y mecánicas. Sin embargo, se pueden hacer algunos avances en el contexto de la modelación actual. Como se mencionó, se puede mejorar la modelación de la curva de polarización. Además, resulta conveniente mejorar la velocidad de convergencia del algoritmo de cálculo con métodos numéricos apropiados, pues el algoritmo de cálculo de potenciales metálicos en electrodos flotantes, presenta una convergencia relativamente lenta comparada con la convergencia de los potenciales en el resto de la celda.

El método determinado para calcular el potencial en electrodos flotantes permite considerar incluir en el software algunas condiciones de operación especiales, las que podrían darse en situaciones de accidentes o que simplemente podrían ser interesantes de analizar. En la implementación del modelo, se definió un electrodo generalizado, el que puede actuar como electrodo bipolar o unipolar. Básicamente, la única diferencia es que en el electrodo unipolar el potencial metálico se asume constante (es un parámetro) y en los electrodos bipolares, por ser flotantes, es una variable dependiente. Quizás podría resultar de interés estudiar situaciones donde; se aplica un voltaje externo a electrodos bipolares o se deja flotantes electrodos unipolares. Para proporcionar la capacidad para simular estas situaciones, bastaría con proporcionar una variable de control por cada electrodo, para que el usuario pueda especificar el estado flotante o energizado de cualquier electrodo. Basado en el principio usado para desarrollar el método que determina el potencial en electrodos flotantes, si tuviera alguna relevancia, resulta factible incorporar fácilmente la capacidad para simular situaciones donde varios electrodos flotantes son conectados entre sí, por medio de conductores de sección despreciable.

# *Bibliografía*

[1] Banks, D.C., (1994). Illumination in Diverse Codimensions, *Computer Graphics Ann. Conf. Series*, pp. 327-334, Julio.

[2] Atkinson, K.E. y G. Birkhoff (1990). *Análisis numérico con pascal*, Addison-Wesley.

[3] Bittner, R., L. Salazar, M. Valenzuela y A. Pagliero, (1998). Modeling the electric field and potential of an electrowinning cell with bipolar floating electrodes, *Proceedings of the IEEE IECON '98*, pp. 365-370.

[4] Bittner, R., A. Pagliero, L. Salazar y M. Valenzuela, (1998). Electric field and potential determination for electrowinning cells with bipolar electrodes by finite difference models, *Conference Record of the IEEE IAS '98 Conference*, pp. 1973-1980.

[5] Bittner, R., (1999). *Modelación y Análisis de Celdas de Electro-obtención de Cobre Basadas en Electrodos Bipolares*, Tesis de Magíster, Universidad de Concepción.

[6] Bittner, R., (1999). Implementación de Modelos de Celdas de Electro-obtención usando la plataforma MATLAB, *Comunicación personal.*

[7] Figueroa, E., (1989). *Introducción al Análisis Lineal*, Texto guía de curso de Cálculo numérico, Dpto. de Matemática, Universidad de Concepción.

[8] Foley, J., A. Van Dam, S. Feiner y J. Hughes, (1997). *Computer Graphics Principles and Practice – Second edition in C*, Addison-Wesley.

[9] Gallagher, R.S., (1995). *Computer Visualization - Graphics Techniques for Scientific and Engineering Analysis*, CRC Press.

[10] Kempf, R., (1997). *Guide to OpenGL on Windows From Silicon Graphics*, Documento 007-3405-001, Silicon Graphics Inc.

[11] Marschner, S. R. y R. Lobb, (1994). An evaluation of Reconstruction Filters for Volume Rendering, *Visualization 94*, IEEE Computer Society Press, pp. 100-107.

[12] Miano, J., T. Cabanski y H. Howe, (1997). *C++ Builder How-To*, Waite Group Press.

[13] Neider, J., T. Davis y M. Woo, (1993). *OpenGL programming Guide, Release 1*, Addison-Wesley.

[14] Petzold, C., (1996). *Programming Windows 95*, Microsoft Press.

[15] Popovic, B. D., (1971). *Introductory Engineering Electromagnetics*, Addison-Wesley.

[16] Protter, M.H. y C.B. Morrey, (1964). *Modern Matematical Analysis*, Addison-Wesley.

[17] Stalling, D., M. Zockler y H.C. Hege, (1995). Fast and Resolution Independent Line Integral Convolution, *Computer Graphics Ann. Conf. Series*, pp. 249-256.

[18] Stalling, D., M. Zockler y H.C. Hege, (1997). Fast Display of Illuminated Field Lines, *IEEE Transaction on Visualization and Computer Graphics*, Vol. 3, Nº 2, pp. 118-127.

[19] Watt, A., (1989). *Fundamentals of Three-Dimensional Computer Graphics*, Addison-Wesley.

# *Distribución homogenea de puntos sobre la superficie de un paralelepípedo*

Se desea resolver el problema de la distribución homogénea de una determinada cantidad de puntos, $N$, sobre la superficie de un paralelepípedo, cuando la cantidad de puntos esta determinada por una densidad de puntos por unidad de superficie $\rho_S$ [puntos/$h^2$] arbitraria, donde $h$ es la unidad de longitud. La solución del problema implica la determinación del número de puntos que se ubicaran equidistantemente sobre las aristas. Estos puntos generaran grillas de muestreo rectangulares en el interior de cada una de las caras.

Considerando que en unidades de discretización $h$, las dimensiones del paralelepípedo son: $I$ en la dirección $x$, $J$ en la dirección $y$, $K$ en la dirección $z$. Se desea determinar el número de puntos sobre las aristas, siendo estos respectivamente $n_i$, $n_j$, y $n_k$, en las direcciones $x, y, z$. Ver Figura 4.3 (página 60).

A mayor dimensión, se desea una mayor cantidad de puntos, para lo que se establecen las siguientes relaciones:

(A.1) $$\frac{I}{J} = \frac{n_i}{n_j}, \quad \frac{J}{K} = \frac{n_j}{n_k}, \quad \frac{I}{K} = \frac{n_i}{n_k}$$

En términos de $n_i$, $n_j$, y $n_k$, el número total de puntos sobre la superficie del paralelepípedo se puede descomponer en los aportes de los vértices ($N_V$), de las aristas sin incluir vértices ($N_A$), y de las caras sin incluir aristas ($N_C$), siendo:

(A.2)
$$\begin{aligned} N_V &= 8 \\ N_A &= 4(n_i - 2) + 4(n_j - 2) + 4(n_k - 2) \\ N_C &= 2(n_i - 2)(n_j - 2) + 2(n_j - 2)(n_k - 2) + 2(n_i - 2)(n_k - 2) \end{aligned}$$

Este planteamiento supone un mínimo de 8 puntos; los vértices del paralelepípedo, por lo que la solución será válida para $n_i \geq 2$, $n_j \geq 2$, y $n_k \geq 2$.

La solución del problema queda determinada entonces por:

$$N = N_C + N_A + N_V = 2\rho_S (IJ + JK + IK)$$

Considerando las relaciones establecidas en (A.1), la ecuación anterior se puede expresar en términos de $n_i$ como única incógnita, resultando el siguiente polinomio de segundo grado, cuyas raíces determina $n_i$

(A.3)
$$p(n_i) = (IJ + JK + IK)n_i^2 - 2I(I + J + K)n_i - I^2\left[(IJ + JK + IK)\rho_S - 4\right] = 0$$

La solución gráfica de la ecuación anterior se ilustra en la Figura A.1.

La densidad que genera 8 puntos es:

(A.4)
$$\rho_8 = \frac{4}{IJ + JK + IK}$$

Analizando el discriminante del polinomio de (A.3), se comprueba que este tiene raíces reales para $\rho_S \geq \rho_{min}$, donde:

(A.5)
$$\rho_{min} = \frac{4(IJ + JK + IK) - (I + J + K)^2}{(IJ + JK + IK)^2} = \rho_8 - \left(\frac{I + J + K}{IJ + JK + IK}\right)^2$$

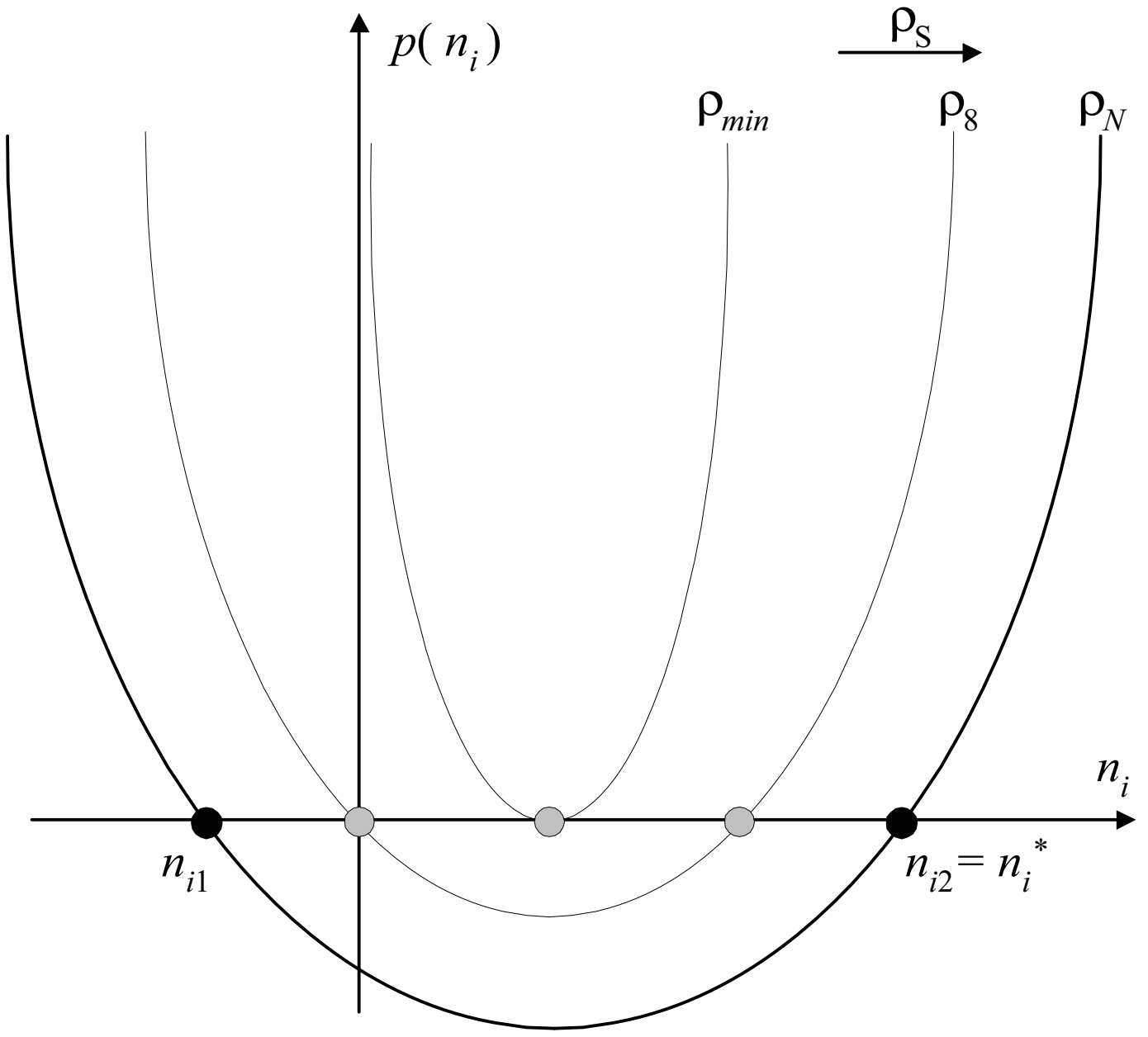

Figura A.1: Solución gráfica de la ecuación (A.3).

Como $\rho_8$ siempre es mayor a $\rho_{\min}$, el problema siempre tiene solución real. Además, se puede comprobar fácilmente que para $\rho_S \geq \rho_8$, sólo una raíz de la ecuación de segundo grado es positiva, siendo esta:

$$n_i^* = \left( (I + J + K)\frac{\rho_8}{4} + \sqrt{\rho_S - \rho_{\min}} \right) I, \quad \rho_S \geq \rho_8 \tag{A.6}$$

Considerando las relaciones de (A.1) y el resultado anterior, se obtienen:

$$n_j^* = \left( (I + J + K)\frac{\rho_8}{4} + \sqrt{\rho_S - \rho_{\min}} \right) J, \quad \rho_S \geq \rho_8 \tag{A.7}$$

$$n_k^* = \left( (I + J + K)\frac{\rho_8}{4} + \sqrt{\rho_S - \rho_{\min}} \right) K, \quad \rho_S \geq \rho_8 \tag{A.8}$$

Como los valores entregados por $n_{\mathrm{i}}^*$, $n_{\mathrm{j}}^*$, y $n_{\mathrm{k}}^*$ no necesariamente son enteros, deben aproximarse al entero más cercano.

Aunque $\rho_S \geq \rho_8$, los resultados anteriores no garantizan que tanto $n_i$, $n_j$, y $n_k$ sean mayores a 1, como supone el problema, debido a que *I*, *J*, y *K* pueden tener magnitudes muy disímiles. Aun más, si la densidad $\rho_S$ es especificada arbitrariamente por el usuario, esta podría ser inferior a $\rho_8$.

Para asegurar que siempre resulten al menos los 8 vértices ($n_i \geq 2$, $n_j \geq 2$, y $n_k \geq 2$), se elige la siguiente solución:

(A.9)
$$\begin{aligned} n_i &= \max\left\{2, \operatorname{int}(n_i^* + 0.5)\right\}, \\ n_j &= \max\left\{2, \operatorname{int}(n_j^* + 0.5)\right\}, \text{ y} \\ n_k &= \max\left\{2, \operatorname{int}(n_k^* + 0.5)\right\} \end{aligned}$$

donde int es la función que retorna la parte entera de un numero real.